\ifx\compilefullpaper\undefined  
\documentclass[pra,longbibliography,twocolumn,showpacs,nofootinbib,superscriptaddress,notitlepage]{revtex4-2} 
\pdfoutput=1	
\usepackage[utf8]{inputenc}
\usepackage[table]{xcolor}

\usepackage{enumitem} 
\usepackage{color}
\usepackage{dsfont}
\usepackage{amsmath}
\usepackage{amssymb,bm}
\usepackage{amsthm}
\usepackage{mathtools}
\usepackage{url}

\usepackage[protrusion=true,expansion=true]{microtype}
\usepackage{booktabs}

\usepackage{array, makecell}
\usepackage{boldline}
\usepackage{enumitem}

\usepackage{color,dsfont}
\usepackage{graphicx}
\usepackage{ragged2e}
\usepackage[colorlinks=true, hyperindex, breaklinks, linkcolor=blue, urlcolor=blue, citecolor=blue]{hyperref} 
\usepackage[normalem]{ulem}
\usepackage[capitalise]{cleveref}
\usepackage{mathrsfs}
\usepackage{keyval}
\usepackage[caption=false]{subfig}
\usepackage{mathtools}
\usepackage{float}
\usepackage{verbatim}
\usepackage{latexsym}
\usepackage{amsmath}
\usepackage{amssymb}
\usepackage{setspace}
\usepackage{amsfonts}
\usepackage{stmaryrd}
\usepackage{xcolor}
\usepackage{enumitem}
\usepackage{lipsum}

\newcommand{\ket}[1]{{|#1\rangle}}

\newcommand{\abs}[1]{{\lvert #1\rvert}}	
\newcommand{\NP}{\ensuremath{\mathsf{NP}}}
\newcommand{\identity}{\ensuremath{\boldsymbol{1}}}
\newcommand{\Id}{\identity} 

\def\llbracket{{[\![}}
\def\rrbracket{{]\!]}}

\definecolor{cardinal}{rgb}{0.827, 0, 0}

\begin{document}

\fi

\vfuzz2pt 

\title{New magic state distillation factories optimized by temporally encoded lattice surgery
}
\author{Prithviraj Prabhu}
\affiliation{University of Southern California, Los Angeles, CA 90089, USA}
\author{Christopher Chamberland}
\affiliation{AWS Center for Quantum Computing, Pasadena, CA 91125, USA}
\affiliation{IQIM, California Institute of Technology, Pasadena, CA 91125, USA}

\begin{abstract}

Fault-tolerant quantum computers, with error correction implemented using topological codes, will most likely require lattice surgery protocols in order to implement a universal gate set. Timelike failures during lattice surgery protocols can result in logical failures during the execution of an algorithm. In addition to the spacelike distance of the topological code used to protect the qubits from errors, there is also the timelike distance which is given by the number of syndrome measurement rounds during a lattice surgery protocol. As such, a larger timelike distance requirement will result in the slowdown of an algorithm's runtime. Temporal encoding of lattice surgery (TELS) is a technique which can be used to reduce the number of syndrome measurement rounds that are required during a lattice surgery protocol. This is done by measuring an over-complete set of mutually commuting multi-qubit Pauli operators (referred to as a parallelizable Pauli set) which form codewords of a classical error correcting code. The results of the over-complete set of Pauli measurements can then be used to detect and possibly correct timelike lattice surgery failures. In this work, we introduce an improved TELS protocol and subsequently augment it with the ability to correct low-weight classical errors, resulting in greater speedups in algorithm runtimes. We also explore large families of classical error correcting codes for a wide range of parallelizable Pauli set sizes. We also apply TELS to magic state distillation protocols in the context of biased noise, where logical qubits are encoded in asymmetric surface codes. Using optimized layouts, we show improvements in the space-time cost of our magic state factories compared to previous protocols. Such improvements are achieved using computations performed in the Clifford frame.

\end{abstract}

\maketitle

\section{Introduction}
\label{sec:Intro}

\begin{figure*}
\hspace{-1.25cm}
\includegraphics[width=1.05\textwidth]{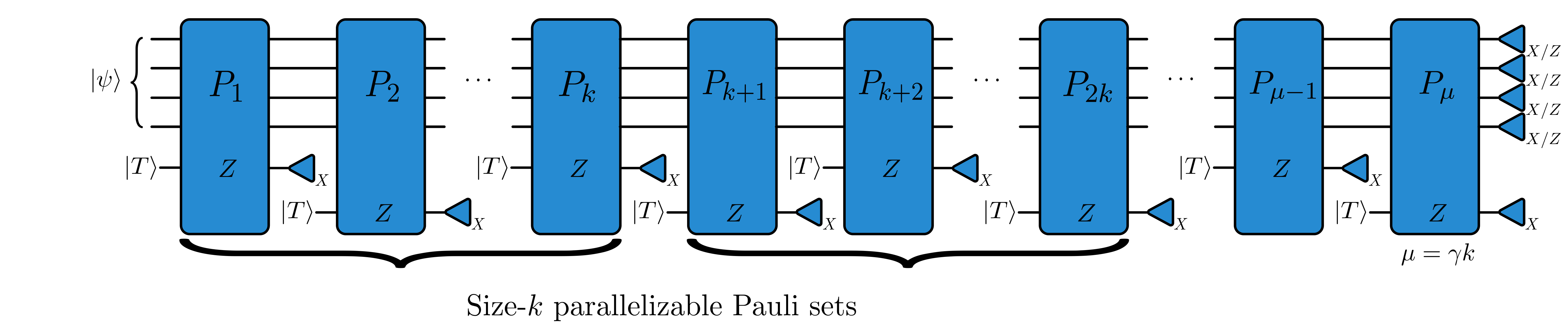}
\caption{General model of Pauli-based computation. A quantum algorithm can be written as a sequence of multi-qubit Pauli measurements which perform both Clifford and non-Clifford operations (here we show the implementation of non-Clifford gates). In general the multi-qubit Pauli operations can be ordered into sets of commuting Pauli operators, where Clifford corrections can be conjugated to the end of each set. Such sets are called parallelizable Pauli (PP) sets. A logical $T$ gate (which is non-Clifford and forms a universal gate set when combined with Clifford operations) can be implemented via a multi-qubit Pauli measurement acting on a set of data qubits and an ancillary magic state $\ket T = (\ket{0} + e^{i \pi / 4}\ket{1}) / \sqrt{2}$. }
\label{fig:PBC}
\end{figure*}

\begin{figure*}
    \centering
    \subfloat[\label{fig:TELSprotocol} ]{\includegraphics[width=.55\textwidth]{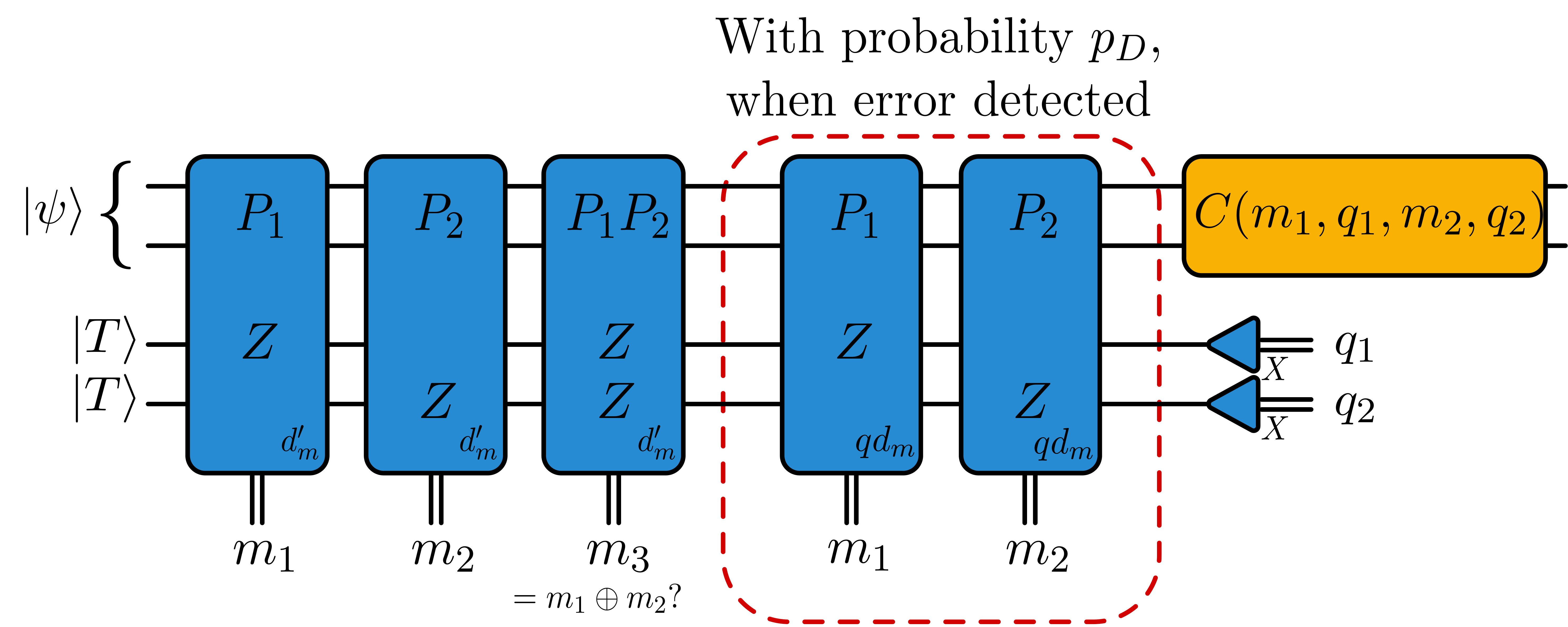}}
\hspace{0.3cm}
\subfloat[\label{fig:TELSprotocolnew} ]{\includegraphics[width=.415\textwidth]{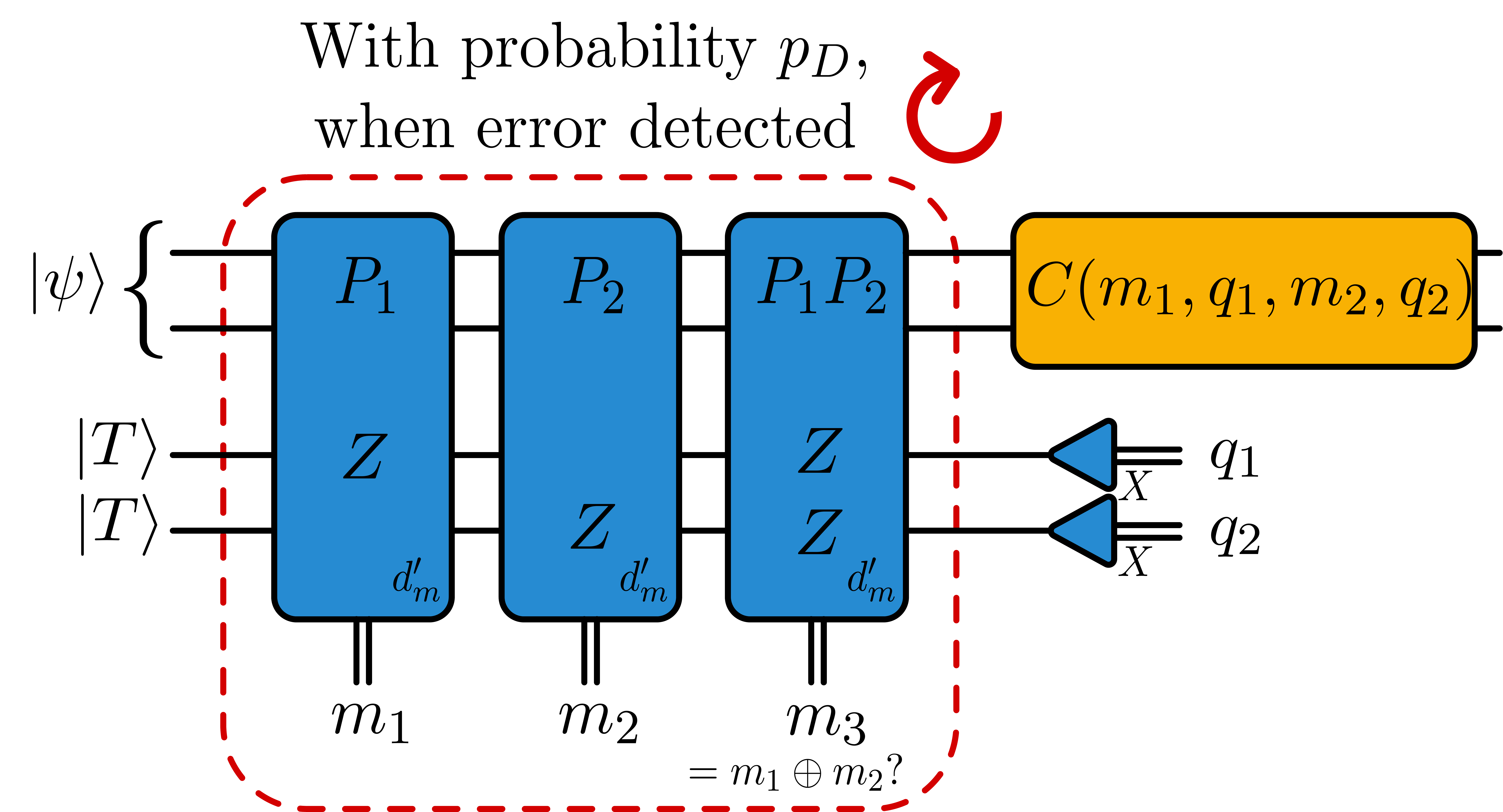}}
    \caption{(a) Old protocol for temporally encoded lattice surgery of a PP set of size-2, where $P_1P_2$ is a redundant measurement which is used to detect failures in the measurements of $P_1$ or $P_2$. If the measurement results of the three multi-qubit Pauli operators are inconsistent, the original multi-qubit Pauli operators $P_1$ and $P_2$ are measured again. Blue boxes correspond to multi-qubit Pauli measurements, and blue triangles correspond to logical single-qubit measurements. Orange boxes correspond to Clifford corrections. (b) New protocol for temporally encoded lattice surgery of a PP set of size-$2$. The operators $P_1$, $P_2$ and $P_1P_2$ are repeatedly measured until no logical timelike failures are detected. In \cref{subsec:RepeatedTELS} we show that such a scheme results in smaller average runtimes for the implementation of a PP set. Orange boxes denote Clifford corrections that result from applying non-Clifford gates.}
\end{figure*}

Universal fault-tolerant quantum computers will be required in order to implement large scale quantum algorithms. However, both the space and time costs due to the implementation of fault-tolerant quantum error correction protocols can be quite high \cite{fowler2012surface,BravsLowOver,JonesToffoli,Yoder2016,chamberland2017overhead,CJO17Reed,BevsUniversal,Chamberland2019Magic1,Litinski19,Litinski19magic,chamberland2020very,ChambsCat,Chamberland22,SC22Modulo}. For fault-tolerant quantum computers where qubits are encoded in topological quantum error-correcting codes, lattice surgery paired with magic state distillation protocols provides an efficient way to implement universal gate sets while being compatible with the locality requirements of two-dimensional planar hardware architectures \cite{BravyiKitaevMagic,BravsLowOver,Knill12,BombinTopoDistill,Campbell2018magicstateparity,Chamberland2019Magic1,chamberland2020very,fowler2018low,Litinski2018latticesurgery,Litinski19,Litinski19magic,Chamberland22,Chamberland22b,PsiQuantumLatticeSurgery}. In addition to the extra space-costs associated with lattice surgery protocols, there are also additional time costs arising from the required protection against timelike failures (which can result in logical parity measurement failures) \cite{Chamberland22,Chamberland22b,Gidney2022stability}. The timelike distance of a lattice surgery protocol is given by the number of syndrome measurement rounds which need to be performed during the measurement of a multi-qubit Pauli operator. 

In Ref.~\cite{Chamberland22}, a protocol called temporal encoding of lattice surgery (TELS) was introduced in order to reduce the required timelike distance of lattice surgery protocols, thus reducing algorithm runtimes. The first step in a TELS protocol is to divide all multi-qubit Pauli measurements $\{ P_1, P_2, \cdots, P_{\mu} \}$ required to run a quantum algorithm into sequences of parallelizable Pauli (PP) sets. For a PP set of size $k$, the multi-qubit Pauli operators in a PP set $P_{[t+1,t+k]} = \{ P_{t+1}, P_{t+2}, \cdots, P_{t+k} \}$ commute, and any necessary Clifford corrections can be conjugated to the end of the sub-sequence. This general model of Pauli-based computation is shown in \cref{fig:PBC}. In this work we consider a universal gate set generated by $\langle T, H, S, \text{CNOT} \rangle$ where $T = \text{diag}(1, e^{i \pi / 4})$ and $H$ and $S$ are the Hadamard and phase gates. For such a universal gate set, a $T$ gate can be implemented using multi-qubit Pauli measurements and the resource magic state $\ket T = (\ket{0} + e^{i \pi / 4}\ket{1}) / \sqrt{2}$. The main idea behind TELS is that for a given PP set, one can measure a larger over-complete set of multi-qubit Pauli operators, where each multi-qubit Pauli operator in the over-complete set is a product of multi-qubit Pauli operators from the original PP set. As was shown in Ref.~\cite{Chamberland22}, each multi-qubit Pauli operator in the new over-complete set is associated with a codeword of a classical $[n,k,d]$ error-correcting code. The measurement results denote the $n$ bits of the classical code. Applying the parity check matrix of the code to these measurement outcomes enables the detection of logical timelike failures. This in turn allows fewer rounds of syndrome measurements for each multi-qubit Pauli, due to the extra protection offered by the overlaying classical~code.

In this work, we introduce new TELS protocols that further reduce the timelike distance required for lattice surgery protocols. In previous TELS protocols, if a lattice surgery failure was detected while measuring the over-complete set, the multi-qubit Pauli operators from the original PP set would be remeasured. We show that better speedups can be obtained if, during the re-measure step, the operators from the over-complete set are repeatedly measured until no logical timelike failures are detected. We also show that in some cases, even larger speedups can be achieved when using the classical error-correcting codes to correct a subset of errors of smaller weight and detect all possible errors of higher weight. We also consider a large number of classical error-correcting codes for various sizes of PP sets. Such considerations enable more efficient TELS protocols to be applied to a wider range of quantum algorithms, where the average PP set sizes depend on the particular algorithm being implemented. 

We then focus on implementing TELS protocols in the context of magic state distillation using Clifford frames. In doing so, we consider a biased circuit-level noise model, where the logical qubits are encoded using asymmetric surface codes. We consider asymmetric surface codes since such codes require fewer qubits to achieve a desired logical failure rate for physical error rates $p = 10^{-3}$ and $p = 10^{-4}$ compared to other topological codes such as $XZZX$ and $XY$ codes \cite{HiggottBP}. By developing new layouts for magic state distillation tiles which are adapted to TELS protocols, we show that the space-time costs of such distillation protocols can be reduced compared to magic state distillation tiles that do not use TELS.

The manuscript is structured as follows. In \cref{subsec:PauliBasedReview} we review the notion of Pauli-based computation implemented via lattice surgery and in \cref{subsec:TELSreview} we review the implementation of previously proposed TELS protocols. Next, in \cref{subsec:RepeatedTELS} we show how repeated encoded multi-qubit Pauli measurements using TELS can result in smaller algorithm runtimes. We then show in \cref{subsec:ErrorDetectCorr} how the simultaneous correction and detection of errors  with the classical codes used in a TELS encoding can lead to reduced runtimes compared to a pure error detection scheme. In \cref{subsec:ManyCodes}, we present the best average runtimes of TELS protocols using a wide range of classical error-correcting codes. In \cref{app:malignantsets}, we provide different methods to count the number of malignant fault sets for a given classical code. This is useful in analyzing the performance of TELS protocols with various classical codes. In \cref{app:codeconstruction}, we show how to construct the different classical codes used in \cref{subsec:ManyCodes}. \cref{app:speedups} contains details about the performance of TELS in additional noise regimes and target logical failure rates. 

Next in \cref{sec:TELSmagicState}, we show how to apply TELS protocols to magic state distillation factories. Specifically, \cref{sec:Cliff} contains a new circuit for executing lattice-surgery-based magic state distillation. In this protocol, magic states are distilled up to a known Clifford correction. In \cref{app:CliffpartTraceproof}, we discuss in detail how Clifford frames are incorporated in our TELS protocols when applied to magic state distillation schemes. In \cref{sec:MStilelayouts}, we analyze the space-time costs of various distillation protocols that use TELS with the distillation circuit of \cref{sec:Cliff}. In particular, for physical error rate $p=10^{-4}$, we use $15$-to-$1$ and $116$-to-$12$ distillation protocols to distill magic states with logical failure probabilities $10^{-10}$ and $10^{-15}$, where $p$ is the noise parameter of a biased circuit-level noise model described in \cref{subsec:PauliBasedReview}. For $p=10^{-3}$, we consider $114$-to-$14$ and $125$-to-$3$ distillation protocols to produce magic states with logical failure rate $10^{-10}$ and $10^{-15}$ respectively. For each distillation protocol, we compute the minimum distances and space time costs using different hardware layouts that are dependent on the chosen lattice surgery mechanism.  \cref{app:golaycodechoice} shows a specific choice of codewords derived from the classical Golay code for use in a $15$-to-$1$ distillation protocol with a TELS implementation. This classical code allows for very low average runtime per Pauli. In \cref{app:algoSTcosts}, we show the general procedure for determining the minimum space-like and time-like distances for different distillation protocols and layouts. \cref{app:constants} contains additional information regarding each of the layouts proposed in this work, which is used in conjunction with \cref{app:algoSTcosts} to determine the minimum distances and space-time costs. We conclude with a note on how to schedule the operation of a distillation factory which contains many distillation tiles. Using a round-robin scheduling algorithm, we optimize the number of tiles that are required to output enough distilled magic states as required for seamless operation of a quantum core.

\section{Review of Pauli-based computation and Temporally Encoded Lattice Surgery}
\label{sec:Review}

\subsection{Pauli-based computation and lattice surgery}
\label{subsec:PauliBasedReview}

There exist many models of computing to execute an algorithm on a quantum computer. The most well-studied is the circuit model of quantum computation~\cite{Deutsch89,Bennett95}. Examples of some other models are adiabatic quantum computation~\cite{Farhi01}, measurement-based models~\cite{Gottesman99,Aliferis04,Briegel09}, and fusion-based computation~\cite{Bartolucci21} which may be more suitable for specific hardware architectures. For a universal quantum computing architecture which uses two-dimensional planar topological codes (such as the surface code), the most natural way to implement a quantum algorithm is to use the Pauli-based model of quantum computation~\cite{Bravyi16}. In a Pauli-based computation (PBC), all logical gates can be applied by measuring multi-qubit Pauli operators (potentially using additional ancillas), as shown in \cref{fig:PBC}. Quantum algorithms are then executed using a pool of purified magic states and a sequence of multi-qubit Pauli measurements. In the interest of speeding up algorithms, we first note that the sequence of multi-qubit measurements can be grouped into subsequences of mutually commuting measurements. In \cref{fig:PBC}, the set $\{ P_1,P_2, \mathellipsis P_k\}$ is such a subsequence, and is called a PP set. An algorithm executing $\mu$ $T$-gates with $T$-depth $\gamma$ can in general be broken down into $\gamma$ PPs of average size $k = \mu / \gamma$. In some algorithms of interest like quantum chemistry simulations~\cite{Kim22}, the size of a PP set is between $9$ and $14$. We also note that for magic state distillation protocols that are expressed as sequences of multi-qubit Pauli measurements, all the non-Clifford gates commute and thus form PP sets. For instance, the $15$-to-$1$ distillation protocol has a PP set of size $11$, and for the $125$-to-$3$ distillation protocol, the PP set is of size $99$.

\begin{figure}
    \centering
    \hspace{-1mm}
    \includegraphics[width=.41\textwidth]{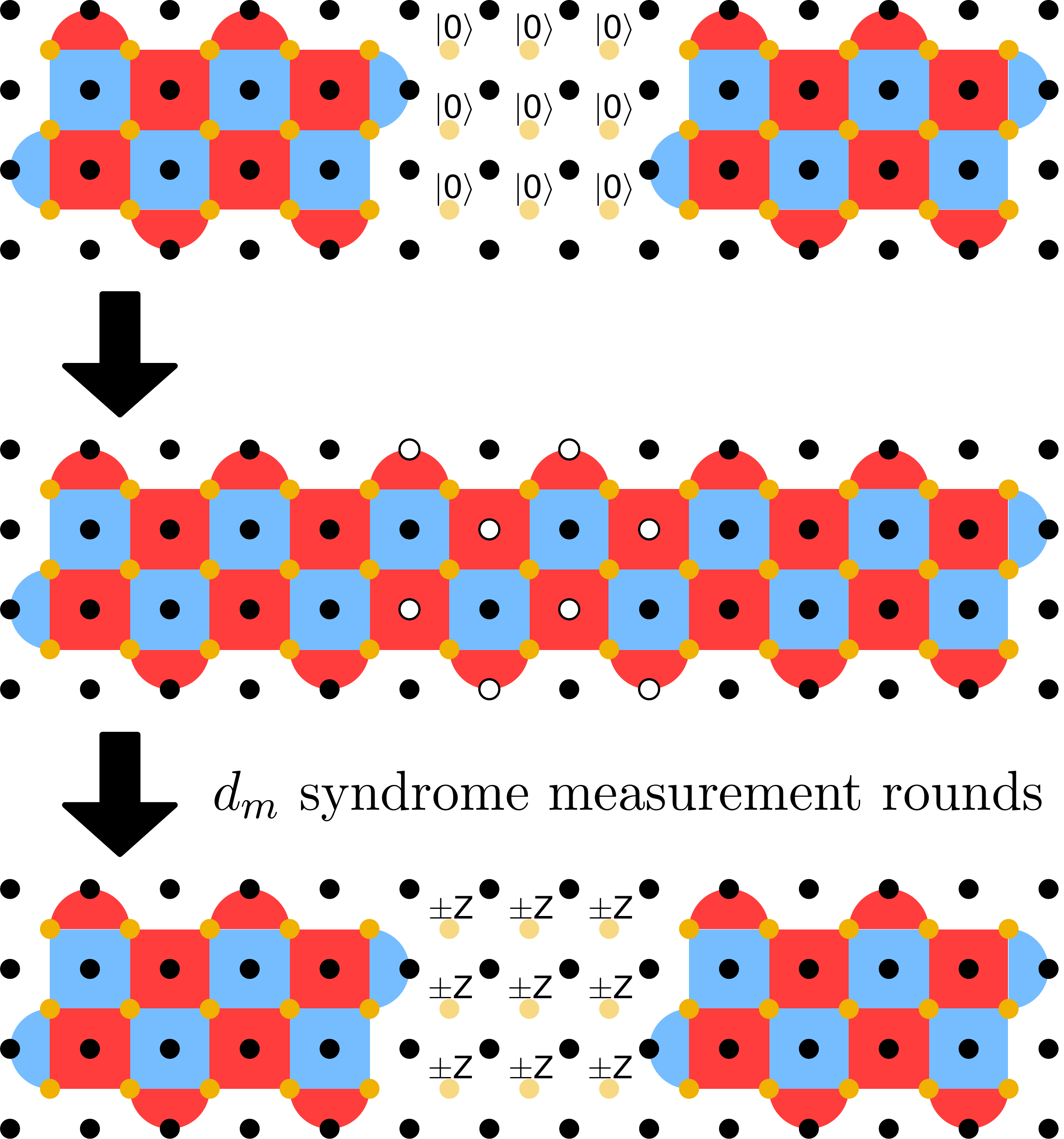}
    \caption{Lattice surgery implementation of an $X \otimes X$ measurement between two logical qubits encoded in $d_x=3$, $d_z=5$ surface code patches. Note that $X$ ($Z$) stabilizers are represented by red (blue) plaquettes. Prior to measuring $X \otimes X$, yellow data qubits in the routing region are prepared in the $\ket 0$ state. The $X \otimes X$ measurement outcome is then obtained by measuring the $X$-stabilizers (shown with white ancillas) in the routing space. The stabilizers of the merged surface code patch are measured for $d_m$ syndrome measurement rounds in order to correct timelike failures which can occur in the first round of the merge resulting in the wrong parity of $X \otimes X$. In the first syndrome measurement round of the merged patch, the individual measurement outcomes of $X$ stabilizers in the routing space region are random, but their product gives the result of the $X \otimes X$ measurement outcome. At the end of the $d_m$ syndrome measurement rounds, the data qubits in the routing space are measured in the $Z$ basis. Since measurement and reset of qubits typically takes a much longer time than the implementation of the physical CNOT gates used to measure the stabilizers, in this work we assume that the qubits in the routing space used to measure $X \otimes X$ are only available one syndrome measurement round after the split. Hence the merge/split operation takes a total of $d_m+1$ syndrome measurement rounds. }
    \label{fig:latsurg}
\end{figure}

In topological codes such as the surface code and color code, lattice surgery is the dominant mechanism used to perform multi-qubit Pauli measurements~\cite{Landahl14,Horsman12}. In \cref{fig:latsurg}, we show how to perform the $X\otimes X$ lattice surgery measurement between two logical qubits encoded in the surface code. Prior to measuring $X \otimes X$, the data qubits in the routing space are initialized in the $\ket 0$ state. Subsequently, the $X \otimes X$ measurement outcome is obtained by measuring $X$ operators in the routing space region shown in \cref{fig:latsurg} (note that a minimum-weight representative of the logical $X$ operator of the surface code is given by the product of $X$ operators along the vertical boundaries of the patch). The measurement outcomes of the $X$ stabilizers in the routing space are random (such stabilizers are illustrated with white ancillas in \cref{fig:latsurg}). However the product of all such stabilizers give the parity of the $X \otimes X$ measurement outcome. The stabilizers of the merged surface code patch are measured for $d_m$ rounds after which the surface code patches are split by measuring the qubits located in the routing space in the $Z$ basis. A logical timelike failure occurs during a lattice surgery protocol when the wrong parity of the multi-qubit Pauli measurement is obtained. Note that the logical timelike failure rate is exponentially suppressed with the number of syndrome measurement rounds $d_m$ (see Ref.~\cite{Chamberland22}).

All numerical results in this work are obtained using the following biased circuit-level noise model:
\begin{enumerate}
    \item Each single-qubit gate location is followed by a Pauli $Z$ error with probability $\frac{p}{3}$ and Pauli $X$ and $Y$ errors each with probability $\frac{p}{3 \eta}$.
	\item Each two-qubit gate is followed by a $\{ Z \otimes I, I \otimes Z, Z \otimes Z \}$ error with probability $p/15$ each, and a $\{ X \otimes I, I \otimes X, X \otimes X, Z \otimes X, Y \otimes I, Y \otimes X, I \otimes Y, Y \otimes Z, X \otimes Z, Z \otimes Y, X \otimes Y, Y \otimes Y \}$ each with probability $\frac{p}{15 \eta}$.
	\item With probability $\frac{2p}{3 \eta}$, the preparation of the $\ket{0}$ state is replaced by $\ket{1}=X\ket{0}$. Similarly, with probability $\frac{2p}{3}$, the preparation of the $\ket{+}$ state is replaced by $\ket{-}=Z\ket{+}$.
	\item With probability $\frac{2p}{3 \eta}$, a single-qubit $Z$ basis measurement outcome is flipped. With probability $\frac{2p}{3}$, a single-qubit $X$-basis measurement outcome is flipped.
	\item Lastly, each idle gate location is followed by a Pauli $Z$ with probability $\frac{p}{3}$, and a $\{ X, Y \}$ error each with probability $\frac{p}{3 \eta}$.
\end{enumerate}
We also set $\eta =100$. Using the above noise model and a minimum-weight perfect matching decoder, in Ref.~\cite{Chamberland22} it was shown that the timelike logical failure rate of an $X \otimes X$ multi-qubit Pauli measurement with $d_m$ syndrome measurement rounds is given by
\begin{equation}
    p_m(d_m) = 0.01634 A (21.93p)^{(d_m+1)/2},
    \label{eq:TimelikePoly}
\end{equation}
when $p$ is below threshold. In \cref{eq:TimelikePoly}, $A$ corresponds to the area of the routing space connecting the various surface code patches that take part in a multi-qubit Pauli measurement. In \cref{sec:NewTELS}, we set $A = 100$ in order to directly compare our results to those in Ref.~\cite{Chamberland22}. Further, we also use \cref{eq:TimelikePoly} when considering multi-qubit Pauli measurements containing $Z$ and $Y$ terms. 

Depending on the accuracy requirements of the algorithm, the target logical error rate per Pauli $\delta$ sets an upper bound on the maximum tolerable noise. This condition is $p_m < \delta$. To achieve low $p_m$, the measurement distance $d_m$ must be accordingly increased. Surface codes with $X$ and $Z$ boundaries can perform $X \otimes X$, $Z \otimes Z$ and $Z \otimes X$ measurements similar to \cref{fig:latsurg}. However to access the $Y$ boundary, twist defects are required. They have been studied extensively in surface codes~\cite{Horsman12,Litinski19,Chamberland22b} and in color codes~\cite{Gowda21}. We use the methods of Ref.~\cite{Chamberland22b} to implement $Y$-type measurements.
As shown in \cref{fig:latsurg}, lattice surgery requires routing space between different logical qubit patches. In the interest of reducing routing space for lattice surgery, Ref.~\cite{Herr19} considered extremely thin data busses that were only one qubit wide. The caveat of this method is that the measurements need to be performed for $d_m^2$ syndrome measurement rounds, instead of $d_m$ in the regular case. In this manuscript we consider routing space of dimensions which are functions of $d_x$ and $d_z$, which are the $X$ and $Z$ distances of the code respectively. For instance, the core of a quantum processor would have a layout given by Fig.~14 (e) in Ref.~\cite{Chamberland22}.

\subsection{Temporally encoded lattice surgery} 
\label{subsec:TELSreview}

It has been shown that encoding the measurement results of lattice surgery in an error-detecting code allows one to effectively reduce the measurement distance $d_m$ of each multi-qubit Pauli measurement~\cite{Chamberland22}. In particular, the Pauli operators of a given PP set $\mathcal{P} = \{ P_{t+1}, P_{t+2}, \cdots, P_{t+k} \}$ can be replaced by a new set $\mathcal{S} = \{Q[\bold{x}^1], Q[\bold{x}^2], \cdots, Q[\bold{x}^n] \}$ where 
\begin{align}
    Q[\bold{x}] = \prod_{j=1}^{k} P_{t+j}^{x_j} \,,
\end{align}
and $\bold{x}$ is a binary vector of length $k$. Such a replacement is allowed since $\langle \mathcal{S} \rangle = \langle \mathcal{P} \rangle$. In this encoding, the vectors $\{\bold{x}^1, \bold{x}^2, \cdots, \bold{x}^n \}$ form the columns of the generator matrix $G$ of some classical code $\mathcal{C}$, where the rows of $G$ are the codewords of $\mathcal{C}$. Note that the encoding of a TELS protocol takes place entirely in the time domain since additional multi-qubit Pauli measurements are performed without requiring additional qubits. By multiplying the measurement outcomes of all operators in $\mathcal{S}$ by the parity check matrix of $\mathcal{C}$, timelike lattice surgery failures will be detected if the result is equal to 1 instead of 0. 

In \cref{fig:TELSprotocol}, we show a TELS protocol used in Ref.~\cite{Chamberland22} for a PP set given by $\mathcal{P} = \{P_1, P_2 \}$, and with $\mathcal{S} = \{P_1, P_2, P_1P_2 \}$ (so that $k=2$ and $n=3$). The generator matrix for the set $\mathcal{S}$ is given by
\begin{align}
    G = \begin{bmatrix} 
101 \\
011
\end{bmatrix},
\end{align}
with the rows of $G$ generating codewords of the classical $[3,2,2]$ code. Note that in a general-purpose quantum computer, the algorithms that are executed may have different sizes of PP sets. Depending on the size of the PP set, a suitable classical code would be chosen (the classical code chosen for a PP set of size-2 in \cref{fig:TELSprotocol} was chosen due to its simplicity). 

When measuring the operators in $\mathcal{S}$, the multi-qubit Pauli operators are measured using $d_m'<d_m$ syndrome measurement rounds during the merge step of the lattice surgery protocol (since the ability to detect logical timelike failures allows for noisier lattice surgery operations). In the example of \cref{fig:TELSprotocol}, if the measurement outcome of $P_1P_2$ is inconsistent with the measurement outcome of $P_1$ and $P_2$, a logical timelike failure has been detected. If timelike logical failures are not detected, Clifford corrections are applied based on the measurement results of the $k$ original Paulis, which are obtained by decoding the $n$ measurement bits according to a classical code (here, the [3,2,2] code). Note however that if two logical timelike failures occur, the protocol in \cref{fig:TELSprotocol} will be unable to detect such a timelike failure (this can also be seen by noting that the distance of the classical code $\mathcal{C}$ is 2). 

If a logical timelike failure during the TELS protocol is detected with probability $p_D$, the original $k$ Paulis are measured with measurement distance $q d_m$, where $q$ is a constant that can be optimized to reduce the overall measurement distance. This offsets the fact that the remeasure round only occurs with probability $p_D$. Note that if the Paulis in the original set $\mathcal{P}$ are measured (due to the detection of a logical timelike failure), timelike failures cannot be detected since only the original $k$ Paulis are measured. The circuits used in the TELS protocol described in this section are performed in the Clifford frame, hence any Clifford corrections that may be required can be conjugated through to the end of the circuit.

To perform a TELS protocol involving Clifford or non-Clifford gates that use resource states,  hardware must be allocated to hold the resource states in memory for the entire TELS protocol. They cannot, in general, be measured out after each Pauli measurement. For example, in \cref{fig:TELSprotocol} the $\ket T$ states used to measure $P_1$ and $P_2$ need to be held in memory until the end of the protocol. Note that a common approach for designing the architecture for a quantum computer is to use a central processing unit (also referred to as a ``core'') and a set of distillation factories. If a quantum computer is built according to this model, the use of TELS results in a small additive factor to the total number of logical qubits needed instead of a multiplicative factor. This does not result in a substantial increase to the space-time cost of implementing algorithms on such an architecture, as in general the algorithms runtime is reduced by a multiplicative factor of between $2 \times$ and $5 \times$, as we show in this manuscript. In contrast, in Ref.~\cite{Litinski19}, Litinski showed that a runtime reduction of approximately $2 \times$ required a $6 \times$ increase in qubit overhead cost. 

We now calculate the total time taken by a TELS protocol to execute a PP set of size $k$, using an $[n,k,d]$ classical code $\mathcal C$. To preface, in a regular lattice surgery protocol, each measurement takes time $(d_m + 1)$, for a total time of $k(d_m + 1)$. Following the TELS protocol described in this section, the total time taken to measure all Paulis in a PP set using TELS is, on average,
\begin{equation}
T_{\text{old}} = n(d_m' + 1) + p_D k (\lceil q d_m \rceil + 1) \: ,
 \label{eq:OldTELStime}
\end{equation}
where the second term is due to the contribution from measuring the Paulis of $\mathcal{P}$ if timelike logical failures are detected. We use the subscript `old' to refer to the TELS protocol described in this section, since in \cref{sec:NewTELS}, the remeasure part of the TELS protocol has a different time cost. Similarly for this protocol, the logical error rate per Pauli is the sum of the logical error rate contributions of the temporally encoded set and the remeasure set,
\begin{align}
p_L = &\frac{1-p_D}{k} \sum_{i=d}^{n} l_i (p_m(d_m'))^i  (1-p_m(d_m'))^{n-i} \nonumber \\
 &\quad+ p_D p_m(\lceil q d_m \rceil) \nonumber \\
 \approx &\frac{1-p_D}{k} l_d (p_m(d_m'))^d (1-p_m(d_m'))^{n-d} \nonumber \\
 &\quad + p_D p_m(\lceil q d_m \rceil) \: ,
 \label{eq:PLfirst}
\end{align}
where $l_i$ is the number of weight-$i$ timelike logical failures that cause trivial syndromes when multiplying the lattice surgery measurement outcomes by the parity check matrix of the classical code $\mathcal{C}$. The variable $p_m(d_m')$ is the probability of a logical timelike failure of a single lattice surgery measurement with measurement distance $d_m'$ (obtained from \cref{eq:TimelikePoly}),
and $p_m(\lceil q d_m \rceil)$ is the probability of a logical timelike failure in the remeasure round, where the measurement distance is $\lceil q d_m \rceil$. Note that the timelike logical error rate is due to wrong Clifford corrections being applied after the non-Clifford gates. Hence if TELS fails without detecting a timelike error, the probability that there is a logical error is scaled by $1-p_D$.
There are various ways that the $l_i$ term in \cref{eq:PLfirst} can be calculated. In \cref{subapp:monte,subapp:bernoulli}, we show how $l_i$ is calculated using sampling methods, where the computational complexity of sampling grows with `$d$`. In \cref{subapp:macwilliams}, we show how the $l_i$ coefficients can be computed determinstically using MacWilliams identities. The advantage of using the MacWilliams identities is that the computational complexity only grows exponentially with $k$ as opposed to $d$.

Finally, the probability that an error is detected during the first stage is 
\begin{equation}
p_D \geq \sum_{i=1}^{d-1} {n \choose i} (p_m(d_m'))^i (1-p_m(d_m'))^{n-i}.
\label{eq:PDgreaterThan}
\end{equation}
Note that in \cref{eq:PDgreaterThan}, we use the $\geq$ sign since some sets of $\geq d$ logical timelike failures will also be detected.

\section{New TELS encoding protocol}
\label{sec:NewTELS}

\subsection{Improvements arising from repeated temporally encoded measurements}
\label{subsec:RepeatedTELS}

In this section we describe an improved TELS protocol compared to the one described in \cref{subsec:TELSreview}. An example of the application of the new protocol is given in \cref{fig:TELSprotocolnew}. The main difference is that if a logical timelike failure is detected when measuring the Paulis in the set $\mathcal{S}$, operators in $\mathcal{S}$ (instead of $\mathcal{P})$ are measured anew. In particular, operators in $\mathcal{S}$ are repeatedly measured until no logical timelike failures are detected. Only at this point can we determine the Clifford corrections that must be applied. 

Although it is clear that the protocol described above is different from the protocol described in \cref{subsec:TELSreview}, it is not clear that the new protocol takes fewer syndrome measurement rounds. First, let us determine the time taken by the protocol described above, which is given by
\begin{align}
    T_{\text{new}} = & n (d_m' + 1) + p_D T_{\text{new}} \nonumber \\
     = & \frac{n (d_m' + 1)}{1-p_D} \: ,
     \label{eq:TEnew}
\end{align}
since the lattice surgery implementation of each multi-qubit Pauli measurement in $\mathcal{S}$ is performed using $d_m'$ syndrome measurement rounds during the merge step.

To compare \cref{eq:TEnew} to the protocol described in \cref{subsec:TELSreview}, we can rewrite \cref{eq:OldTELStime} as
\begin{align}
     T_{\text{old}} = & n (d_m' + 1) + p_D u n (d_m' + 1) \nonumber \\
      = & n (d_m' + 1) (1 + p_D u) \: ,
\end{align}
where $u = \frac{k (\lceil q d_m \rceil + 1)}{n (d_m' + 1)}$.

If $\frac{1}{1-p_D} < (1 + p_D u)$, the revised protocol is more efficient than the one described in \cref{subsec:TELSreview}.
Simplifying this, we can get a constraint on $u$,
\begin{equation}
    u > \frac{1}{1-p_D} \: .
    \label{eq:ucheck}
\end{equation}

We computed the value of $u$ for all the classical codes considered in this paper that were used in a TELS protocol. If the condition in \cref{eq:ucheck} was satisfied (which was true for all classical codes considered in the work except the distance-$2$ Single Error Detect code (see \cref{subsubsec:SED})), we used the new protocol for TELS. An interesting note is that the time difference between the old protocol and the new is only exacerbated when $p_D$ is larger. 

The probability of a logical error in this new protocol is the probability that an undetectable set of measurement errors occurs during the last (or successful) iteration of temporally encoded lattice surgery. Let $p'_L \equiv \frac{1}{k} \sum_{i=d}^{n} l_i \, (p_m(d_m'))^i (1-p_m(d_m'))^{n-i}$, which corresponds to the probability (per Pauli) that a series of timelike logical failures during the execution of the measurements in $\mathcal{S}$ results in a trivial syndrome when multiplied by the parity check matrix of the classical code $\mathcal{C}$. The TELS protocol failure probability is then given by the following equation
\begin{align}
p_L &= (1-p_D)p'_L + p_D(1-p_D)p'_L + p_D^2(1-p_D)p'_L + \cdots \nonumber \\
 &= (1-p_D)p'_L(1 + p_D + p_D^2 + \cdots) \nonumber \\
 &= p'_L \nonumber \\
 &= \frac{1}{k} \sum_{i=d}^{n} l_i \, (p_m(d_m'))^i (1-p_m(d_m'))^{n-i} \: .
\end{align}

\vspace{-.5cm} 
\subsection{Error Correct + Error Detect}
\label{subsec:ErrorDetectCorr}

In Ref.~\cite{Chamberland22}, the classical codes used in a TELS protocol were for a pure error detection scheme (as described in \cref{subsec:TELSreview}). Further, it was argued that performing error correction using TELS would result in worse speedups compared to a pure error detection scheme. In this section we show that performing a hybrid scheme using TELS, where errors of low weight are corrected and errors of higher weight are detected, can result in further performance improvements compared to a pure error detection scheme. The overall effect is to reduce the average time per Pauli measurement, while staying under the logical error rate threshold set by $\delta$. This effect is more dominant when using classical codes with high $k$ and $d$, and at higher tolerable noise rates $\delta$.  At larger $\delta$, it is possible to correct classical errors of higher weight than for smaller $\delta$ since there is more of an error budget and $p_L$ can be increased to match $\delta$. In addition, $p_D$ tends to be higher when $d_m$ is small, and $d_m$ is smallest at large delta. As we show in \cref{tab:ECED}, the benefit of using error correction is that  $p_D$ can be made smaller and that in turn allows smaller average runtimes per Pauli.

When using a distance-$d$ classical code in an error detection scheme, an error is detected with probability $p_D \approx O(p_m(d_m'))$. When using classical codes of high distance, the target logical timelike failure rate of a lattice surgery protocol may be achieved using measurement distances $d_m$ close to 1. However, with such small measurement distances, $p_m(d_m')$ is inevitably large, and so~is~$p_D$.

\pagebreak 

If we instead use the classical code to correct all errors up to some weight $c$, a detection event is only triggered by errors of weight at least $c+1$. In other words, $p_D \approx O\big ((p_m(d_m'))^{c+1} \big)$. A lower value of $p_D$ thus requires fewer repeated measurements of the Paulis in the set $\mathcal{S}$. Although time is now saved by performing fewer remeasure rounds, the logical error rate increases relative to a pure error detection scheme. In a pure error detection scheme, the logical error rate scales as $O \big ( (p_m(d_m'))^d \big )$. However, some errors of weight $d-c$ will have the same syndrome as errors of weight-$c$, since they may differ by a logical operator. The most probable correction for this syndrome is the weight-$c$ error, but applying this correction yields a logical error with probability $O\big ((p_m(d_m'))^{d-c}\big )$. Consequently, using the classical code in a TELS protocol to correct low-weight errors leads to a tradeoff between the logical error rate of the encoded measurements and the time saved due to fewer remeasure rounds. 
Such tradeoffs have also been considered for magic state distillation schemes (see for instance Ref.~\cite{Haah18}).

By incorporating the ability to correct errors up to weight $c$, the probability of observing an uncorrectable but detectable error becomes
\begin{equation}
p_D \geq \sum_{i=c+1}^{d-c-1} {n \choose i} (p_m(d_m'))^i (1-p_m(d_m'))^{n-i} \: ,
\end{equation}
which is significantly smaller than in \cref{eq:PDgreaterThan}.
The probability of a logical error per Pauli due to the TELS protocol described in this section is
\begin{align}
p_L =& \frac{1}{k} ( \sum_{i=d-c}^{d-1} l_i ( p_m(d_m'))^i (1-p_m(d_m'))^{n-i} + \nonumber \\ 
& \sum_{i=d}^{n} l_i (p_m(d_m'))^i (1-p_m(d_m'))^{n-i}) \nonumber \\
 \approx & \frac{1}{k} l_{d-c} ( p_m(d_m'))^{d-c} (1-p_m(d_m'))^{n-d+c} \: .
\label{eq:p_LtotECED}
\end{align}
Note that in \cref{eq:p_LtotECED}, $l_i$ in the second term of the sum includes contributions from both undetectable errors and errors of weight greater than $d$ which have the same syndrome as correctable errors. We only include the leading order term since higher order terms have very small contributions. Note however that for classical codes with very large values of $d$, a larger value of $p_m(d_m')$ may be tolerated. In such cases, including higher order terms in \cref{eq:p_LtotECED} may be required.

To show the benefits of including error correction in a TELS protocol, we consider TELS with a classical $[127,92,11]$ BCH code (see \cref{subsubsec:BCH}) at a physical error rate of $10^{-4}$ and a target logical error rate of $\delta = 10^{-15}$ per Pauli. In \cref{tab:ECED}, we show that by correcting errors up to weight three, it is possible to reduce the average time per Pauli measurement by $36 \%$ .
 
\setlength{\tabcolsep}{8pt}
\begin{table}[]
    \centering
    \begin{tabular}{c c c c c c}
    $d_m$ & $c$ & $p_{L}$ & $p_D$ & $T$  & $T/k$ \\
    \hline
    $1$ & $0$ & $1 \times 10^{-24}$ & $0.37$ & $400.71$  & $4.36$  \\[.2cm]
    $1$ & $1$ & $2.5 \times 10^{-21}$ & $0.076$ & $275.08$ & $3$  \\
    $1$ & $2$ & $1 \times 10^{-18}$ & $0.011$ & $256.83$ & $2.79$  \\
    $1$ & $3$ & $2.5 \times 10^{-16}$ & $0.0012$ & $254.3$  & $2.76$  \\[.2cm]
    $11$ & $-$ & $1.8 \times 10^{-16}$ & $-$ & $1104$ & $12$
    \end{tabular}
    \caption{Comparison between the performance of a TELS protocol implemented using pure error detection versus protocols implemented using combined error detection and correction with the $[127,92,11]$ BCH code. The last line of the table shows the performance of unencoded lattice surgery. Here, $d_m$ is the measurement used when measuring multi-qubit Pauli operators using lattice surgery, and $c$ is the maximum weight of errors that can be corrected by the classical code used in the TELS protocol. The objective is to minimize the average time taken per Pauli measurement (last column) while ensuring the logical error rate is less than $10^{-15}$ per Pauli. The logical error rate per Pauli $p_L$ is calculated using \cref{eq:p_LtotECED}, where the routing space area is $A=100$. The results in the first row are for the TELS protocol implemented using pure error detection. By correcting weight-one, -two and -three errors, the average measurement time is reduced to $2.76$ syndrome measurement rounds, as opposed to $4.36$ when using TELS with pure error detection, or $12$ without TELS. Note that a pure error detection scheme with $d_m \ge 3$ results in a larger runtime than those obtained in the first four rows of this table since the total number of syndrome measurement rounds will be at least $127 \times 4 = 508$.}
    \label{tab:ECED}
\end{table}

\pagebreak 
\subsection{Protocols for PP sets of size up to $100$}
\label{subsec:ManyCodes}

When performing temporally encoded lattice surgery, it is unclear which classical $[n,k,d]$ code will give the best speedup. \cref{eq:TEnew} shows that the total time to measure all the multi-qubit Pauli operators using a TELS protocol is clearly proportional to $n$. Also, the time is inversely proportional to the distance of the classical code used since a higher value of $d$ results in a lower value of $d_m'$. However the contribution arising from $p_D$ needs to be determined, as well as the tradeoffs between a pure error detection scheme and an correction + detection scheme. 

Ref.~\cite{Chamberland22} evaluated the speedups arising from using TELS protocols at a physical error rate of $p=10^{-3}$, $\delta = 10^{-15}$ and $A=100$ (see \cref{eq:TimelikePoly}). Three classical codes were considered, and the speedups arising from each code were computed. In the analysis, the distance-$4$ Extended Hamming code resulted in the best performance improvements.
For the TELS protocols considered in this work, it was unclear at the outset which classical codes gave the best performance improvements. As such, we collected results for an expanded list of classical codes, which are listed in \cref{tab:codes}. Note that in our analysis, we only consider odd values of $d_m'$, as \cref{eq:TimelikePoly} only applies to odd measurement distances. Since the calculation of the number of malignant fault sets is computationally expensive, we could not consider classical codes of distances higher than those shown in \cref{tab:codes}. The code constructions for the codes in \cref{tab:codes} are provided in \cref{app:codeconstruction}.

\begin{table}[]
    \centering
    \begin{tabular}{c c c}
        Code / Code family  &  Distances \\
        \hline
        Single Error Detect(SED)  & $2$\\
        Hamming (Hamm) & $3$ \\
        Concatenated SED (CSED) & $4$ \\
        Extended Hamming (EHamm) & $4$ \\
        Golay (Gola) & $7$ \\
        Extended Golay (EGol) & $8$ \\
        Doubly Concatenated SED (DCSED) & $8$ \\
        Reed-Muller (RM) & $8$ \\
        Polar (Pol) & $4,8$ \\
        Zetterberg (Zett) & $5,6$ \\
        Bose–Chaudhuri–Hocquenghem (BCH) & $3,5,7,9,11$ 
    \end{tabular}
    \caption{Classical codes (and their associated distances) used in the TELS protocols considered in this work. In different noise regimes, some codes perform better than others, as we show in \cref{fig:delt10}, \cref{fig:p3delt152025} and \cref{fig:p4delt1520}. We provide explicit constructions of the best performing codes in \cref{app:codeconstruction}.}
    \label{tab:codes}
\end{table}

\begin{figure*}
    \centering
    \includegraphics[width=.98\textwidth]{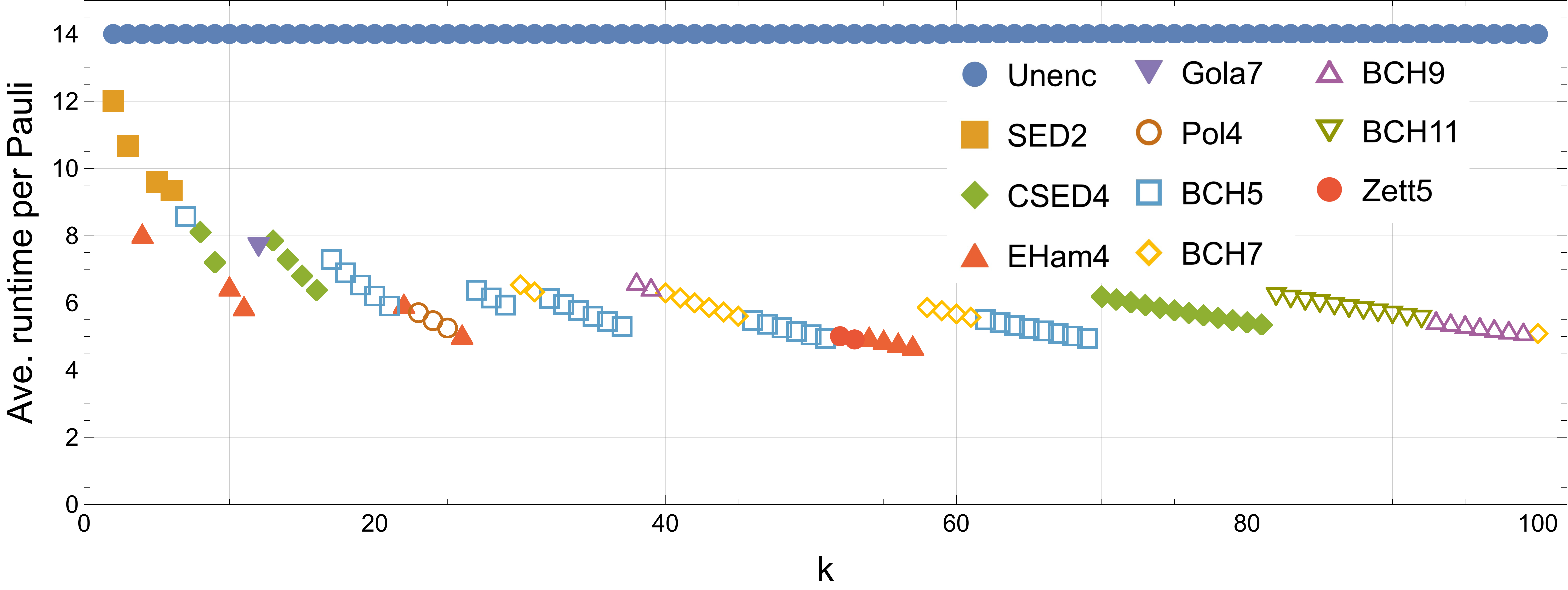}
    \caption{The best average runtime per Pauli (in units of syndrome measurement rounds) for all classical codes considered in this work when using the TELS protocol of \cref{sec:NewTELS} for PP sets of size $k \in \{ 2,3, \mathellipsis, 100\}$ at $p=10^{-3}$, $\delta = 10^{-10}$ and with maximum routing space $A = 100$. For example, for PP sets of size $k=20$, a distance-$5$ BCH code achieves the lowest average runtime per Pauli among all the classical codes considered. To calculate $p_L$, we used \cref{eq:p_LtotECED}, with $p_m$ given by \cref{eq:TimelikePoly} and $d_m$ chosen to minimize the runtime while keeping $p_L<\delta $. We compare the results of TELS with un-encoded lattice surgery, which is shown here to take $14$ syndrome measurement rounds per Pauli. The legend labels correspond to codes from \cref{tab:codes}. Low distance codes perform better for small values of $k$, whereas for larger values of $k$, the high rate of larger-distance codes enables smaller measurement distances.}
    \label{fig:delt10}
\end{figure*}

For $k \in \{ 2,3, \mathellipsis 100\}$, the lowest achievable average time per Pauli measurement for TELS protocols described in this section and which use the codes of \cref{tab:codes} are shown in \cref{fig:delt10}. The average runtime per Pauli is $T_{\text{new}} / k$ where $T_{\text{new}}$ is given by \cref{eq:TEnew}. Such a calculation allows us to compare the performance of a TELS protocol relative to the time taken to measure multi-qubit Pauli operators in the original PP set $\mathcal{P}$ without using TELS. Our results are obtained using the parameters $p=10^{-3}$ and $\delta = 10^{-10}$. Other regimes are considered in \cref{app:speedups}. In particular, results are obtained for the parameters $\delta \in \{ 10^{-15}, 10^{-20}, 10^{-25}\}$ with $p=10^{-3}$ and $\delta \in \{  10^{-15}, 10^{-20}\}$ with $p=10^{-4}$. 
We also show the average runtimes per Pauli for unencoded lattice surgery protocols. 

With a classical code that is defined to encode $k$ logical bits, we can implement TELS protocols for PP sets of any size \textbf{up to} $k$. This allows us to use good classical codes for many different sizes of PP sets. If the PP set is of size $k-j$, the Paulis associated with the remaining $j$ logical bits of the code are simply set to the identity. When decoding the temporally encoded measurement results, information corresponding to the extra logical bits is thrown away.

The biggest insight from our findings is that small codes with low distances perform well at small values of $k$, whereas at larger values of $k$, the larger and higher distance codes perform better. Moreover, we notice that for values of $k$ higher than $30$, codes from the BCH family generally give the largest speedups. Of course, the list of codes considered in \cref{tab:codes} is not all-encompassing. There may be many codes that perform better than the codes considered here. To find codes that work well for a specific value of $k$, the primary target should be to ensure that the rate of the code ($\tfrac{k}{n}$) is not too low. In our observations, codes with rate less than $1/2$ generally did not give the best speedups. The second biggest consideration is the distance of the classical code. High-distance classical codes admit very low values of $d_m'$ at the cost of much higher probabilities of detecting a logical timelike failure. Error correction for smaller weight errors can be used to reduce the detection rate as long as the logical failure rate per Pauli is below the target rate $\delta$.

\section{TELS for magic state distillation}
\label{sec:TELSmagicState}

Quantum gates can be classified into those that can be efficiently simulated by classical computers (such as Clifford gates), and those that cannot~\cite{Aaronson04}. As was discussed in \cref{sec:Intro}, a universal gate set can be achieved by combining Clifford gates with at least one non-Clifford gate. However, the implementation of logical non-Clifford gates with topological codes is not as straightforward as implementing gates from the Clifford group. Examples of non-Clifford gates include rotations about one of the Bloch sphere axes by angles of $j\pi/8$ where $j\in \{ 1,1/2,1/4, \mathellipsis\}$  or multiply-controlled $Z$ gates $C^jZ$ where $j\geq 2$. One common method used to implement non-Clifford gates is to use magic states as resource states along with stabilizer operations to perform gate teleportation.

A logical magic state can be created by first preparing a physical magic state, followed by a gauge fixing step which encodes the state into a logical qubit patch such as the surface code~\cite{Vuillot19}. However such operations are not fault-tolerant and lead to encoded magic states with unacceptably high physical noise rates. To get high fidelity magic states, the prepared magic states are then injected into a magic state distillation protocol where a quantum error-correcting code uses stabilizer operations to detect logical failures present on the injected magic states. Such protocols can be concatenated to achieve any desired target logical failure rate.

\subsection{Magic state injection}
\label{subsec:MagicInject}

In this section, we briefly describe various magic state injection methods used to prepared magic states prior to performing a distillation protocol. 

Ref.~\cite{Lao22} considers a state injection protocol on the rotated surface code afflicted by a standard depolarizing noise model. On the other hand, Ref.~\cite{Singh22} considers an injection protocol under a biased depolarizing noise model with $\eta = 10^3$ and $\eta = 10^4$. Further, the magic states are prepared in the $XZZX$ code rather than the rotated surface code. In the implementation of the protocol, the magic states are initialized into an effective two-qubit repetition code using low-error $ZZ$ rotations, and the stabilizers of this code are measured twice. Afterwards, the error detecting code is merged into the final $XZZX$ code, where $d_m$ syndrome measurement rounds are performed. The prepared magic states are then shown to have failure probability $O(p^2)$. In this work, we conjecture that a similar injection protocol can be devised using rectangular surface codes in the presence of biased noise. However, we consider the entire protocol to require only two syndrome measurement rounds since the $d_m$ syndrome measurement used after merging the error detecting code with the final code can be part of the lattice surgery operations used in the magic state distillation schemes described in \cref{subsec:MagicDist}. 

Recall that the noise model used in this work has bias $\eta = 10^2$, as opposed to $\eta = 10^3$ or $\eta = 10^4$ used in Ref.~\cite{Singh22}. Further, in Fig. 3 of Ref.~\cite{Singh22}, it can be seen that for values of the physical noise rate parameter $p \le 10^{-3}$, the injected magic states have logical error rate less than $p$ (with a quadratic scaling as a function of $p$). Since we use a rectangular surface code with bias $\eta = 10^2$, we take the injected magic states to have a logical error rate 
\begin{align}
    \epsilon_{\text{L,X}} = \epsilon_{\text{L,Y}} = \frac{p}{3 \eta},
    \epsilon_{\text{L,Z}} = \frac{p}{3},
    \label{eq:InjectProbs}
\end{align}
where $\epsilon_{\text{L,P}}$ is the probability of a logical Pauli $P$ error when injecting the prepared magic state. Note that the previous expression may be optimistic depending on the hardware implementation of the $ZZ(\theta)$ rotation used in Ref.~\cite{Singh22} and given that our noise bias $\eta = 100$ is lower than what was considered in Ref.~\cite{Singh22}. On the other hand, if the hardware architecture allows for the implementation of the $ZZ(\theta)$ as in Ref.~\cite{Singh22} and values of $p$ in the range $10^{-4} \le p \le 10^{-3}$ are below threshold at $\eta = 100$, then the expressions for $\epsilon_{\text{L,P}}$ in \cref{eq:InjectProbs} can be quite pessimistic. Note that if the underlying hardware architectures enables the implementation of the methods described in Refs.~\cite{chamberland2020very,SC22Modulo}, \cref{eq:InjectProbs} may be improved even further\footnote{Implementing the schemes described in Refs.~\cite{chamberland2020very,SC22Modulo} would result in a higher space-time cost compared to the other injection protocols described in this section, since color code patches would be used, and $\mathcal{O}(d)$ rounds of error detection would be required. However the injected magic states would have failure probabilities that scale as $\mathcal{O}(p^{(d+1)/2})$ instead of $\mathcal{O}(p)$ or $\mathcal{O}(p^2)$. Such low failure rates could then result in much smaller magic state distillation factories (see for instance Ref.~\cite{ChambsCat}). }. In what follows, we define 
\begin{align}
    \epsilon_L = \epsilon_{\text{L,X}} + \epsilon_{\text{L,Y}} + \epsilon_{\text{L,Z}} = \frac{p}{3} + \frac{2p}{3 \eta}.
    \label{eq:epsL}
\end{align}

Using the results of Fig.~$3$ of Ref.~\cite{Singh22}, we approximate the success probability of the state injection protocol to be greater than $99 \%$ for $p \leq 10^{-3}$. Hence, we take the time taken to prepare a magic state to be approximately $T_{\text{inj}} = 2/0.99$ syndrome measurement rounds for $p=10^{-3}$ and $T_{\text{inj}} = 2/0.995$ syndrome measurement rounds for $p=10^{-4}$. It is especially important to consider the time taken for injection since in TELS protocols, the time for each multi-qubit Pauli measurement can be as low as two syndrome measurement rounds (if we include the time required for ancilla reset). Note that the precise success probabilities require a more careful analysis, however we don't expect such analysis to have a significant effect~on~$T_{\text{inj}}$.

\begin{figure*}
    \centering
    \includegraphics[width=0.98\textwidth]{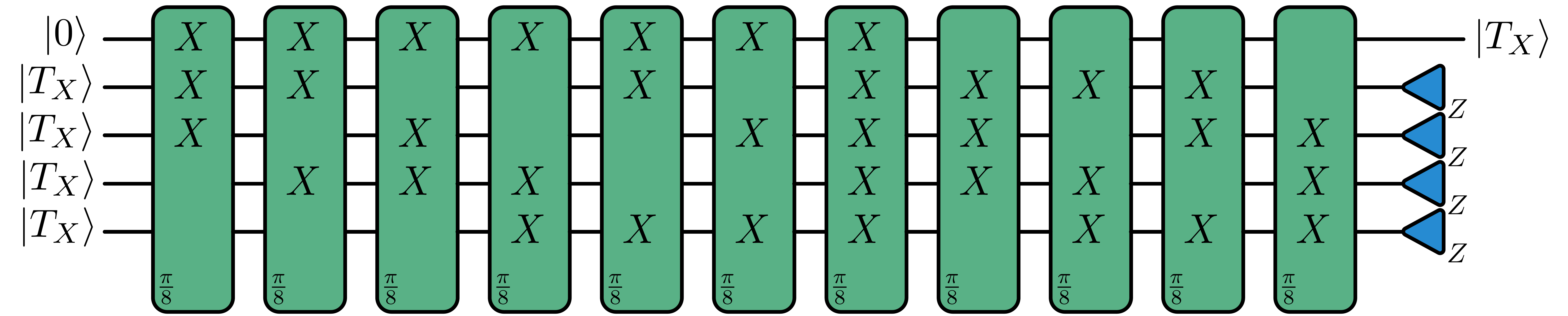}
    \caption{Circuit used in a $15$-to-$1$ magic state distillation protocol expressed as a sequence of multi-qubit non-Clifford gates. The circuit above is the Hadamard-transformed version of Fig.~15 of Ref.~\cite{Litinski19}, which produces the state $H \ket T$. This state can be used in the same way as the regular magic state, with only a change of Pauli basis while measuring.}
    \label{fig:Haddist}
\end{figure*}

\pagebreak 
\subsection{Magic state distillation}
\label{subsec:MagicDist}

A magic state distillation protocol takes several noisy magic states as input (which are encoded in some code such as the surface code), and by performing stabilizer operations which are part of an error detection protocol, yields fewer magic states of much higher fidelity. The output yield and magic state error probability depend entirely on the quantum code used for the error detection protocol. Bravyi and Haah suggested a class of distance-2 quantum codes for distilling $k$ magic states from $3k+8$ input magic states~\cite{Bravyi12}. These codes offer minimal protection, but have good yield. One of the protocols we consider in this work is the popular $15$-to-$1$ distillation protocol~\cite{Bravyi05}, which can be transformed into a series of $11$ commuting multi-qubit Pauli measurements~\cite{Litinski19}. These Pauli measurements form a PP set of size $11$, and thus a TELS protocol can be used to reduce the runtime required to measure each multi-qubit Pauli operator. In search of magic state distillation protocols with good output-to-input ratio, we found infinite families of protocols with near constant rate, where the focus is not on concatenating protocols but instead on measuring the stabilizers of an outer code using operations that are transversal for an inner code~\cite{Haah17}. Other techniques construct more complex codes for distillation by generalizing triorthogonal codes, or by puncturing Reed-Muller codes~\cite{Haah18}. In addition to the $\llbracket 15,1,3\rrbracket$ code, in this work we also consider several triorthogonal quantum codes that are constructed by puncturing a $[128,29,32]$ classical Reed Muller code, as described in Ref.~\cite{Haah18}. In particular, the distillation protocols we consider are derived from $\llbracket 114,14,3\rrbracket$, $\llbracket 116,12,4\rrbracket$ and a $\llbracket 125,3,5\rrbracket$ quantum codes. It is beneficial to consider quantum codes of different distances as this allows distilling magic states across a range of target logical error~probabilities.

Since our focus of applying TELS to magic state distillation is to reduce space-time costs, we comment on some previous work by Litinski~\cite{Litinski19,Litinski19magic}. First it was shown that the time costs of distillation algorithms can be reduced, however this leads to a disproportionate space increase, leading to overall larger space-time costs~\cite{Litinski19}. In further work, it was shown how to reduce space-time costs by performing the distillation using surface code patches of reduced distance, and using faulty $T$ measurements instead of $T$ state injection~\cite{Litinski19magic}. 
Improvements due to TELS protocols may also be applied to the work in Ref.~\cite{Litinski19magic}, allowing even smaller space-time costs than shown in \cref{tab:STcosts} of \cref{sec:MStilelayouts} (where we apply TELS to various distillation protocols and compute the space-time costs). The main objective of the results given in \cref{tab:STcosts} is to show that space-time costs can be reduced when using TELS as opposed to when there is no temporal encoding. In addition, we make a careful assessment of the time and space required for magic state injection, which makes the results in \cref{tab:STcosts} look more pessimistic when compared to results such as in Ref.~\cite{Litinski19,Litinski19magic}.

Traditional distillation algorithms, such as those in Ref.~\cite{Litinski19} perform only pure multi-qubit Pauli $Z$ measurements. In the remainder of this work, we apply Hadamard transformations to the distillation circuits to produce circuits consisting of pure multi-qubit $X$ measurements. Such a transformation reduces the space-time costs of the distillation factories, as only the shorter logical $X$ boundary of the asymmetric surface code patch will need to be accessed. The Hadamard-transformed version of the $15$-to-$1$ distillation circuit of Ref.~\cite{Litinski19} is shown in \cref{fig:Haddist}.

\subsection{Clifford frame distillation circuit}
\label{sec:Cliff}

To obtain the space-time costs of various magic state distillation protocols implemented with TELS, we first design an appropriate distillation protocol which outputs the desired magic state up to a Clifford correction. For simplicity, the general protocol can be separated into two steps. First the non-Clifford gates are applied using temporally encoded lattice surgery. If a non-trivial lattice surgery measurement failure is detected, all the physical qubits in the distillation tile are reset and the protocol restarts. We allow for the TELS protocol to also use the developments of \cref{subsec:ErrorDetectCorr}, where classical errors of low weight may be corrected before signaling a lattice surgery measurement failure. Alternatively, if we follow the TELS protocol of \cref{subsec:RepeatedTELS}, more magic states would need to be simultaneously held in memory, and hence the hardware requirements would be larger. If TELS was successful, we are left with a distilled magic state up to a Clifford frame, prior to performing the single-qubit measurements. In the second part of the protocol, the Clifford frame is conjugated through the final single-qubit measurements. This changes the single-qubit measurements into multi-qubit $\pi/2$ Pauli measurements implemented via lattice surgery. Note that these measurements may now be tensor products of arbitrary Paulis and not just $\Id$ and $X$. These multi-qubit measurements may also be sped up using TELS, since they all commute. In particular, we use the TELS protocol of \cref{subsec:RepeatedTELS} for these final multi-qubit Pauli measurements.

\begin{figure}
    \centering
    \subfloat[\label{fig:Pnoncliff} ]{\includegraphics[width=.46\textwidth]{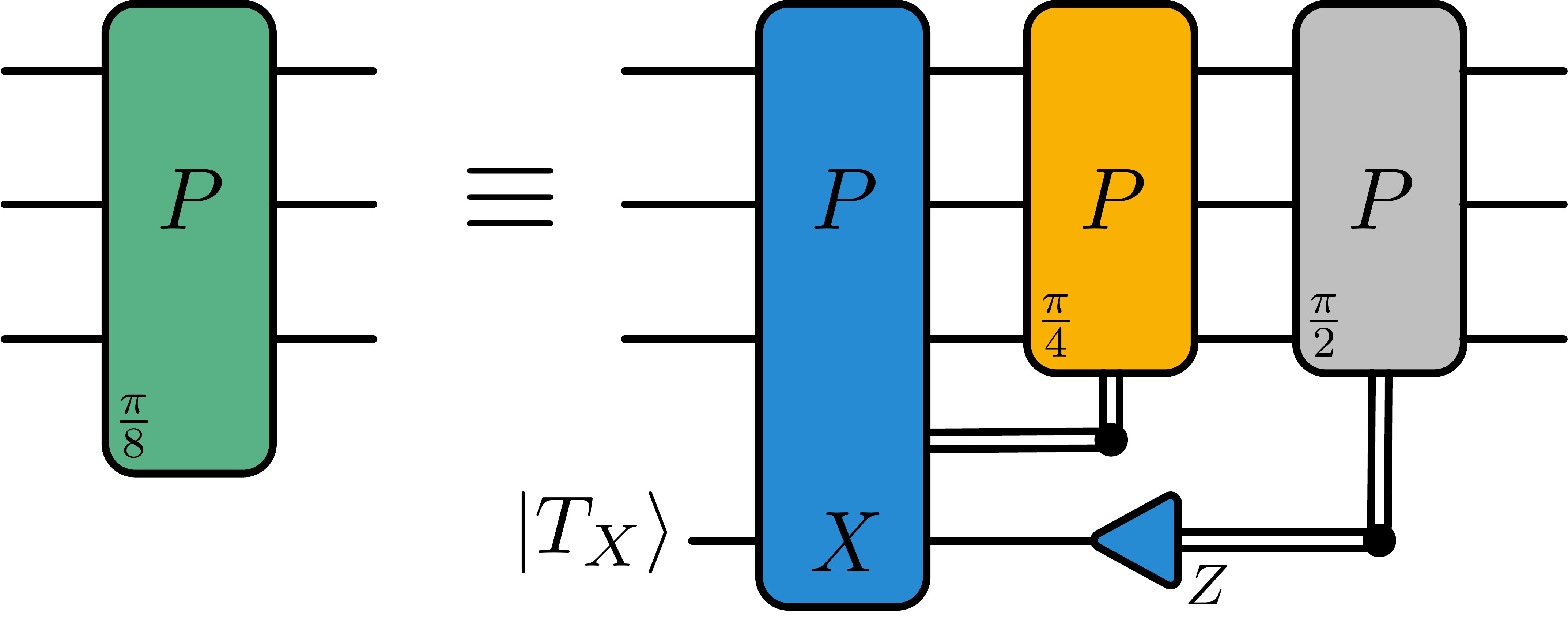}}
    \hspace{0.2cm}
\subfloat[\label{fig:Pcliff} ]{\includegraphics[width=.42\textwidth]{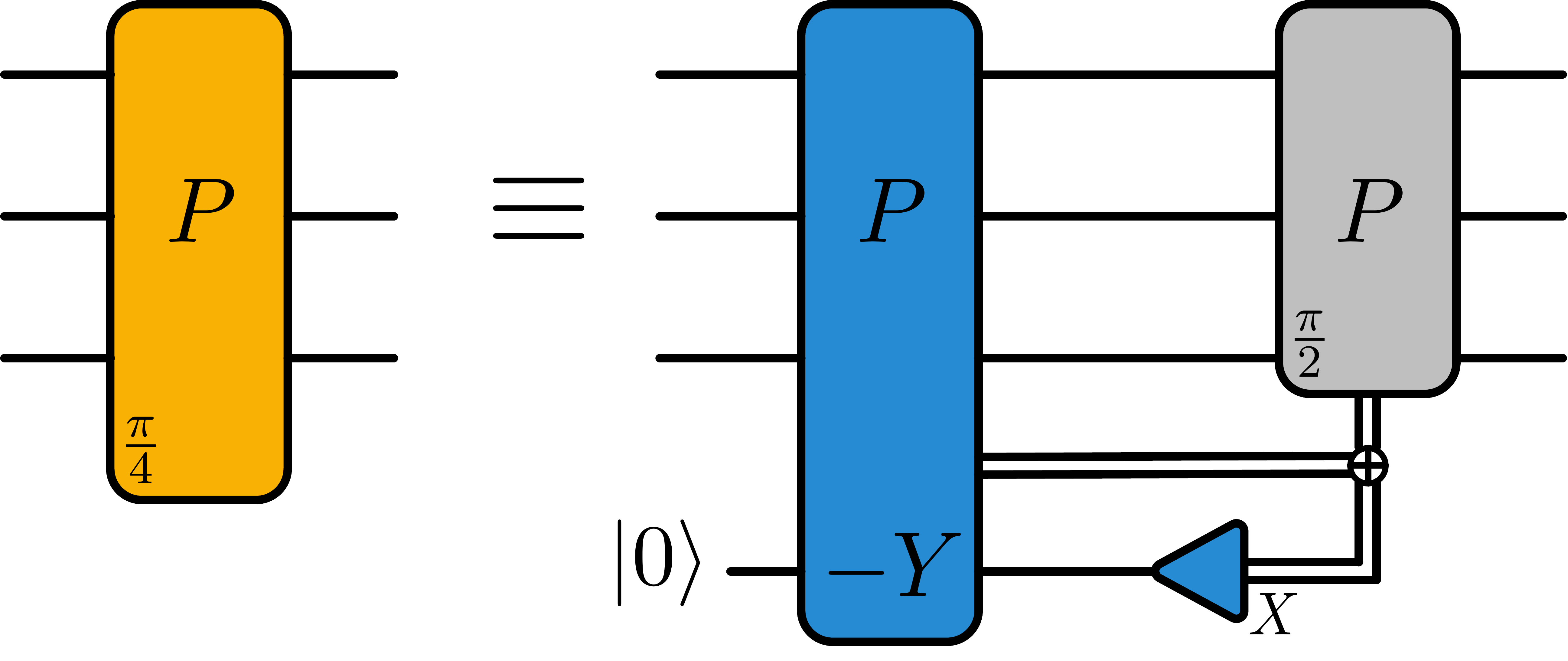}}
    \caption{(a) Circuit for performing a $\pi / 8$ multi-qubit Pauli measurement. The circuit requires a $\ket{T_X} = H \ket{T}$ resource state, and a Clifford correction may be required depending on the $P \otimes X$ measurement outcome. (b) Circuit for performing a Clifford gate using an ancilla prepared in $\ket{0}$. Both circuits are adapted from Ref.~\cite{Litinski19}.}
    \label{fig:Cliffgadgets}
\end{figure}

After the Clifford frame is conjugated through the single-qubit measurements, the output distilled magic states are correct up to a Clifford correction. In fact, for the example in \cref{fig:Haddist}, the resulting state is exactly one of $\ket{T_X},X_{\pi/4}\ket{T_X},X_{\pi/2}\ket{T_X}$ ,or $X_{3\pi/4}\ket{T_X}$, as we prove in \cref{app:CliffpartTraceproof}.  When using the distilled state in an algorithm, the final magic state measurement axis is modified depending on the Clifford frame. If magic states were prepared in the Pauli frame (see \cref{fig:PauliAutocorr}) rather than Clifford frames, such magic states would be measured in the $Z$ basis using transversal single-qubit measurements (see \cref{fig:Pnoncliff}). However, the measurement basis may now be $-Y, -Z, Y$ given that
\begin{align}
    X_{\pi/4} Z & X^{\dagger}_{\pi/4} = -Y \: , \nonumber \\
    X_{\pi/2} Z & X^{\dagger}_{\pi/2} = -Z \: , \nonumber \\
    X_{3\pi/4} Z & X^{\dagger}_{3\pi/4} = Y \: .
\end{align}
For protocols that distill multiple magic states, the Clifford frame of the distilled states may contain multi-qubit operators. After the distilled states are used by non-Clifford gates in a core of a quantum computer, the Clifford frame must be further conjugated through the remaining single-qubit $Z$ measurements. This may result in further multi-qubit Pauli measurements and additional routing space area in order to access the $Y$ and $Z$ logical boundary of the distilled states. A caveat from using the Clifford frame is that $Y$ basis measurements may require an extra ancilla (depending on the chosen hardware implementation). As such, the design of magic state distillation factories may require additional routing space to store the ancillas needed for $Y$ measurements.

We now address the additive space cost of TELS and how it may be minimized in a distillation protocol. Distillation tiles are essential building blocks of fault-tolerant universal quantum computers, so it is worth finding the smallest, most optimal qubit layouts for them. Consider the distillation circuit in \cref{fig:Haddist}. Each of the $11$ non-Clifford gates requires an input $\ket{T_X}$ state, as shown in the non-Clifford circuit gadget of \cref{fig:Pnoncliff}. If we use TELS to perform the $11$ measurements, $11$ magic states will need to be held in memory, as indicated in \cref{fig:TELSprotocolnew}. In contrast, when performing lattice surgery without temporal encoding, only one cell is assigned to repeatedly prepare magic states for non-Clifford gates~\cite{Litinski19, Litinski19magic}. Upon close inspection of the multi-qubit Pauli measurements being performed in a TELS protocol, we notice that a magic state is stored only for as long as the Pauli it was associated with from the original PP set $\mathcal P$ appears in the sequence of new measurements $\mathcal S$.  Consequently, magic states do not need to be stored for the entire protocol. Consider performing TELS for a PP set of size $3$, where each Pauli measurement consumes a magic state. We use the $[4,3,2]$ error-detect code with the cyclic codeword matrix
\begin{equation}
    G = \begin{bmatrix} 
1100\\
0110 \\
0011
\end{bmatrix}.
\label{eq:Gmat432}
\end{equation}
Here, the choice of a cyclic representation is what allows us to reduce space requirements. We may read off the new multi-qubit Pauli measurements from  $\mathcal S$ as $\{ P_1, P_1 P_2, P_2 P_3, P_3\}$. Notice that after the measurement $P_1 P_2$, the magic state associated with $P_1$ does not need to be accessed again. At this point the  hardware holding the magic state used for the $P_1$ multi-qubit Pauli measurement can be reset to prepare the magic state required for the $P_3$ measurements. Since $P_1$ does not appear in any of the measurements after the first occurrence of $P_3$, the magic states associated with $P_1$ and $P_3$ are never simultaneously accessed. Continuing with this argument, it can be seen that space for only two magic states is required to perform all the measurements given by \cref{eq:Gmat432}.

\pagebreak 
Hence for every code, a particular choice and ordering of codewords can result in a smaller quantity of $\ket{T_X}$ states that need to be stored. For classical codes that are cyclic, a cyclic description of the codeword generator matrix allows for a reduced number of magic states needed to be held in memory. To determine exactly how many, note that for any column in this matrix (corresponding to a Pauli measurement from $\mathcal S$), the number of rows between the first $1$ and the last $1$ denotes the number of magic states that need to be held in memory for that Pauli measurement. The maximum number of $\ket{T_X}$ states required for any of the columns of the codeword generator matrix is the maximum for the entire PP set. For codes that do not have a natural cyclic set of codeword generators, the codewords must be chosen and ordered very carefully. In general, finding a sequence that minimizes the space requirements of magic states is an $\NP$ problem as there are exponentially many orderings of codewords. For instance, in the 15-to-1 distillation protocol of \cref{fig:Haddist}, one of the TELS protocols we considered (see \cref{sec:MStilelayouts}) uses the classical Golay code. For this code, there is a natural cyclic representation of codewords as we show in \cref{subsubsec:Golay}. According to this construction, $12$ magic states will need to be held in memory. However in \cref{app:golaycodechoice}, we show a specific choice of codewords that can minimize the space-requirements to $10$ magic states.

To execute the non-Clifford and Clifford gates in \cref{fig:Haddist}, we use the gate gadgets in \cref{fig:Cliffgadgets}. Since we conjugate the Clifford frame through to the final single-qubit measurements, we do not actually need to perform any $\pi/4$ Clifford gates, since such gates are converted to $\pi/2$ multi-qubit Pauli measurements~\cite{Litinski19}.

\subsubsection{Time cost analysis.}
\label{subsubsec:clifftimecost}

The time taken to successfully complete one round of distillation is calculated as follows. Let $T_1$ be the time taken to implement TELS on the non-Clifford gates and $T_2$ be the variable time associated with the final multi-qubit $\pi/2$ Pauli measurements. $T_2=0$ if there are no updates to the Clifford frame. If a lattice surgery logical timelike failure is detected during TELS with probability $p_D$,
\begin{align}
    T_1 = & T_{\text{inj}} + n (d_m'+1) + p_D \Big( T_{\text{inj}} + n (d_m'+1) \Big) + \nonumber \\
    &p_D^2 \Big(T_{\text{inj}} + n (d_m'+1) \Big) + p_D^3 \mathellipsis \nonumber \\
    =& (T_{\text{inj}} + n (d_m'+1))(1 + p_D +p_D^2 + \cdots) \nonumber \\
    =& \frac{T_{\text{inj}} + n (d_m'+1)}{1-p_D} \: .
\end{align}
Here, $T_{\text{inj}}$ is the time taken to inject a magic state into a cell, calculated using the analysis in \cref{subsec:MagicInject}. Note that in the above equation, we assume that all the Pauli measurements after the first one do not wait any extra time for newly injected magic states. This assumption is validated by the fact that, for $p=10^{-3}$, magic state injection requires two syndrome measurement rounds (with probability $99\%$), or four (with probability $99.99\%$). However for all the distillation protocols we consider, TELS requires at least four syndrome measurement rounds per Pauli measurement.

The final multi-qubit $\pi/2$ Pauli measurements are non-deterministic, implying we may need to perform $k'$ Pauli measurements, where $0 \leq k' \leq \kappa$.  In the previous inequality, $\kappa$ is the number of single-qubit measurements on the input $\ket{T_X}$ states of the distillation protocol (not the ones used for the $\pi / 8$ measurements) before the Clifford frame is conjugated through (for instance, in \cref{fig:Haddist}, $\kappa=4$). To execute these Pauli measurements, we perform TELS according to the method of \cref{sec:NewTELS}. If there are $k'$ measurements to perform, this takes time $T_2 = k' (d_m + 1)$ without TELS. If we use TELS with measurement distance $d_m''$ and an $[n', k', d']$ code with detection probability $p_D'$,
\begin{align}
    T_2 &= n' (d_{m}''+1) + p_{D}' (n' (d_{m}''+1) ) + (p'_{D})^2 ... \nonumber \\
    &= \frac{ n' (d_{m}''+1) }{1-p_{D}' } \: .
\end{align}
Note that when measuring the final multi-qubit Paulis, if a detection event is observed, the entire protocol does not need to be restarted. Instead, it is sufficient to just redo the Pauli measurements associated with the TELS protocol, as is done in \cref{fig:TELSprotocolnew}. 

The time to successfully distill the magic state also relies on whether the distillation protocol itself detected an error in any of the input magic states. This is modeled by the probability that the magic state protocol detects an error on an input magic state, which we denote $p_D^{(M)}$. Thus the total time required to successfully distill a magic state is
\begin{align}
\label{eq:timecostCliff}
    T &= T_1 + T_2 + p_D^{(M)} T \nonumber \\
      &= \frac{T_1 + T_2}{1- p_D^{(M)}} \: .
\end{align}

\subsection{Challenges of extending TELS protocols to Pauli frames for magic state distillation}
\label{sec:pauliMSD}

\begin{figure}
    \centering
    \hspace{-0.2cm}
    \includegraphics[width=0.48\textwidth]{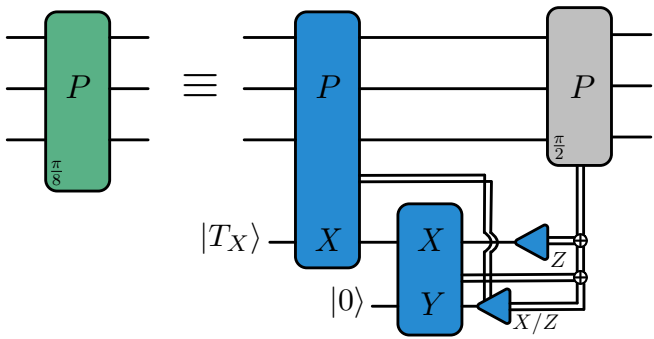}
    \caption{Circuit gadget for an auto-corrected non-Clifford gate. The circuit does not require the application of conditional Clifford gates to the logical data qubits. However, an extra ancilla prepared in $\ket{0}$ is required. }
    \label{fig:PauliAutocorr}
\end{figure}

The multi-qubit Paulis associated with the encoded TELS measurements in a Clifford-frame distillation circuit are performed using the circuit shown in \cref{fig:Pnoncliff}. However this results in a Clifford frame which eventually must be implemented using the Clifford gate gadgets in \cref{fig:Pcliff}. Keeping track of Clifford frames can be avoided by using the auto-corrected $T$ gadgets shown in \cref{fig:PauliAutocorr}. In such an implementation, the time associated with the Clifford correction can be traded for the extra space used by the additional $\ket 0$ ancilla. However, using auto-corrected $T$ gadgets in a TELS protocol leads to additional challenges. When the $k$ $P \otimes X$ measurements are performed using TELS, there will be an additional space cost associated with holding some magic state cells in memory. In order to benefit from the time speedups provided by TELS, a TELS protocol may need to also be performed on the $\ket 0$ ancilla states. Such considerations would result in the space cost being tripled (relative to the Clifford frame scheme). To see this, note that the $P \otimes X$ and $X \otimes Y$ measurements occur simultaneously since they have the same measurement distance and both $X$ boundaries of the $\ket{T_X}$ state can be accessed simultaneously. In such a protocol, the number of $\ket{0}$ and $\ket{T_X}$ states are identical. Furthermore, each $\ket{0}$ state requires an additional cell in order to access its $Y$ boundary. 

\setlength{\tabcolsep}{7pt}
\begin{table*}
    \centering
    \vspace{0.5cm} 
    \begin{tabular}{c c c c c c c c c c c}
        $p$ &  $\delta^{(M)}$ & Dist.  &  Circuit type & $d_x$ & $d_z$ & Space & Time & Space-time  & Space-time\\
         &  & code  &  &  &  &  &  & (NS)  &per o/p state \\
        \hline
        $10^{-4}$  & $10^{-10}$ & $\llbracket 15, 1, 3 \rrbracket$  & No encoding & $7$ & $9$ & $1360$ & $110.06$ & $1.5 \times 10^5$ \\
         &  &  &No enc., Par & $7$ & $9$ & $3300$ & $60.03$ & $1.98 \times 10^5$ \\
         &  &  & \textbf{Cliff-SED}& $7$ & $9$  & $1120$ & $104.05$ & $\mathbf{1.17 \times 10^5}$ & $1.17 \times 10^5$\\
         &  &  & Cliff-SED, Par& $7$ & $9$ & $2352$ & $68.03$ & $1.6 \times 10^5$ \\
         &  &   &  Cliff-BCH& $7$ & $9$ & $1344$ & $92.08$ & $1.24 \times 10^5$ \\
         &  &  &  Cliff-BCH, Par& $7$ & $9$ & $2688$ & $64.06$ & $1.72 \times 10^5$ \\
         &  &  &  Cliff-Golay& $7$ & $9$ & $1792$ & $124.06$ & $ 2.23 \times 10^5$ \\
         &  &  &  Cliff-Golay, Par& $7$ & $9$ & $3024$ & $80.04$ & $2.42 \times 10^5$ \\[0.25cm]
         
        $10^{-4}$ &  $ 10^{-15}$ & $\llbracket 116,12,4 \rrbracket$  & No encoding & $9$ & $15$ & $9440$ & $1391.48$ & $1.31 \times 10^7$  \\
         &  &   & No enc., Par & $9$ & $15$ & $14934$ & $702.77$ & $1.05 \times 10^7$ \\
         &  &  & Cliff-BCH9, Par & $9$ & $15$ & $22800$ & $431.82$ & $9.85 \times 10^6$ \\
         &  &   & \textbf{Cliff-Zett5, Par} & $9$ & $15$ & $18600$ & $442.41$ & $\mathbf{8.23 \times 10^6}$ & $6.86 \times 10^5$\\[.25cm]
         
        $10^{-3}$ & $10^{-10}$ & $\llbracket 114,14,3 \rrbracket$  &  No encoding & $9$ & $19$ & $11750$ & $1646.61$ & $1.93 \times 10^7$  \\
         &  &   & No enc., Par &  $9$ & $17$ & $16836$ & $831.62$ & $1.4 \times 10^7$ \\
         &  & & Cliff-BCH7, Par &  $9$ & $17$ & $22400$ & $595.31$ & $1.33 \times 10^7$ \\
         &  &  & \textbf{Cliff-Zett5, Par} &  $9$ & $17$ & $20480$ & $623.27$ & $\mathbf{1.28 \times 10^7}$  & $9.14 \times 10^5$\\[.25cm]
         
        $10^{-3}$  & $10^{-15}$ & $\llbracket 125, 3, 5 \rrbracket$  & No encoding & $13$ & $25$  & $17514$ & $2479.17$ & $4.34 \times 10^7$  \\
         &  &  & No enc., Par &  $13$ & $25$ & $29548$ & $1252.11$ & $3.7 \times 10^7$ \\
         &  &  & Cliff-BCH7, Par&  $13$ & $25$ & $38640$ & $859.8$ & $3.32 \times 10^7$ \\
         &  &  & \textbf{Cliff-BCH9, Par}&  $13$ & $25$ & $42504$ & $729.66$ & $\mathbf{3.1 \times 10^7}$  & $1.03 \times 10^7$
    \end{tabular}
    \vspace{0.3cm} 
    \caption{Space-time costs of different distillation protocols on a biased-noise planar surface code. $\delta^{(M)}$ is the target logical error rate per output magic state. TELS protocols are labeled ``Cliff-xxx'', with ``Par'' implying that measurements are performed two at a time (i.e., with lattice surgery measurements which can access the two $X$ logical boundaries of surface code patches simultaneously). The number of physical qubits is two times the space cost, since the space cost counts only the number of data qubits of the surface code. The probability that a distillation algorithm rejects due to an error in an injected magic state is  $p_D^{(M)} = 1-(1-\epsilon_L)^n$ where $\epsilon_L$ is given by \cref{eq:epsL}. For the $15$-to-$1$ distillation protocol, the space time cost of a protocol using TELS is approximately $30\%$ smaller ($1.17 \times 10^5$) than a protocol that does not use TELS ($1.5 \times 10^5$). For the $125$-to-$3$ distillation protocol, the space time cost is decreased by approximately $20\%$ with TELS. The label NS refers to the number of syndrome measurement rounds required for the entire distillation protocol.}
    \label{tab:STcosts}
\end{table*}

If instead, we do not perform TELS on the $X \otimes Y$ measurements, there are two options. The lattice surgery operations (with large measurement distance) can either be performed sequentially, which will result in a speed mismatch between the $X \otimes Y$ measurements and the $P \otimes X$ measurements, leading to a backlog of $\ket{T_X}$ states and $\ket{0}$ states that will need to be held in memory (and so there would be no time improvement due to TELS). Another option is to perform the slow lattice surgery operations in parallel, but this also admits an additional space cost to hold all the $\ket 0$ cells.

Given the above considerations, and the challenges associated with the design of distillation tiles for the inclusion of $\ket{0}$ ancillas, we leave the analysis of TELS protocols applied to magic state distillation protocols in the Pauli frame to future work.

\section{Magic state distillation tiles and space-time cost analysis}
\label{sec:MStilelayouts}

In this section we analyze space-time costs of various distillation protocols in different noise regimes. At a physical error rate of $p=10^{-4}$, one round of $15$-to-$1$ distillation with robust lattice surgery operations is sufficient to distill magic states with final error probability $\delta^{(M)} \le 10^{-10}$. For $\delta^{(M)}= 10^{-15}$ (which is relevant for larger algorithms), or for distillation protocols with $p=10^{-3}$, we considered $100+$ qubit quantum codes, as suggested in Ref.~\cite{Haah18}. To the best of our knowledge, our work is the first to analyze space-time costs of distillation protocols using these larger codes. For the noise rate regimes $p=10^{-4}$ and $p=10^{-3}$, we estimate the space-time costs of the various distillation protocols using different implementations of lattice surgery. We first consider protocols that do not use TELS. These protocols will execute non-Clifford gates using the auto-corrected non-Clifford gates of \cref{fig:PauliAutocorr}. Subsequently, we consider distillation protocols that perform TELS, using the methods developed in \cref{sec:Cliff}. In contrast to Ref.~\cite{Litinski19}, the distillation tiles developed in this paper are all rectangular, minimizing wasted space when tiled on a $2$D grid of qubits. Note also that in Ref.~\cite{Litinski19magic}, the logical qubits are designed to have different space-like distances $d_x$ and $d_z$ even without a biased noise model. This improvement is permitted due to the specific function of each qubit in the distillation protocol. When applying these improvements to the distillation protocols and layouts in this work, the space-time costs may be further reduced. 

The time-like distance of lattice surgery $d_m$ and the space-like distances $d_x$ and $d_z$ of the logical qubits and routing regions are computed using a procedure detailed in \cref{app:algoSTcosts}. Essentially, a set of distances $\{ d_x,d_z,d_m\}$ must be determined that minimize the overall space-time cost, while ensuring the output magic states have logical errors with probability at most $\delta^{(M)}$. In solving \cref{eq:conditiondelta}, we must first determine certain constants related to each hardware layout. This includes the area of the distillation tile used in each protocol (which we denote as the space cost), the number of logical qubits used in a tile $N$, the worst-case routing space area $A$ and maximum area used during any lattice surgery measurement. We develop $20$ different layouts in this work, and the associated constants for each of them are tabulated in \cref{app:constants}. 

In \cref{tab:STcosts}, we display the space-time costs of the various distillation protocols considered in this work. Using TELS protocols, it is possible to achieve lower space-time costs than using protocols without any temporal encoding. Although there are only minor improvements to the space-time cost of TELS-assisted distillation tiles, many of these tiles will be needed in each distillation factory. As a result, the improvements add up and the quantum computer as a whole will have a lower space-time volume. Moreover with reduced time costs, fewer distillation tiles may be required altogether, as we show in \cref{subsec:scheduling}. This in turn further reduces the space-time cost of distillation factories. Interestingly, for the $15$-to-$1$ distillation protocol, circuits that perform TELS do not produce tiles that have smaller time costs. Instead, TELS-assisted tiles require fewer qubits, and this can be attributed to the use of non-Clifford gate gadgets as shown in \cref{fig:Pnoncliff}. 

\subsection{$15$-to-$1$ distillation}
\label{sec:15to1}

The $15$-to-$1$ magic state distillation protocol is one of the most widely known protocols for distilling $\ket T$ states (see for instance Refs.~\cite{Bravyi05, BevsUniversal, fowler2018low, Litinski19, Litinski19magic}). The protocol originates from a $\llbracket 15,1,3\rrbracket$ triorthogonal CSS quantum code. The code has the property that the application of $T$ gates on all of the physical qubits of the code implements a logical $T^{\dagger}$ gate. Since we perform distillation with $\ket{T_X}$ states, we define this code to contain $1$ logical qubit,  $10$ $X$-type stabilizers and $4$ $Z$-type stabilizers. A distilled magic state is produced by encoding a logical $\ket{0}$ state, applying the transversal $T$ gates, decoding  and then performing measurements. By propagating the the Clifford gates past the transversal $T$ gates and removing the redundant parts of the  circuit, we are left with the circuit in \cref{fig:Haddist} (see Ref.~\cite{Litinski19} for a more detailed derivation). The circuit contains $11$ commuting Pauli measurements on $5$ logical qubits (four of which are logical $\ket{T_X}$ states). Since the $11$ Pauli measurements commute, they form a size-$11$ PP set. There exist many choices of classical codes to be used in TELS protocols with size-$11$ PP sets. In this paper we will focus on using a Single Error Detect code of distance $2$ ($[12,11,2]$), a BCH code of distance $3$ ($[15,11,3]$) and the Golay code of distance $7$ ($[23,12,7]$).

We will consider using this distillation protocol in a regime where the physical error rate is $p=10^{-4}$ and the target logical error rate per magic state is  $\delta^{(M)}=10^{-10}$. Using the noise model described in \cref{subsec:MagicInject} for the injection of magic states, we apply the analysis in Ref.~\cite{Litinski19magic} to determine the logical failure probability per output magic state for one round of a $15$-to-$1$ distillation scheme which is given by
\begin{align}
p^{(M)}_L = &35 \Big ( (\epsilon_{\text{L,Z}})^3+\frac{1}{2} 6 (\epsilon_{\text{L,Z}})^2 \epsilon_{\text{L,X}} \nonumber \\
&+\frac{1}{4} 12 \epsilon_{\text{L,Z}} (\epsilon_{\text{L,X}})^2 + \frac{1}{8} 8 (\epsilon_{\text{L,X}})^3  \Big ) \nonumber \\
=& \frac{35(1+\eta)^3}{27 \eta^3} p^3 \: .
\end{align}

For $p=10^{-4}$ and $\eta=100$, the probability that the distillation succeeds is $1 - p_D^{(M)} = (1-\epsilon_L)^{15} = 0.999$ and $p_L^{(M)} = 1.33 \times 10^{-12}$. As this is sufficiently below $\delta^{(M)}$, the lattice surgery measurements used to execute the distillation protocol must be modeled with measurement distance large enough to allow for distilled magic states of logical error rate at most $\delta^{(M)}$. Using the procedure in \cref{app:algoSTcosts}, we determined that the minimum spacelike distances required are $d_x=7$ and $d_z=9$. 

In the subsequent subsections, we detail the specifics of the hardware layouts that are used for the various distillation protocols, both with and without TELS. Arranging the logical qubits according to these layouts minimizes the space requirements of distillation blocks. In addition, TELS is used to minimize the time costs. Overall we observe that protocols that use TELS can achieve lower space-time costs than those that do not.

\begin{figure}
    \centering
\subfloat[\label{fig:15ue} ]{\includegraphics[width=.104\textwidth]{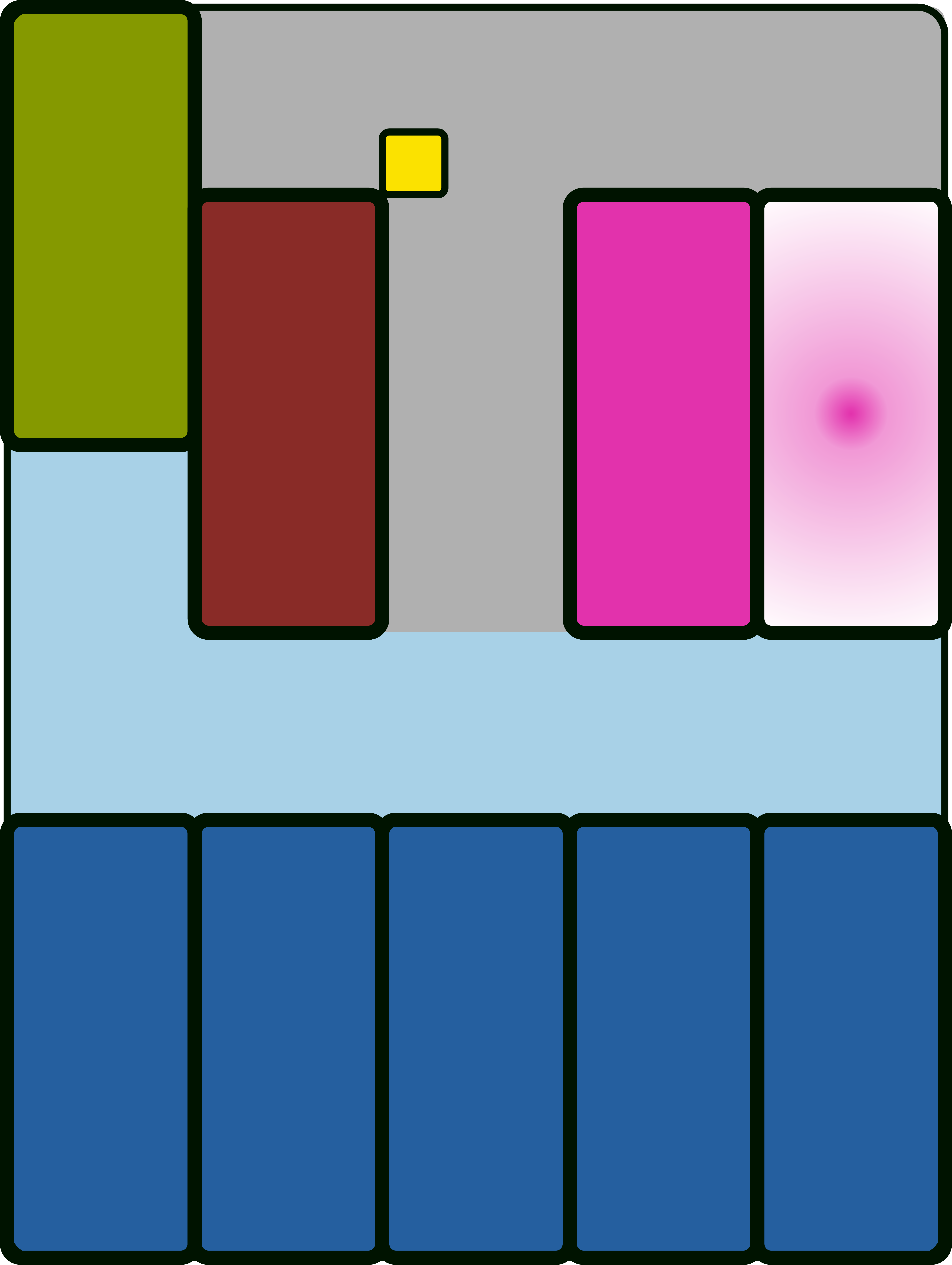}}
    \hspace{0.35cm}
    \subfloat[\label{fig:15sed} ]{\includegraphics[width=.104\textwidth]{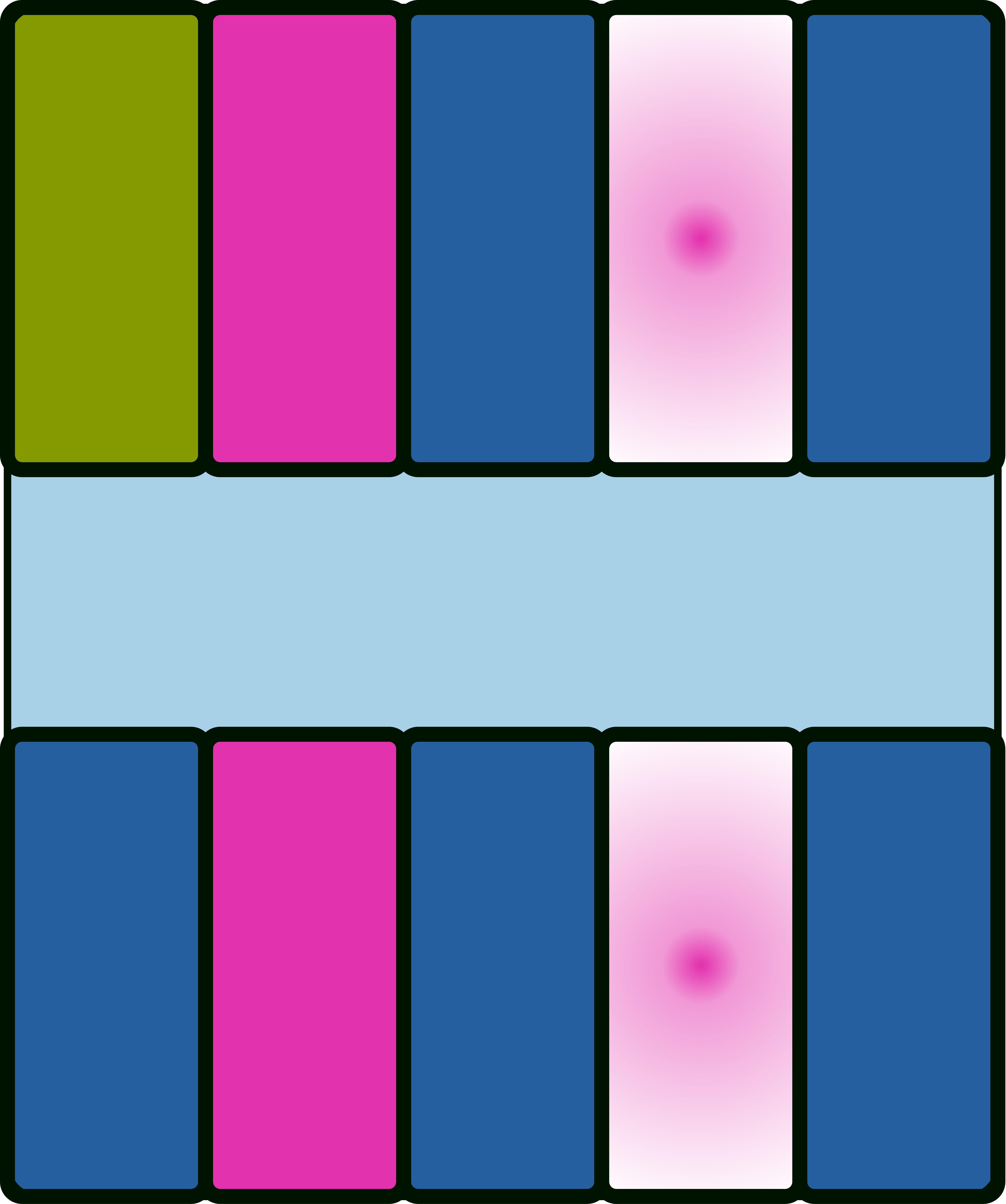}}
    \hspace{2.5cm}
    \subfloat[\label{fig:15bch3} ]{\includegraphics[width=.125\textwidth]{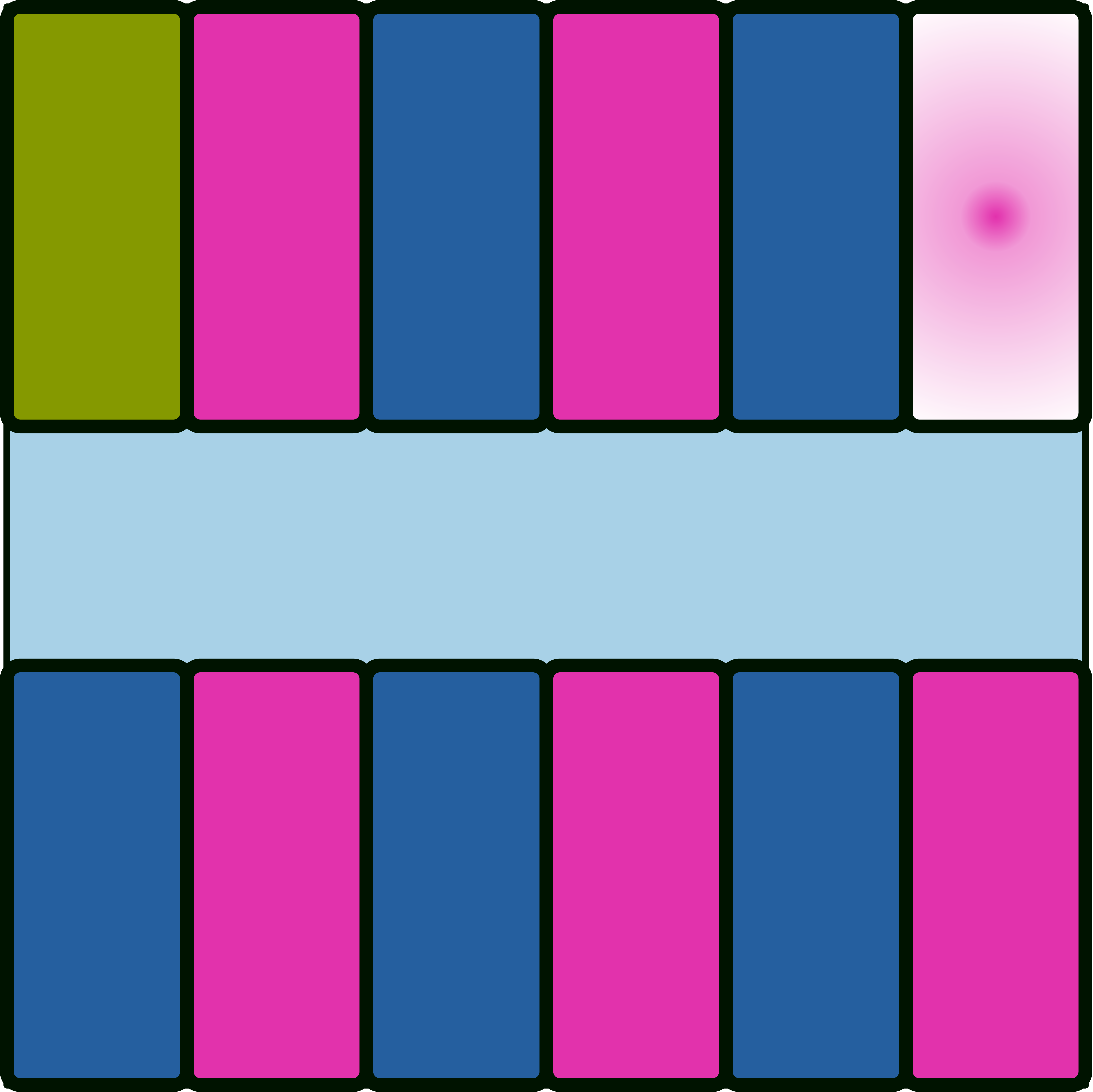}}
    \hspace{0.35cm}
    \subfloat[\label{fig:15golay} ]{\includegraphics[width=.167\textwidth]{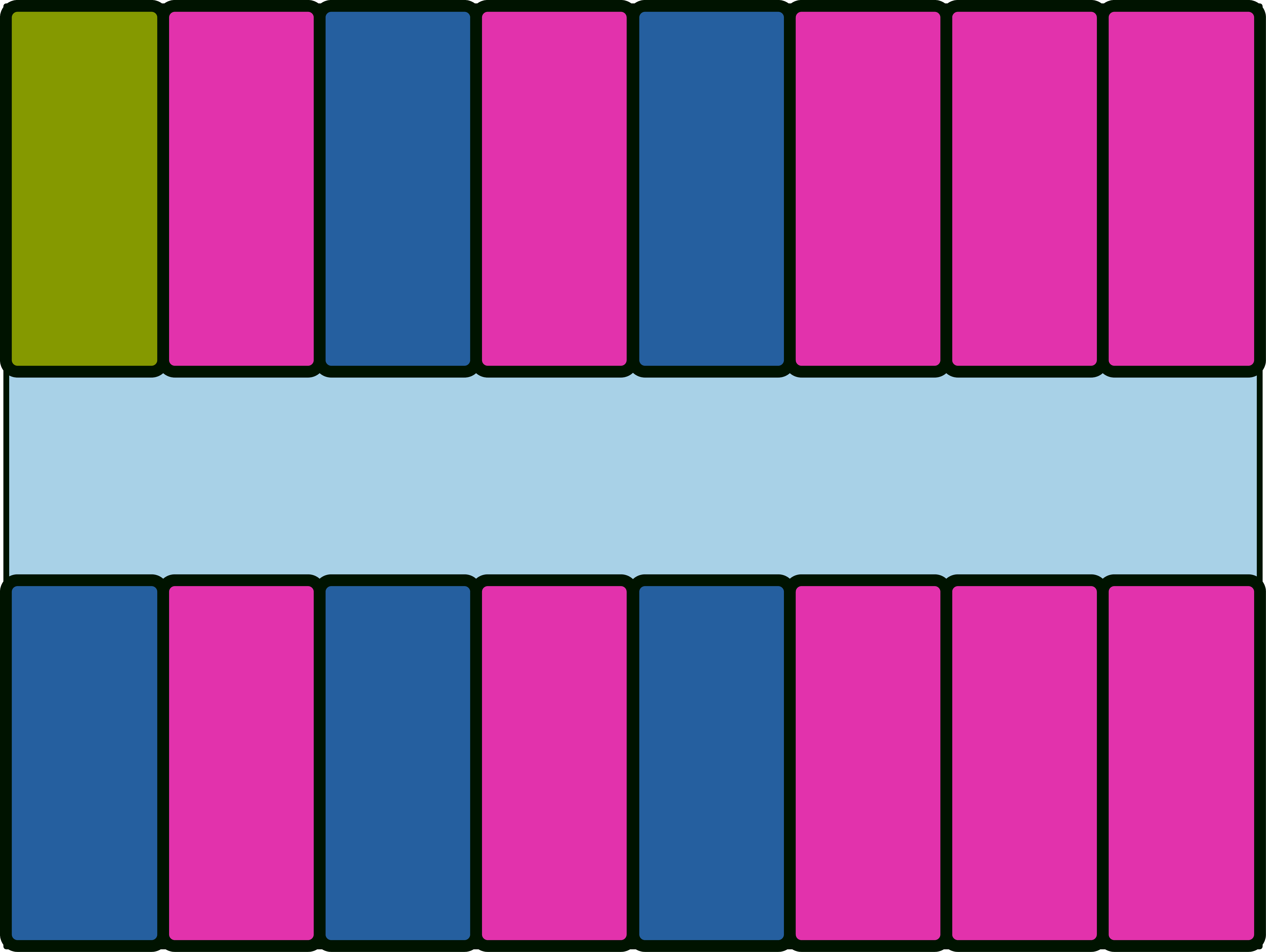}}
    \label{fig:15to1layouts}
    \caption{Layouts of logical qubits for TELS-assisted $15$-to-$1$ state distillation. Data qubits are placed in blue cells. Magic states are in pink cells, where cells with a radial shading are extra cells used to prepare new magic states in parallel with the Pauli measurements. $\ket 0$ ancillas for autocorrected gadgets are placed in the brown cells adjacent to the yellow squares used for twists. Green cells are used to store distilled magic states for use by the core while the next round of distillation occurs. Additional green cells may be required if a distillation tile produces magic states faster than the core consumes them (alternatively, the magic states can be transported to additional tiles surrounding the core). Routing regions between cells are split into grey and blue to show that the relevant lattice surgery operations will not clash. (a) Layout for un-encoded lattice surgery using autocorrected non-Clifford gate gadgets of \cref{fig:PauliAutocorr}. The grey routing region handles the $X \otimes Y$ measurements and the blue routing regions performs $X$-boundary measurements between different logical qubits. (b) Layout for $15$-to-$1$ distillation with TELS, using the $[12,11,2]$ Single Error Detect code. Note that we only need one radial pink cell. However given the geometry of the entire tile, we use the remaining space for another pink radial tile. (c) Layout using the $[15,11,3]$ BCH code, and, (d)~using the $[23,12,7]$ Golay code.}
    \label{fig:15to1layouts}
\end{figure}

\begin{figure}
    \centering
    \subfloat[\label{fig:15uepar} ]{\includegraphics[width=.104\textwidth]{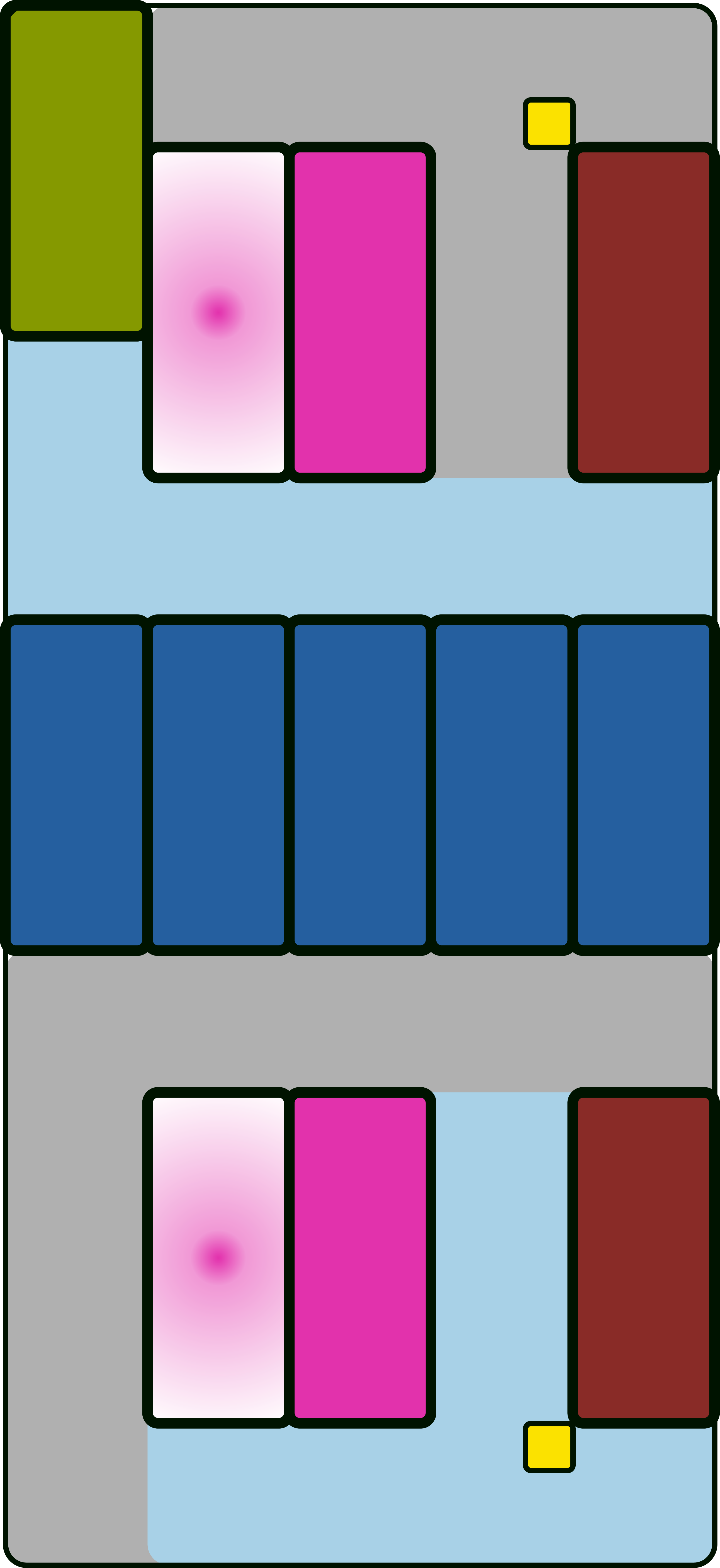}}
    \hspace{25cm}
    \subfloat[\label{fig:15sedparstag1}
    ]{\includegraphics[width=.146\textwidth]{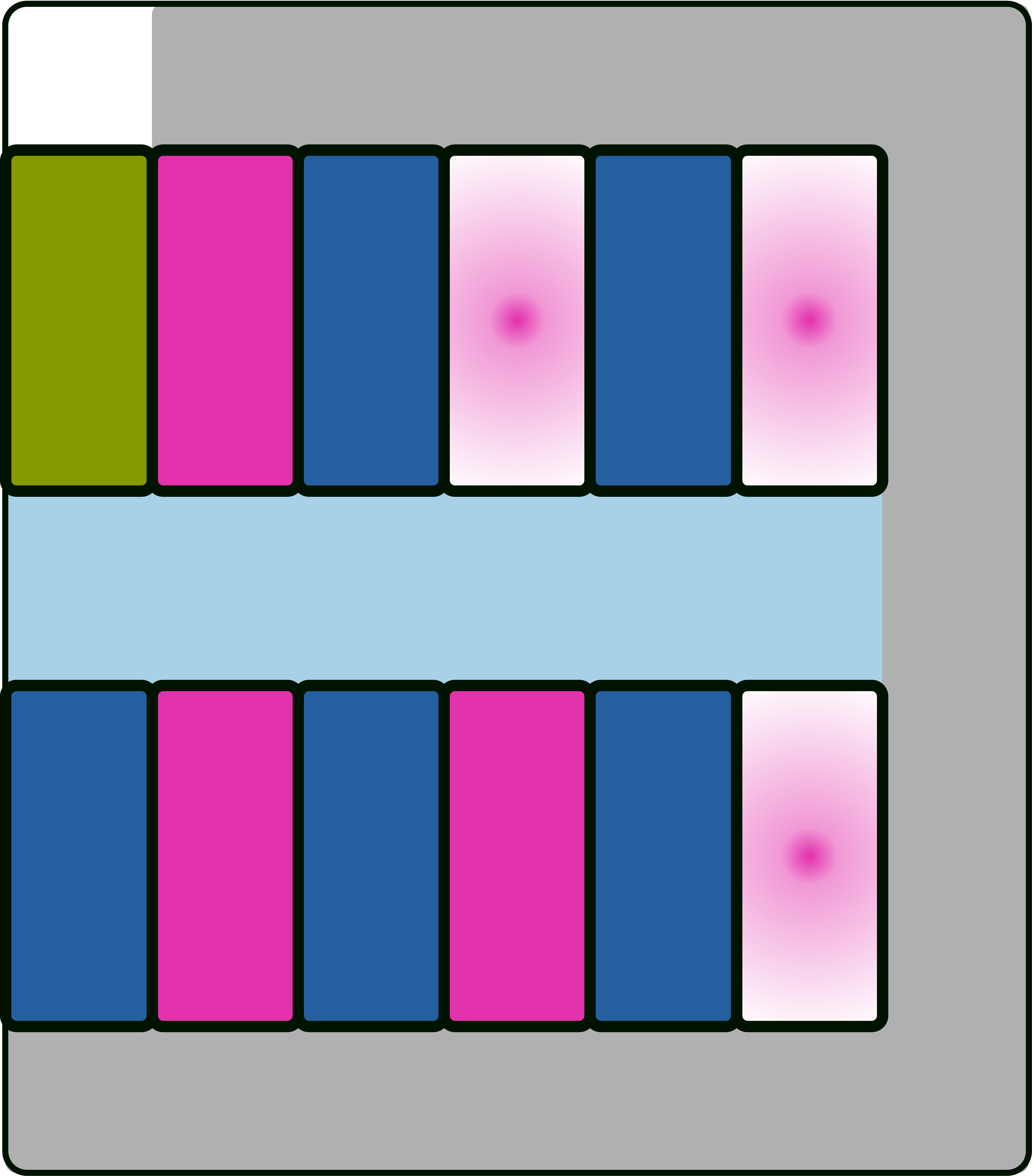}}
    \hspace{0.4cm}
    \subfloat[\label{fig:15sedparstage2} ]{\includegraphics[width=.146\textwidth]{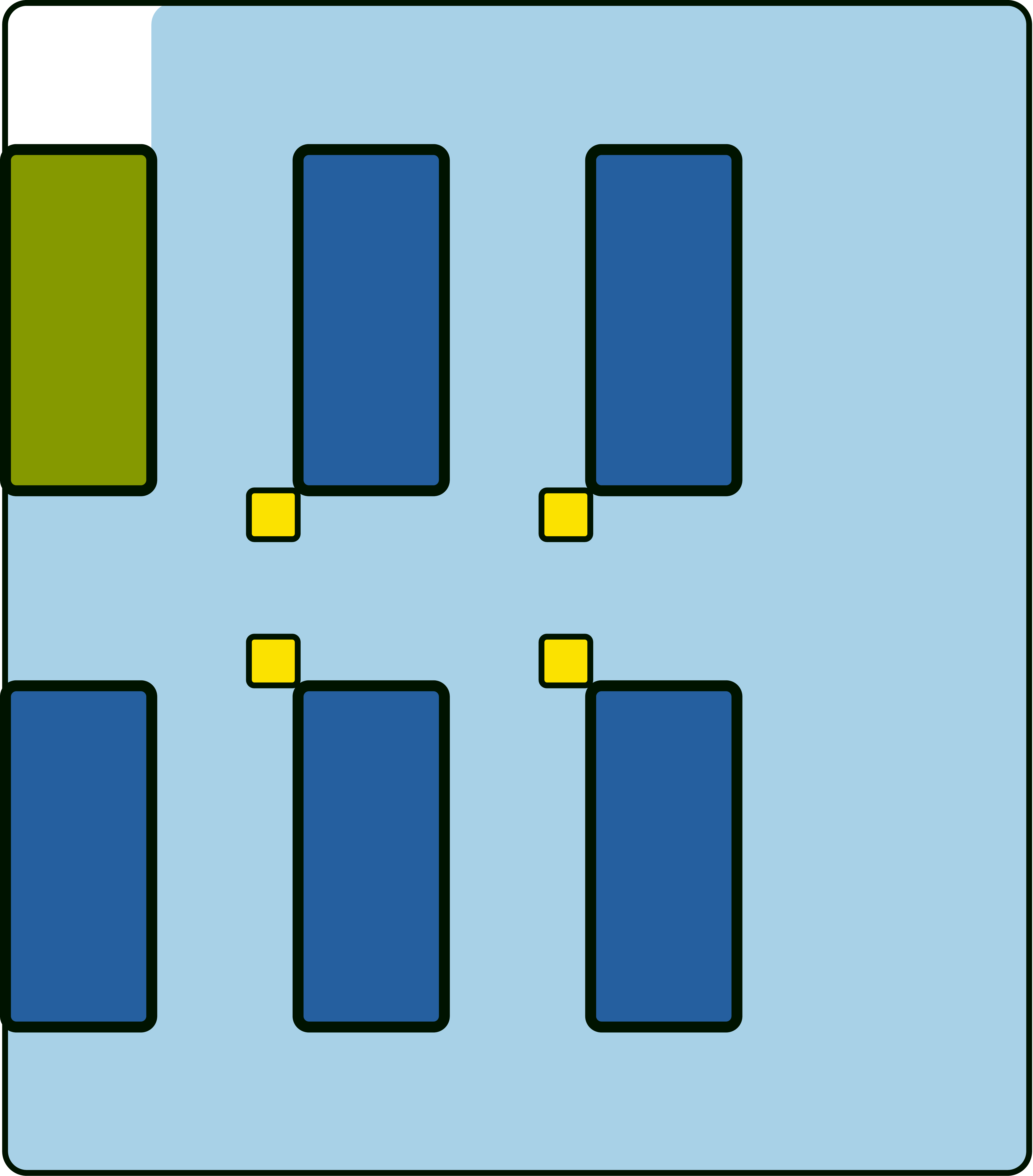}}
    \hspace{1cm}
    \subfloat[\label{fig:15bch3par} ]{\includegraphics[width=.167\textwidth]{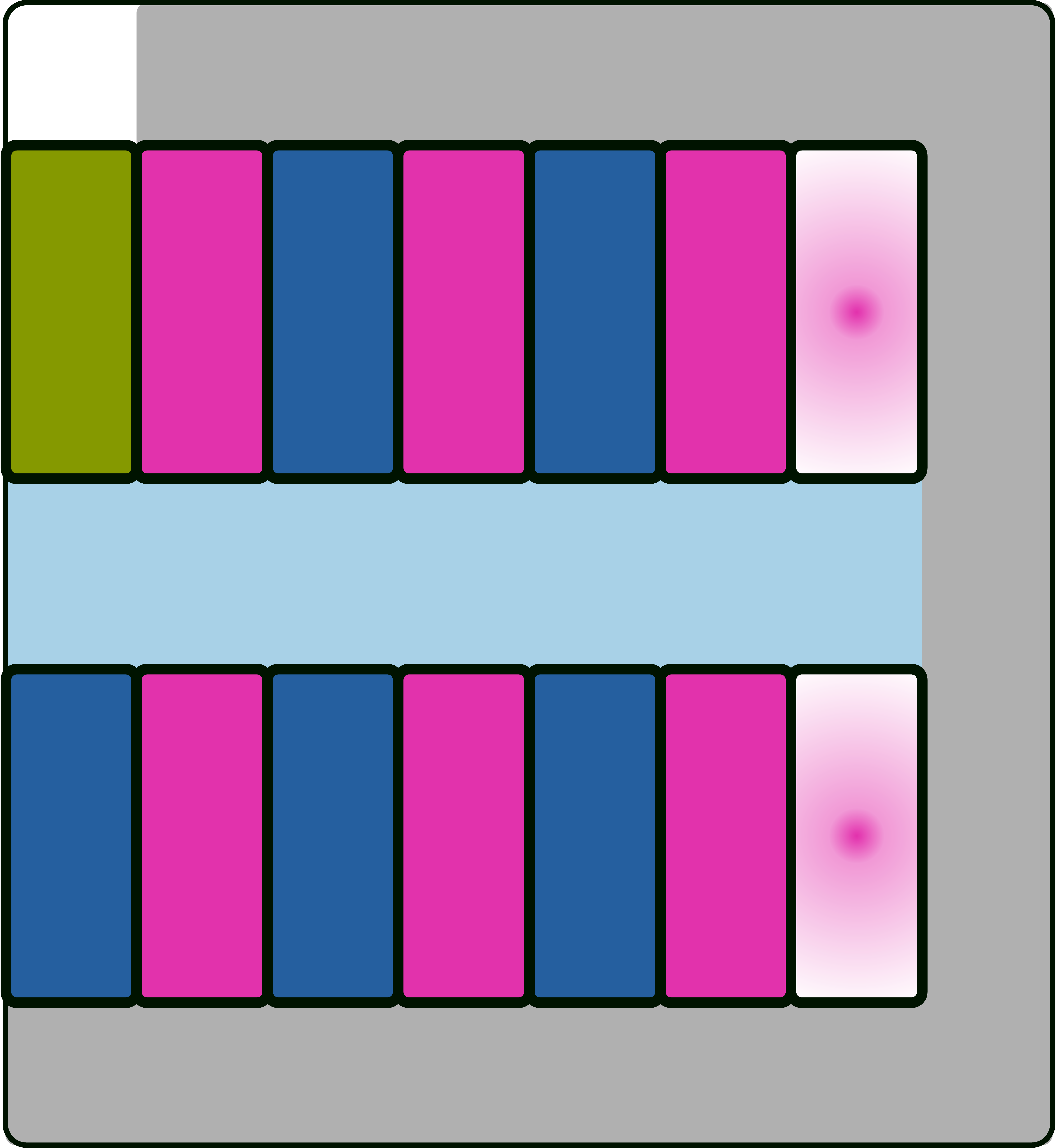}}
    \hspace{0.3cm}
    \subfloat[\label{fig:15golaypar} ]{\includegraphics[width=.205\textwidth]{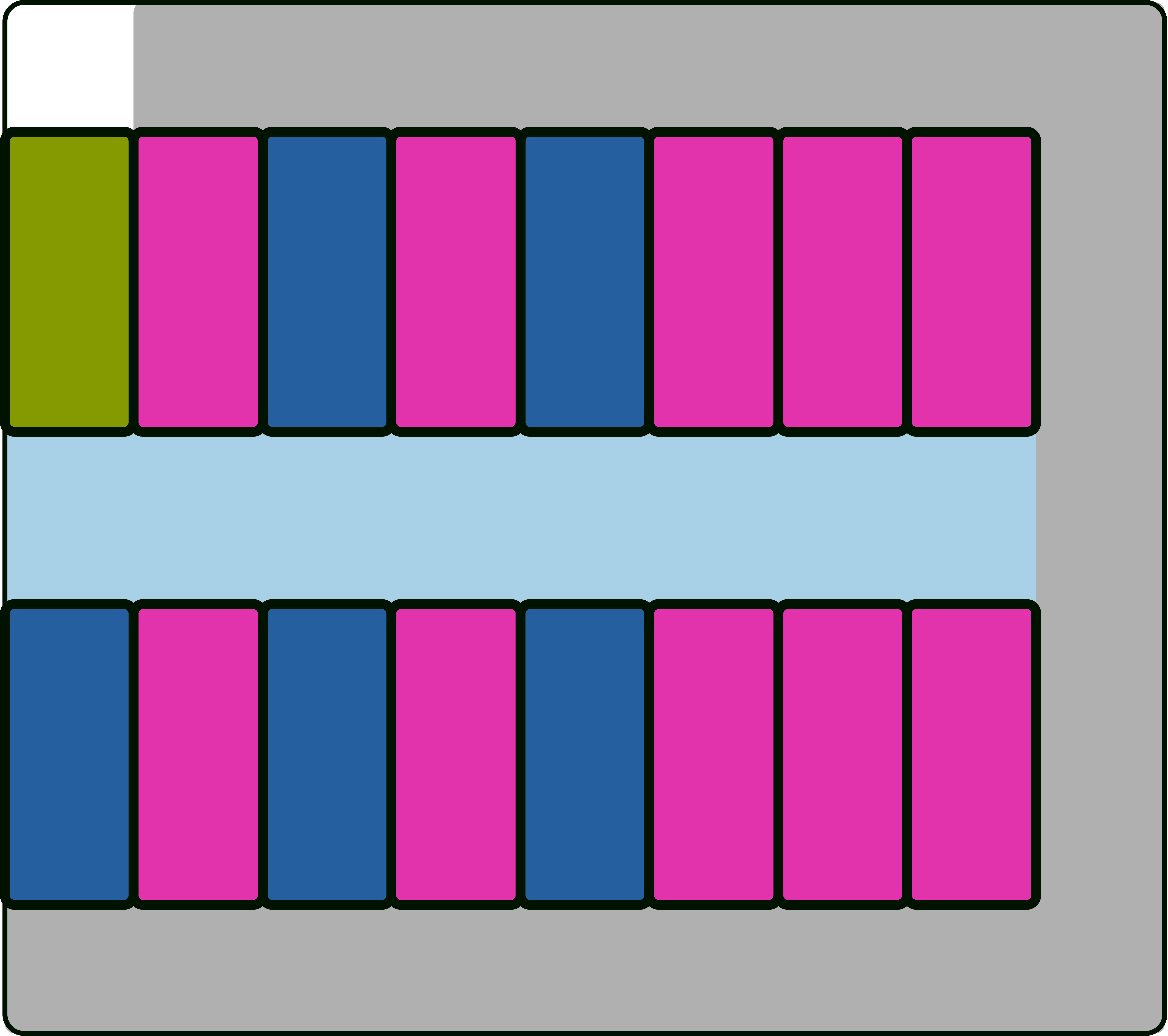}}
    \caption{Layouts of logical qubits for parallelized TELS-based $15$-to-$1$ state distillation protocols. The meaning of each color is described in the caption of \cref{fig:15to1layouts}. (a) Layout for un-encoded lattice surgery, with two routing regions, each accessing one $X$ boundary of the logical qubits. Each routing region has access to a separate magic state and a $\ket 0$ ancilla used in the circuit of \cref{fig:PauliAutocorr}. (b)~Layout for distillation with TELS, using the $[12,11,2]$ Single Error Detect code. Three magic state tiles are held in memory for each pair of parallel Pauli measurements. Then two are discarded and two prepared magic states on other pink cells are used in the following round. (c)~Layout and routing region used for the final multi-qubit Pauli measurements in the Clifford frame distillation protocol. (d)~Layout for parallelized distillation with TELS, using the $[15,11,3]$ BCH code. (e)~Layout for parallelized distillation with TELS using the $[23,12,7]$ Golay code. For (d) and (e), the layouts used to perform the final multi-qubit Pauli operations required by the  Clifford frame can be found in an analogous way from going from  (b) to (c). }
    \label{fig:15to1layoutsparallelized}
\end{figure}

\subsubsection{No temporal encoding} 

\begin{figure*}
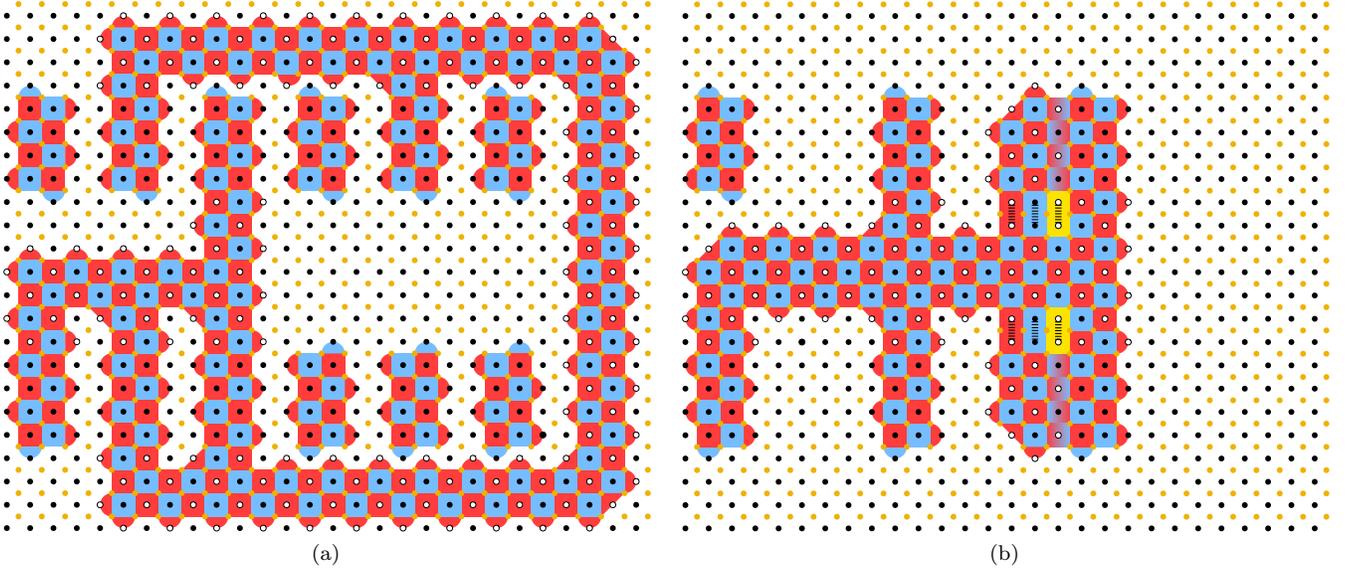

    \centering
    \subfloat[\label{fig:layoutSpec} ]{\includegraphics[width=0.48\textwidth]{images/latsurg1.pdf}}
    \hspace{0.3cm}
    \subfloat[\label{fig:layoutSpecPaulis} ]{\includegraphics[width=0.48\textwidth]{images/latsurg2.pdf}}
    \caption{(a) On the layout of \cref{fig:15sedparstag1}, we show how two separate routing spaces can be used to perform parallel lattice surgery measurements. The logical measurements are $X_1 \otimes X_2 \otimes X_3 \otimes X_{T_{X,1}}$ (in the equatorial routing space) and $X_3 \otimes X_4 \otimes X_{T_{X,1}} \otimes X_{T_{X,2}}$ (in the circumferential routing space). These are the first and second measurements respectively when performing TELS-assisted distillation using the $[12,11,2]$ SED code (see \cref{eq:CyclicG12} of \cref{app:codeconstruction} for the codeword generator matrix). Alternatively, they correspond to the first measurement and the product of the first and second measurements from \cref{fig:Haddist}. The logical patches have code distances $d_x=3$, $d_z=5$. $X$ stabilizers are in red, and $Z$ stabilizers are in blue. The product of the $X$ stabilizers indicated by white vertices gives the parity for the multi-qubit Pauli measurement outcomes. (b) On the same layout, we show how to perform Pauli measurements which are tensor products of $X,Y$, or $Z$ on the data qubits. These measurements are performed after the non-Clifford gates of the distillation protocol. The example in the figure measures $X \otimes Y \otimes X \otimes X \otimes Y$ on the five data qubits. The yellow stabilizers are twist defects that are used to access $Y$ boundaries of logical qubits that are originally defined with only $X$ and $Z$ boundaries, using the techniques shown in Ref.~\cite{Chamberland22b}. Note that the size of the routing space area separating the top and bottom rows of data qubits is taken to be large enough to allow for $Y$ measurements requiring twists. }
\end{figure*}

First, we calculate the space-time cost of the distillation circuit of \cref{fig:Haddist} without temporally encoded lattice surgery. For this, we consider a modified version of the layout used by Litisnki (see Fig. $18$ of Ref.~\cite{Litinski19}), as shown in \cref{fig:15ue}. In this figure, the five blue cells at the bottom correspond to the data qubits of \cref{fig:Haddist}; the pink cells are used to store magic states for performing the $\pi / 8$ multi-qubit Pauli measurements; radial pink cells are used to inject new magic states for subsequent Pauli measurements (thus preventing time delays due to state injection); brown cells with an adjacent yellow square for twists are $\ket 0$ ancillas used in the autocorrected non-Clifford gate gadget of \cref{fig:PauliAutocorr}; and green cells are used to store the distilled magic state from the previous round of distillation, so that it may be accessed by the core of the quantum computer.  

There are two cells assigned for magic states. Without TELS, only one magic state is used per non-Clifford gate (note that one $X$ boundary has access to the data qubits, and the other boundary has access to the $Y$ boundary of the $\ket 0$ ancilla, hence these lattice surgery measurements may be performed in parallel). However, at any given time, one magic state cell will take part in a non-Clifford gate, and the other will be used to prepare a noisy magic state for the subsequent non-Clifford gate. This is required as it takes a non-trivial amount of time to prepare a noisy magic state (roughly 2 syndrome measurement rounds as described in \cref{subsec:MagicInject}). In \cref{fig:15to1layouts,fig:15to1layoutsparallelized,fig:125to1layouts,fig:116to12layouts,fig:114to14layouts}, we use the radial-shaded pink cell to denote the extra cells needed for this simultaneous magic state preparation.  

Since we assume classical processing is instantaneous, only one qubit cell is assigned for the $\ket 0$ ancilla used in the autocorrected gadget. Note however, that in Ref.~\cite{Litinski19} and Ref.~\cite{EarlParallel}, it was shown that for finite decoding times, additional $\ket{0}$ ancilla qubits can be used to offset the extra time cost associated with decoding all syndrome measurement rounds associated with the previous lattice surgery operations. However with classical parallelization and pre-decoders \cite{ChambsLocalNN22,BrownLocalPre22,EarlParallel,ChaoPrallel22}, such additional ancillas may be unnecessary. 

Note that we do not need to shuttle the distilled magic state from a blue cell to a green cell. We design the distillation tile such that the output magic state is in the right most blue cell. In the next round of distillation, the layout is mirrored about the equator and the magic state cell now becomes a green cell with core access. In this way, the distillation protocol may be restarted without any shuttling delays. Using the procedure detailed in \cref{app:algoSTcosts}, we determined that the lattice surgery measurement distance must be $d_m=9$ to obtain magic states with logical failure rate at most $10^{-10}$.

For every data qubit cell, there are two accessible $X$ boundaries. We can almost trivially speed up the protocol by a factor of two by assigning new routing space and ancilla cells that access the second $X$ boundary of the data qubits. This new hardware layout allows us to perform two multi-qubit Pauli measurements in parallel. This idea was originally proposed in Ref.~\cite{Litinski19}. We show a layout that performs lattice surgery measurements two at a time without temporal encoding in \cref{fig:15uepar}. Note that with this layout, a distilled magic state present on a blue data cell must be shuttled to a green storage cell. There is an extra time cost associated with the shuttling operation. We do not include the time cost of shuttling in \cref{tab:STcosts} as it may still be possible to eliminate shuttling using a more clever layout. In any case, the layout without parallel measurements yields a smaller space-time cost.

\subsubsection{TELS-Single Error Detect $\mathbf{[12,11,2]}$} Next, we consider a distillation protocol that uses TELS to execute the non-Clifford gates, with the protocol described in \cref{sec:Cliff}. We first determine the space-time cost using the Single Error Detect $[12,11,2]$ code. The codeword generator matrix for this code is given in \cref{eq:CyclicG12} of \cref{app:codeconstruction}. If the measurements are performed sequentially, as in the layout of \cref{fig:15sed}, one routing space with access to the $X$ boundaries of all the qubits will suffice. For a faster distillation tile that performs measurements two at a time, the measurements may be performed using extra routing space as shown in \cref{fig:15sedparstag1}. Note that we now use the non-Clifford gadget of \cref{fig:Pnoncliff} to perform the $\pi / 8$ rotations since we perform distillation in the Clifford frame. This frees up space, as we do not need to allocate qubits for a $\ket{0}$ ancilla with $Y$ boundary access. This can reduce the routing space and the number of logical qubit cells needed. On the other hand, TELS incurs a larger space cost as more magic states need to be held in memory.

In \cref{fig:15sedparstag1}, we separate the routing space into grey and light blue regions to show non-intersecting routing areas for the two parallel multi-qubit Pauli measurements. Each of these routing spaces has access to the $X$ boundaries of all the data and magic state cells involved in the distillation. In \cref{fig:layoutSpec}, we show the routing space regions that are used to perform the lattice surgery measurements corresponding to the first two parallelizable Paulis when the $[12,11,2]$ code is used for TELS. In the first syndrome measurement round of the lattice surgery measurement, stabilizers with white vertices yield random outcomes due to the gauge fixing step~\cite{Vuillot19}. The lattice surgery measurement outcomes are then the error-corrected measurement values corresponding to the $X$ stabilizers in the respective routing regions. Using \cref{app:algoSTcosts}, we determined that $d_m = 5$ is sufficient for $\delta^{(M)} = 10^{-10}$ when using the $[12,11,2]$ code for TELS.

Access to only the $X$ boundaries of the data cells is sufficient for the first stage of the protocol, which is the temporally encoded measurements for the non-Clifford gates. In the second stage, we must perform multi-qubit Pauli measurements which are tensor products of $X$, $Y$ and $Z$. In \cref{fig:15sedparstage2}, we show how to perform these measurements on the same layout without the shuffling around of surface code patches. On the distilled magic state (bottom left blue cell), the multi-qubit measurements only need $X$ boundary access. The remaining data qubits will need at least one accessible $Z$ and $Y$ boundary. For the $15$-to-$1$ distillation protocol, there are at most $4$ multi-qubit $\pi/2$ Pauli measurements. Since these measurement may require access to different types of boundaries on each data cell, they cannot in general be performed in parallel. Hence the entire routing space (in light blue) is used to perform these measurements. As shown in \cref{sec:Cliff}, this set of measurements corresponds to the second PP set. Since there are four single-qubit measurements at the end of \cref{fig:Haddist}, there are at most $4$ measurements in this second PP set. We perform these measurements using TELS with a $[5,4,2]$ Single Error Detect code and with measurement distance $d_m = 5$.

When using the $[12,11,2]$ code for the lattice surgery measurements of the non-Clifford gates, we define $G$ as a cyclic code where at most two magic states need to be accessed simultaneously for each Pauli measurement. After each measurement, a magic state cell can be reset and reused for a future non-Clifford gate. If Pauli measurements are performed two at a time, three magic states will need to be concurrently held in memory. These are the solid pink tiles of \cref{fig:15sedparstag1}. In addition, for each subsequent pair of measurements, at least two injected magic states are required, which is why there is not just one additional magic state tile associated for injection but three. One of them may be removed, but then distillation tiles will either not be rectangular or will contain wasted physical qubits. In any case the extra cell for injection can ensure there is always a magic state injected and ready for a non-Clifford gate. Note however that since keeping track of Clifford frames requires occasionally performing $Y$ measurements when using the distilled magic states in the core, the extra pink radial cell could also be used to store the ancilla needed to perform the $Y$ measurement\footnote{Performing a $Y$ measurement on a surface code patch can be done in various ways. For instance, one can perform a logical phase gate, followed by measuring all the data qubits in the $X$ basis. However, performing a logical phase gate on a two-dimensional planar architecture with the surface code requires additional routing space and measurements involving twists (see for instance Fig. 23 of Ref.~\cite{PsiQLaticceSurgery21}. Alternatively, one could use an ancilla prepared in the logical $\ket{0}$ state, and perform a $Y \otimes Z$ measurement to get the parity of the $Y$ measurement outcome.}.

\subsubsection{Other TELS protocols}

In addition to the Single Error Detect code, we considered distillation with a distance-$3$ BCH code and the distance-$7$ classical Golay code. In \cref{fig:15bch3}, we show a layout for a distillation tile that performs TELS with a classical $[15,11,3]$ BCH code. The time cost can be decreased by performing the lattice surgery measurements corresponding to the non-Clifford gates two at a time. A layout with sufficient routing space for this is shown in \cref{fig:15bch3par}. Note that since there are two disjoint routing spaces (in blue and grey), the set of $n=15$ Pauli measurements can be performed in the time required for $8$ sequential measurements. To obtain the time cost shown in \cref{tab:STcosts}, we used $d_m = 3$.  Similarly, in \cref{fig:15golay,fig:15golaypar}, we show layouts for $15$-to-$1$ distillation tiles that perform TELS with a classical $[23,12,7]$ Golay code. Here, we used the parameters $d_m = 3$ and $c=2$ (where $c$ is the maximum weight of classical errors that are corrected in a TELS protocol) to obtain the time costs shown in \cref{tab:STcosts}. For the $15$-to-$1$ distillation protocol, implementing TELS using the classical Golay code does not allow for smaller space or time costs. However, for a small enough physical error rate, it is sufficient to consider a measurement distance $d_m=1$, which allows for a smaller time cost than any other lattice surgery protocol.

The space requirements of all the above layouts are described as functions of $d_x$ and $d_z$ in \cref{app:constants}. Additional constants in \cref{app:constants} can be used with the procedure of \cref{app:algoSTcosts} to determine all the minimum distances (spacelike and timelike) for the distillation protocols.

\subsection{$125$-to-$3$ distillation}
\label{sec:125to3}

The $125$-to-$3$ magic state distillation protocol is obtained from a triorthogonal CSS quantum $\llbracket 125,3,5\rrbracket$ code. This code is constructed by puncturing the $[128,29,32]$ Reed-Muller code at any three locations~\cite{Haah18}. As a result, the quantum code will contain $3$ logical qubits, $96$ $X$-type stabilizers and $26$ $Z$-type stabilizers. After applying the circuit transformation from a gate-based model to the PBC model~\cite{Litinski19}, we are left with a sequence of $99$ commuting Pauli measurements on $29$ logical qubits, three of which will finally become the distilled magic states. These $99$ measurements form a size-$99$ PP set.

In this paper, we will consider using this protocol in a regime where the physical error rate is $p=10^{-3}$ and our target is to distill magic states with logical error probability at most $\delta^{(M)}=10^{-15}$. Using the noise model described in \cref{subsec:MagicInject} for the injection of magic states, we apply the analysis in Ref.~\cite{Litinski19magic} to determine the logical failure probability per output magic state for one round of a $125$-to-$3$ distillation scheme,
\begin{align}
    p_L^{(M)}  = & \frac{1}{3} 31 \Big ( (\epsilon_{\text{L,Z}})^5 + \frac{1}{2} 10(\epsilon_{\text{L,Z}})^4 \epsilon_{\text{L,X}} \nonumber \\
    & \quad + \frac{1}{4} 40 (\epsilon_{\text{L,Z}})^3 (\epsilon_{\text{L,X}})^2
    + \frac{1}{8} 80(\epsilon_{\text{L,Z}})^2 (\epsilon_{\text{L,X}})^3 \nonumber \\
    &\quad + \frac{1}{16} 80 \epsilon_{\text{L,Z}} (\epsilon_{\text{L,X}})^4
    + \frac{1}{32} 32(\epsilon_{\text{L,X}})^5 \Big ) \nonumber \\
    = &\frac{31(1+\eta)^5}{729 \eta^5} p^5 \: .
\end{align}
For $p=10^{-3}$ and $\eta=100$, the probability that the distillation succeeds is $1 - p_D^{(M)} = (1-\epsilon_L)^{125} = 0.9584$ and $p_L^{(M)} = 4.47 \times 10^{-17}$. As this is sufficiently below $\delta^{(M)}$, the lattice surgery measurements used to execute the distillation protocol must be modeled with measurement distance large enough to allow for distilled magic states of logical error rate at most $\delta{(M)}$. Using the procedure in \cref{app:algoSTcosts}, we determined that the minimum spacelike distances are $d_x=13$ and $d_z=25$. 

\begin{figure}
    \centering
    \subfloat[\label{fig:125ue} 
    ]{\includegraphics[width=.2568\textwidth]{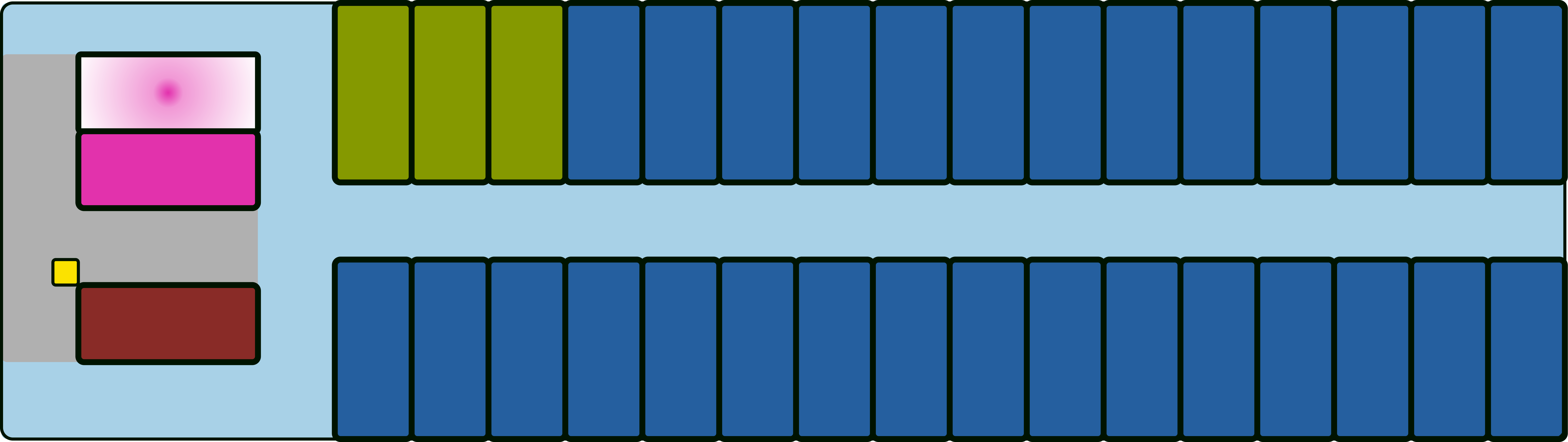}}
    \hspace{0.1cm}
    \subfloat[\label{fig:125uepar}
    ]{\includegraphics[width=.31\textwidth]{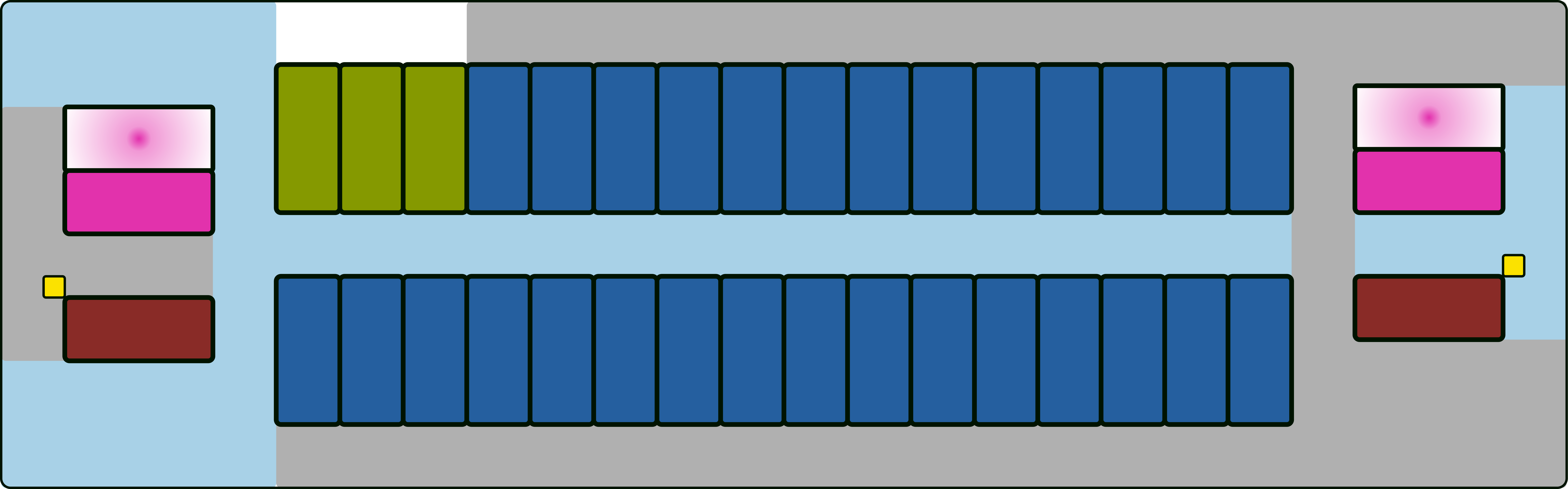}}
    \hspace{0.1cm}
    \subfloat[\label{fig:125bch7par}
    ]{\includegraphics[width=.379\textwidth]{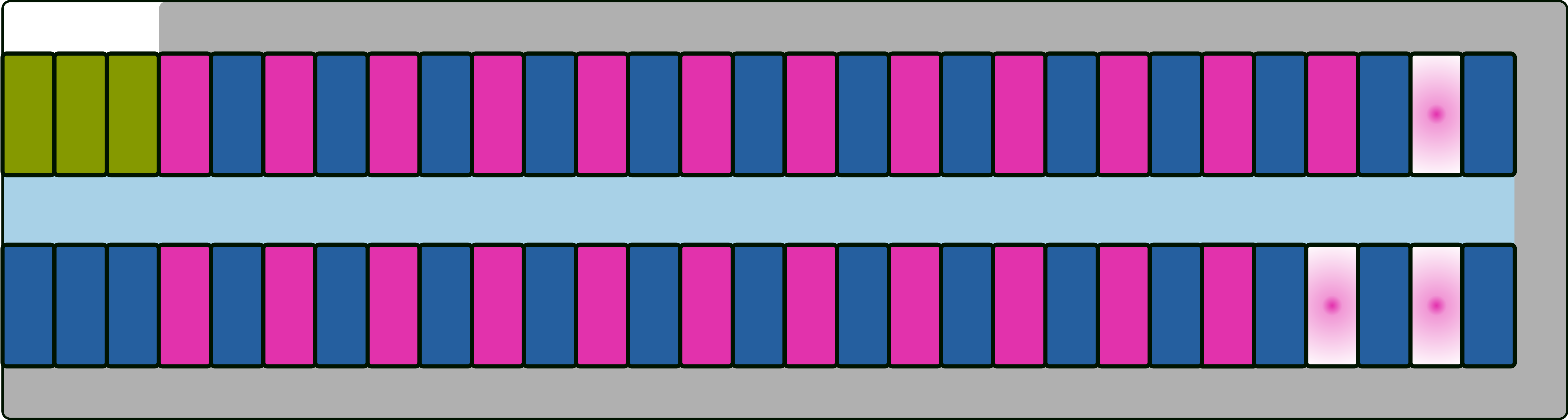}}
    \hspace{0.1cm}
    \subfloat[\label{fig:125bch9par} ]{\includegraphics[width=.417\textwidth]{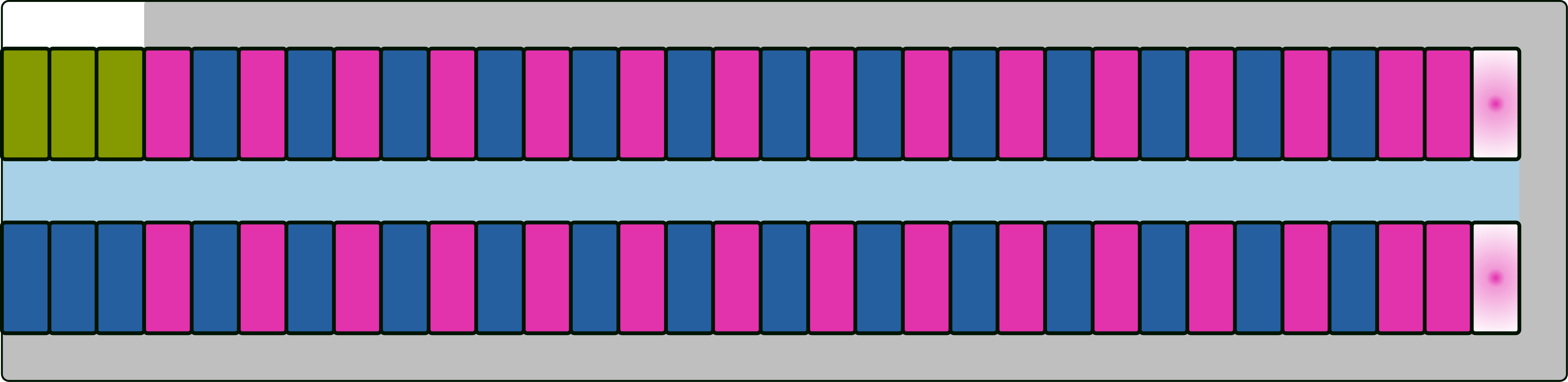}}
    \caption{Layouts of logical qubits for lattice-surgery-based $125$-to-$3$ magic state distillation. Cell color legend in caption of \cref{fig:15to1layouts}.~(a) Layout for a distillation tile without temporally encoded lattice surgery measurements, using an auto-corrected non-Clifford gate gadget. The blue routing space allows Pauli $X$-type measurements as there is access to the $X$ boundaries of all the cells. The gray routing region allows access to a $\ket 0$ ancilla with $Y$ boundary access. This region contains hardware to execute a non-Clifford gate gadget.~(b) Layout for a distillation tile that performs Pauli measurements two at a time, without temporal encoding. The long blue routing space performs one set of measurement with the auto-corrected non-Clifford gadget hardware at the left, and the large grey routing space uses the gadget hardware on the right.~(c) Layout for a distillation tile performing temporally encoded lattice surgery with the $[127,106,7]$ BCH code. Non-Clifford gates are performed two at a time, using separate routing spaces shown in gray and blue.~(d) Layout for a TELS-assisted distillation tile using the $[127,99,9]$ BCH code, performing non-Clifford gates two at a time.}
    \label{fig:125to1layouts}
\end{figure}

We show two layouts for distillation tiles that do not use TELS in \cref{fig:125ue} and \cref{fig:125uepar}. These layouts perform distillation in the Pauli frame as described in Ref.~\cite{Litinski19} using auto-corrected non-Clifford gadgets (\cref{fig:PauliAutocorr}). On the layout of \cref{fig:125ue}, $99$ Pauli measurements are performed, each with measurement distance $d_m = 23$ (also derived using \cref{app:algoSTcosts}). On the layout of \cref{fig:125uepar}, Pauli measurements can be performed two at a time. Hence the time required is only the time for $50$ sequential lattice surgery measurements with measurement distance $d_m = 23$. 

In \cref{fig:125bch7par}, we show a layout for a distillation tile that performs TELS with a classical $[127,106,7]$ BCH code. Note that since there are two disjoint routing spaces (in blue and grey), the set of $n=127$ Pauli measurements can be performed in the time required for $64$ sequential measurements. This way, the time cost is nearly halved, with only a minor increase to the height of the distillation tile (two rows of routing space of height $d_x$). To obtain the time cost shown in \cref{tab:STcosts}, we used $d_m = 7$ and $c=2$, where $c$ is the maximum weight of classical errors that are corrected in a TELS protocol. Only classical errors of weight greater than or equal to three triggered detection events. Similarly, in \cref{fig:125bch9par}, we show a layout for a $125$-to-$3$ distillation tile that performs TELS with a classical $[127,99,9]$ BCH code. Here, we used the parameters $d_m = 5$ and $c=1$ to obtain the time cost shown in \cref{tab:STcosts}. For both of the layouts that use TELS, we have only discussed the TELS code used to execute the non-Clifford gates of \cref{sec:Cliff}. For the $125$-to-$3$ distillation protocol, there may be at most $26$ additional Pauli measurements to perform due to the conditional Clifford corrections. The results of these measurements are used to detect if there are errors in the final distilled magic states. These Pauli measurements form a PP set of size at most $26$. For the worst case where the PP set is of size 26, the TELS protocol  uses the $[31,26,3]$ BCH code with lattice surgery measurement distance $d_m = 9$.

The space requirements of all the above layouts are described as functions of $d_x$ and $d_z$ in \cref{app:constants}. Additional constants in \cref{app:constants} can be used with the procedure of \cref{app:algoSTcosts} to determine all the minimum distances (spacelike and timelike) for the distillation protocols.

\subsection{$116$-to-$12$ distillation}
\label{sec:116to12}

The $116$-to-$12$ magic state distillation protocol is obtained from a triorthogonal CSS quantum $\llbracket 116,12,4\rrbracket$ code. This code is constructed by puncturing the $[128,29,32]$ Reed-Muller code at a specific set of $12$ locations as shown in Ref.~\cite{Haah18}. As a result, the quantum code will contain $12$ logical qubits, $87$ $X$-type stabilizers and $17$ $Z$-type stabilizers. After applying the circuit transformation from a gate-based model to the PBC model~\cite{Litinski19}, we are left with a sequence of $99$ commuting Pauli measurements on $29$ logical qubits, twelve of which will finally become the distilled magic states. These $99$ measurements form a size-$99$ PP set.

In this paper, we will consider using this protocol in a regime where the physical error rate is $p=10^{-4}$ and our target is to distill magic states with logical error probability at most $\delta^{(M)}=10^{-15}$. Using the noise model described in \cref{subsec:MagicInject} for the injection of magic states, we apply the analysis in Ref.~\cite{Litinski19magic} to determine the logical failure probability per output magic state for one round of a $116$-to-$12$ distillation scheme,
\begin{align}
    p_L^{(M)}  = & \frac{1}{12} 495 \Big ( (\epsilon_{\text{L,Z}})^4 + \frac{1}{2} 8(\epsilon_{\text{L,Z}})^3 \epsilon_{\text{L,X}} \nonumber \\
    & \quad + \frac{1}{4} 24 (\epsilon_{\text{L,Z}})^2 (\epsilon_{\text{L,X}})^2
    + \frac{1}{8} 32(\epsilon_{\text{L,Z}}) (\epsilon_{\text{L,X}})^3 \nonumber \\
    &\quad + \frac{1}{16} 16 (\epsilon_{\text{L,X}})^4
     \Big ) \nonumber \\
    = &\frac{495(1+\eta)^4}{972 \eta^4} p^4 \: .
\end{align}
For $p=10^{-4}$ and $\eta=100$, the probability that the distillation succeeds is $1 - p_D^{(M)} = (1-\epsilon_L)^{116} = 0.9961$ and $p_L^{(M)} = 5.3 \times 10^{-17}$. As this is sufficiently below $\delta^{(M)}$, the lattice surgery measurements used to execute the distillation protocol must be modeled with measurement distance large enough to allow for distilled magic states of logical error rate at most $\delta{(M)}$. Using the procedure in \cref{app:algoSTcosts}, we determined that the minimum spacelike distances are $d_x=9$ and $d_z=15$. 

\begin{figure}
    \centering
    \subfloat[\label{fig:116ue}
    ]{\includegraphics[width=.32\textwidth]{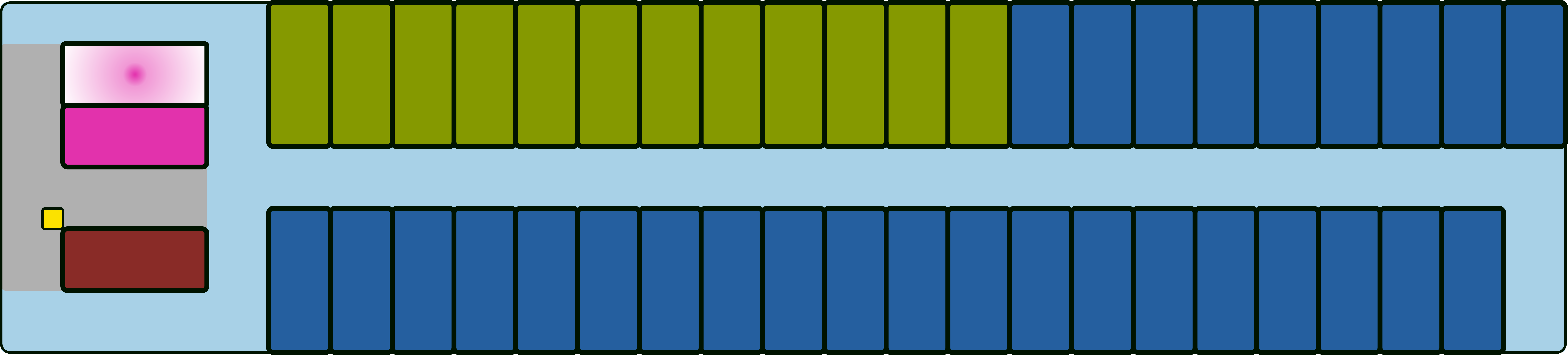}}
    \hspace{0.1cm}
    \subfloat[\label{fig:116uepar}
     ]{\includegraphics[width=.3746\textwidth]{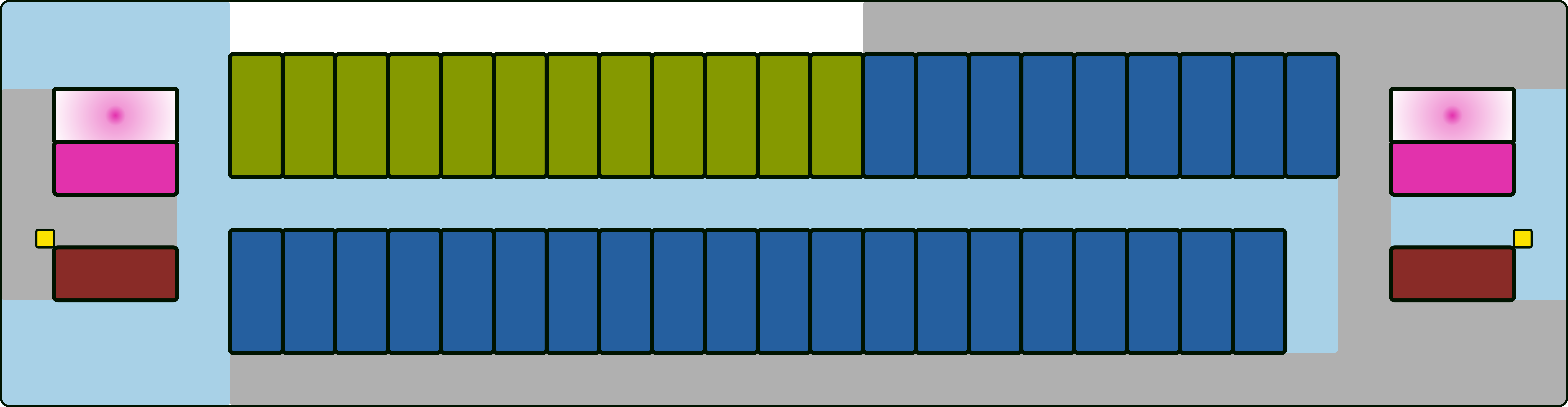}}
    \hspace{0.1cm}
    \subfloat[\label{fig:116zett5par}
    ]{\includegraphics[width=.3915\textwidth]{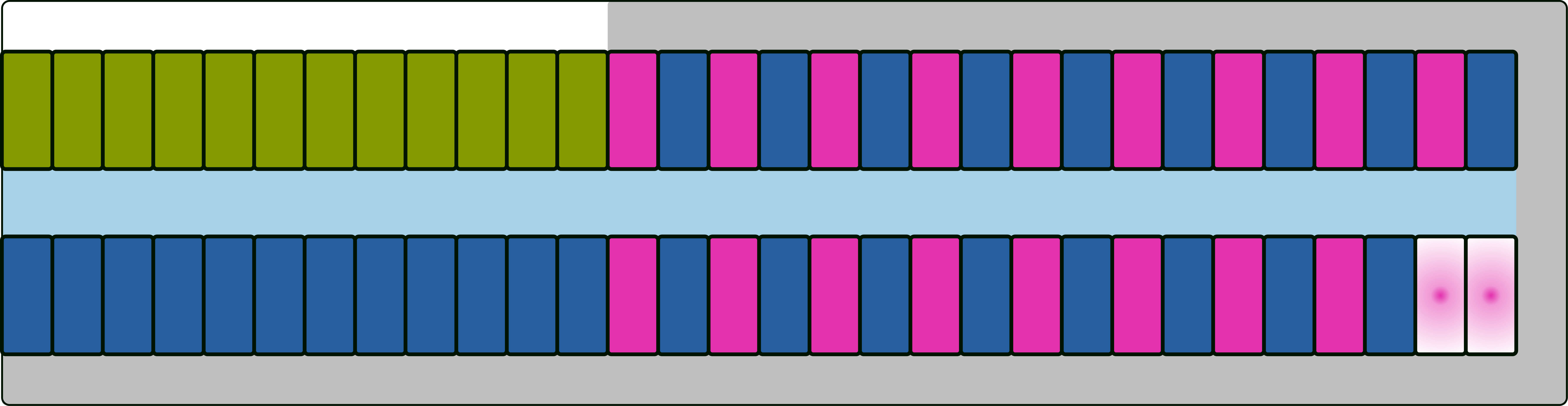}}
    \hspace{0.1cm}
    \subfloat[\label{fig:116bch9par} ]{\includegraphics[width=.48\textwidth]{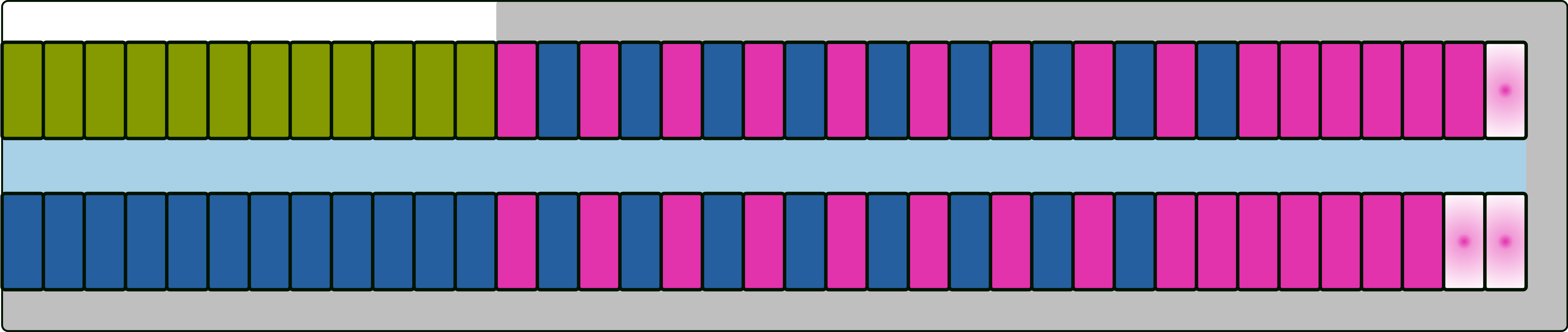}}
    \caption{Layouts of logical qubits for lattice-surgery-based $116$-to-$12$ magic state distillation. Cell color legend in caption of \cref{fig:15to1layouts}.~(a) Layout for a distillation tile without temporally encoded lattice surgery measurements, using an auto-corrected non-Clifford gate gadget.~(b) Layout for a distillation tile that performs Pauli measurements two at a time, without temporal encoding.~(c) Layout for a distillation tile performing temporally encoded lattice surgery with the $[129,114,6]$ Zetterberg code. Non-Clifford gates are performed two at a time, using separate routing spaces shown in gray and blue.~(d) Layout for a TELS-assisted distillation tile using the $[127,99,9]$ BCH code, performing non-Clifford gates two at a time.}
    \label{fig:116to12layouts}
\end{figure}

We show two layouts for distillation tiles that do not use TELS in \cref{fig:116ue} and \cref{fig:116uepar}. These layouts perform distillation in the Pauli frame as described in Ref.~\cite{Litinski19} using auto-corrected non-Clifford gadgets (\cref{fig:PauliAutocorr}). On the layout of \cref{fig:116ue}, $99$ Pauli measurements are performed, each with measurement distance $d_m = 13$. On the layout of \cref{fig:125uepar}, Pauli measurements can be performed two at a time. Hence the time required is only the time for $50$ sequential lattice surgery measurements with measurement distance $d_m = 13$. 

In \cref{fig:116zett5par}, we show a layout for a distillation tile that performs TELS with a classical $[129,114,6]$ Zetterberg code. Note that since there are two disjoint routing spaces (in blue and grey), the set of $n=129$ Pauli measurements can be performed in the time required for $65$ sequential measurements. To obtain the time cost shown in \cref{tab:STcosts}, we used $d_m = 3$ and $c=0$. Similarly, in \cref{fig:116bch9par}, we show a layout for a $116$-to-$12$ distillation tile that performs TELS with a classical $[127,99,9]$ BCH code. Here, we used the parameters $d_m = 3$ and $c=2$ to obtain the time cost shown in \cref{tab:STcosts}. As discussed in \cref{sec:Cliff}, we must also execute a second PP set, now of maximum size $17$, due to the conditional Clifford corrections. The results of these measurements are used to detect if there are errors in the final distilled magic states. These Pauli measurements form a PP set of maximum size $17$, and in the worst case, the TELS protocol used is a $[43,17,7]$ BCH code with lattice surgery measurement distance $d_m = 3$ and $c=2$.

The space requirements of all the above layouts are described as functions of $d_x$ and $d_z$ in \cref{app:constants}. Additional constants in \cref{app:constants} can be used with the procedure of \cref{app:algoSTcosts} to determine all the minimum distances (spacelike and timelike) for the distillation protocols.

\subsection{$114$-to-$14$ distillation}
\label{sec:114to4}

The $114$-to-$14$ magic state distillation protocol is obtained from a triorthogonal CSS quantum $\llbracket 114,14,3\rrbracket$ code. This code is constructed by puncturing the $[128,29,32]$ Reed-Muller code at a specific set of $14$ locations as shown in Ref.~\cite{Haah18}. As a result, the quantum code will contain $14$ logical qubits, $85$ $X$-type stabilizers and $15$ $Z$-type stabilizers. After applying the circuit transformation from a gate-based model to the PBC model~\cite{Litinski19}, we are left with a sequence of $99$ commuting Pauli measurements on $29$ logical qubits, fourteen of which will finally become the distilled magic states. These $99$ measurements form a size-$99$ PP set.

In this paper, we will consider using this protocol in a regime where the physical error rate is $p=10^{-3}$ and our target is to distill magic states with logical error probability at most $\delta^{(M)}=10^{-10}$. Using the noise model described in \cref{subsec:MagicInject} for the injection of magic states, we apply the analysis in Ref.~\cite{Litinski19magic} to determine the logical failure probability per output magic state for one round of a $114$-to-$14$ distillation scheme,
\begin{align}
    p_L^{(M)}  = & \frac{1}{14} 30 \Big ( (\epsilon_{\text{L,Z}})^3 + \frac{1}{2} 6(\epsilon_{\text{L,Z}})^2 \epsilon_{\text{L,X}} \nonumber \\
    & \quad + \frac{1}{4} 12 (\epsilon_{\text{L,Z}}) (\epsilon_{\text{L,X}})^2
    + \frac{1}{8} 8 (\epsilon_{\text{L,X}})^3 \Big ) \nonumber \\
    = &\frac{30(1+\eta)^3}{378 \eta^3} p^3 \: .
\end{align}
For $p=10^{-3}$ and $\eta=100$, the probability that the distillation succeeds is $1 - p_D^{(M)} = (1-\epsilon_L)^{114} = 0.962$ and $p_L^{(M)} = 8.18 \times 10^{-11}$. Now, the lattice surgery measurements used to execute the distillation protocol must be modeled with measurement distance large enough to allow for distilled magic states of logical error rate at most $\delta{(M)}$. 

\begin{figure}
    \centering
    \subfloat[\label{fig:114ue} 
    ]{\includegraphics[width=.333\textwidth]{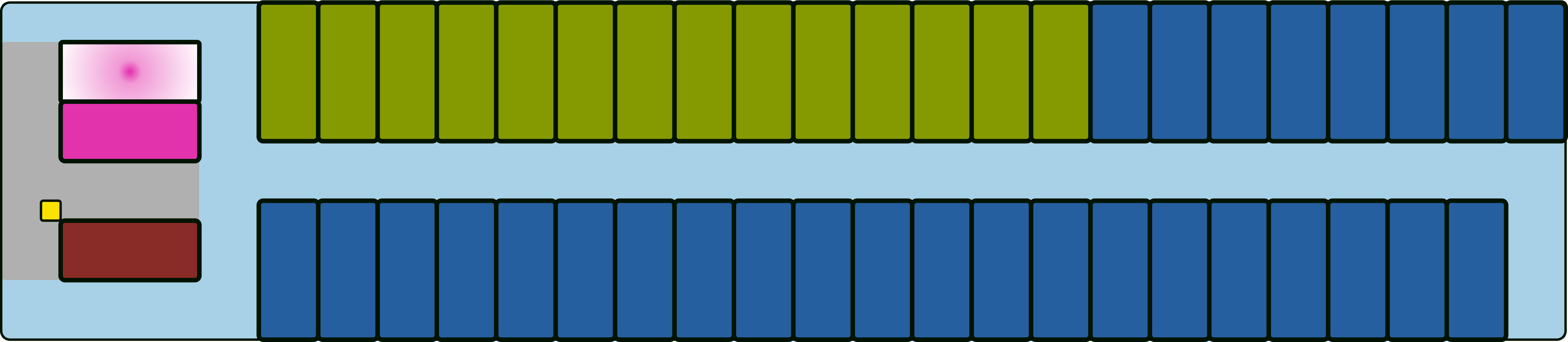}}
    \hspace{0.1cm}
    \subfloat[\label{fig:114uepar}
    ]{\includegraphics[width=.387\textwidth]{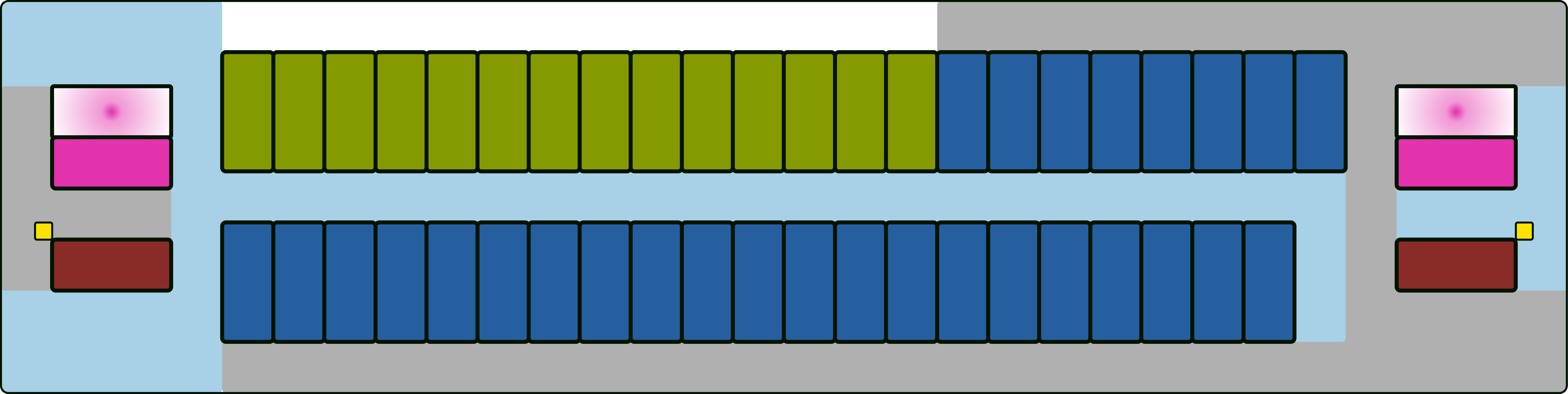}}
    \hspace{0.1cm}
    \subfloat[\label{fig:114zett5par}
    ]{\includegraphics[width=.404\textwidth]{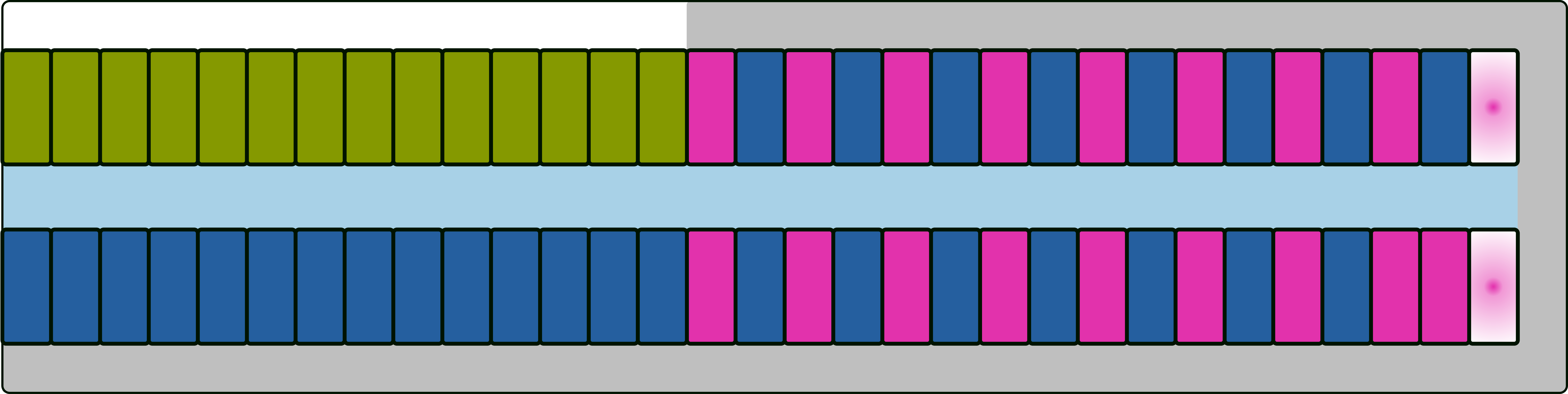}}
    \hspace{0.1cm}
    \subfloat[\label{fig:114bch7par} ]{\includegraphics[width=.442\textwidth]{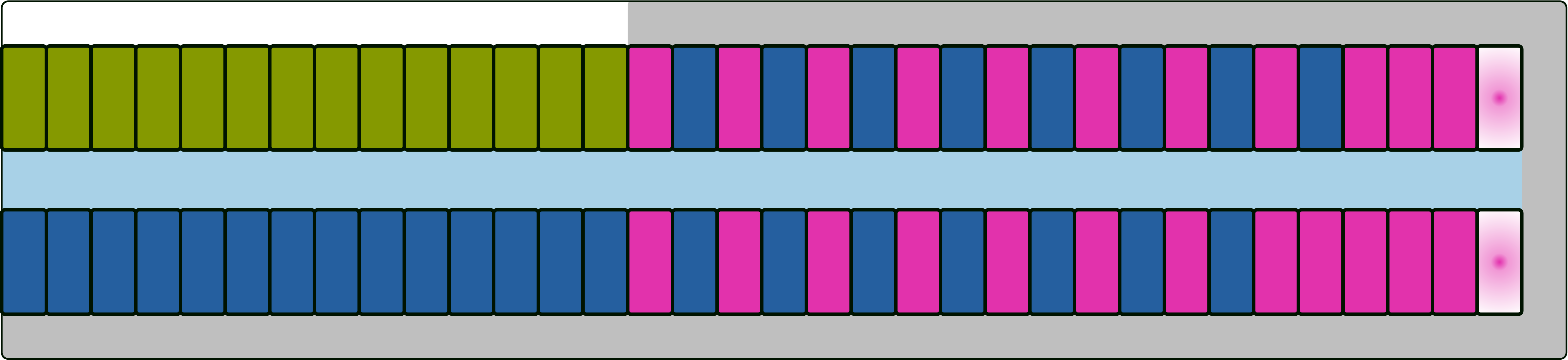}}
    \caption{Layouts of logical qubits for lattice-surgery-based $114$-to-$14$ magic state distillation with a $\llbracket 114,14,3\rrbracket$ quantum code. Cell color legend in caption of \cref{fig:15to1layouts}.~(a) Layout for a distillation tile without temporally encoded lattice surgery measurements, using an auto-corrected non-Clifford gate gadget.~(b) Layout for a distillation tile that performs Pauli measurements two at a time, without temporal encoding.~(c) Layout for a distillation tile performing temporally encoded lattice surgery with the $[129,114,6]$ Zetterberg code. Non-Clifford gates are performed two at a time, using separate routing spaces shown in gray and blue.~(d) Layout for a TELS-assisted distillation tile using the $[127,106,7]$ BCH code, performing non-Clifford gates two at a time.}
    \label{fig:114to14layouts}
\end{figure}

We show two layouts for distillation tiles that do not use TELS in \cref{fig:114ue} and \cref{fig:114uepar}. These layouts perform distillation in the Pauli frame as described in Ref.~\cite{Litinski19} using auto-corrected non-Clifford gadgets (\cref{fig:PauliAutocorr}). On the layout of \cref{fig:114ue}, $99$ Pauli measurements are performed, each with measurement distance $d_m = 15$. Using the procedure in \cref{app:algoSTcosts}, we determined that the minimum spacelike distances for this layout are $d_x=9$ and $d_z=19$. On the layout of \cref{fig:114uepar}, Pauli measurements can be performed two at a time. Hence the time required is only the time for $50$ sequential lattice surgery measurements with measurement distance $d_m = 15$. However in this case, when we calculated the minimum spacelike distances, the $Z$-distance could be dropped by two. This can be attributed to the fact that the distillation protocol finished in nearly half the time, and the probability of a logical $Z$-type error (see \cref{app:algoSTcosts}) scales linearly with time. Hence $d_x=9$ and $d_z=17$.

In \cref{fig:116zett5par}, we show a layout for a distillation tile that performs TELS with a classical $[129,114,6]$ Zetterberg code. Note that since there are two disjoint routing spaces (in blue and grey), the set of $n=129$ Pauli measurements can be performed in the time required for $65$ sequential measurements. To obtain the time cost shown in \cref{tab:STcosts}, we used $d_m = 5$ and $c=0$. Similarly, in \cref{fig:116bch9par}, we show a layout for a $114$-to-$14$ distillation tile that performs TELS with a classical $[127,106,7]$ BCH code. Here, we used the parameters $d_m = 5$ and $c=1$ to obtain the time cost shown in \cref{tab:STcosts}. As discussed in \cref{sec:Cliff}, we also execute a second PP set, now of maximum size $15$, due to the conditional Clifford corrections. The results of these measurements are used to detect if there are errors in the final distilled magic states. These Pauli measurements form a PP set of maximum size $15$, and in the worst case, the TELS protocol used is a $[31,16,7]$ BCH code with lattice surgery measurement distance $d_m = 5$ and $c=2$.

The space requirements of all the above layouts are described as functions of $d_x$ and $d_z$ in \cref{app:constants}. Additional constants in \cref{app:constants} can be used with the procedure of \cref{app:algoSTcosts} to determine all the minimum distances (spacelike and timelike) for the distillation protocols.

\subsection{Scheduling distillation tiles in a factory for magic states used by the core}
\label{subsec:scheduling}

\begin{figure*}
    \centering
    \includegraphics[width=0.85\textwidth]{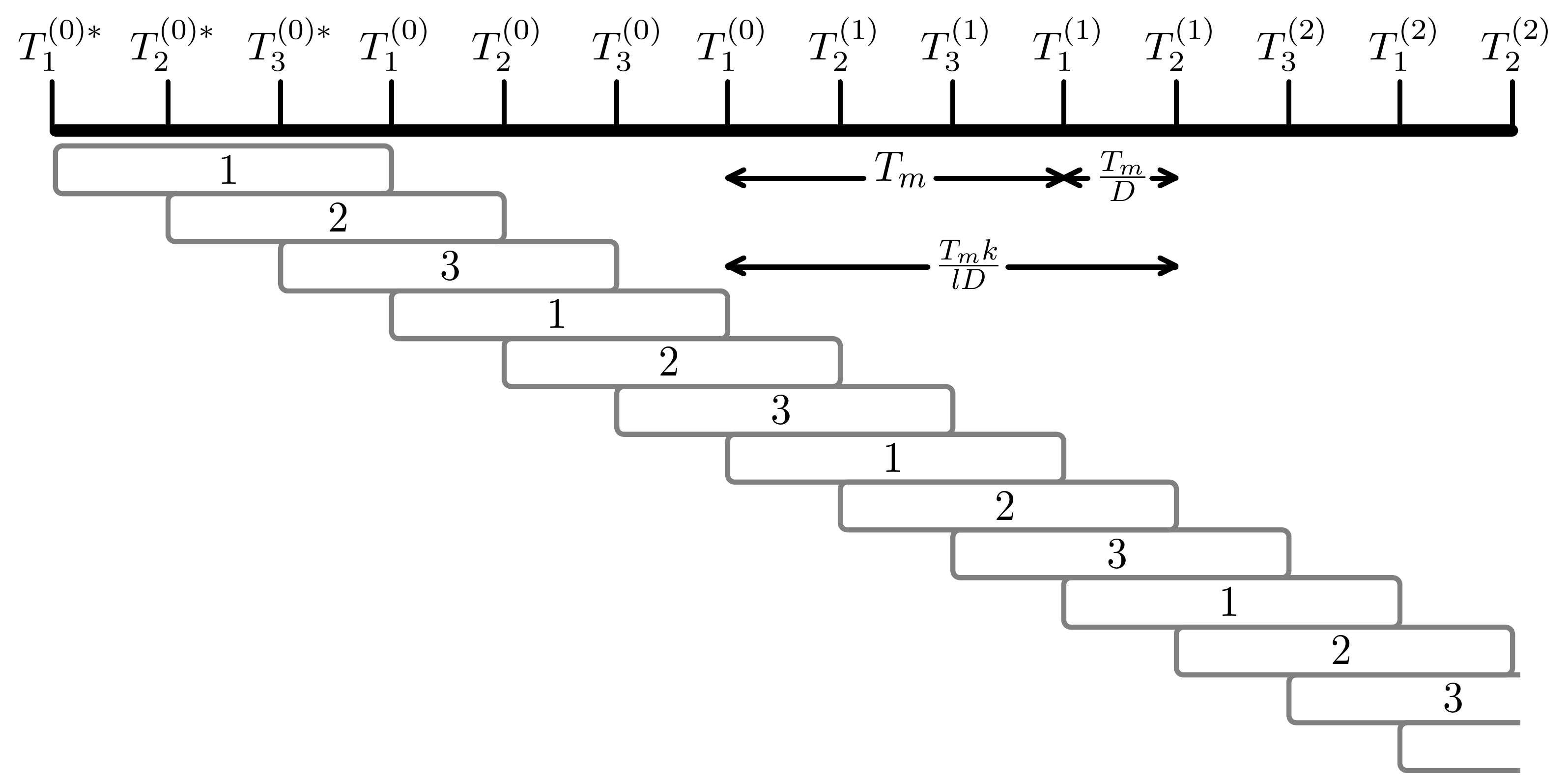}
    \caption{Round robin scheduling of deterministic-time distillation tiles in a factory. This scheduling method allows minimizing core wait time if a distillation tile rejects. The horizontal axis is time and the labels above are timestamps. Timestamp $T_j(i)$ indicates the end of the $j$th distillation tile while accumulating magic states for the $i$th PP set executed in the core. A * indicates the start of the distillation tile for the first time. In this example, there are $D=3$ distillation tiles, each producing $l=2$ distilled magic states in time $T_m$, for a core that executes PP sets of size $k=8$. }
    \label{fig:factoryscheduling}
\end{figure*}

Quantum computer architectures that perform Pauli-based computation generally contain two parts: a core and a magic state distillation factory. The core contains the data qubits taking part in the logical computation. It is known that TELS can speed up the runtime of PP sets executed in the core~\cite{Chamberland22}. In this paper, we also applied TELS to distillation circuits and observed reduced space-time costs. However distillation tiles are just modules that are used to construct a complete distillation factory, where many distillation tiles must be arranged with sufficient routing space to access the core. When merging a factory with a core, we are faced with additional scheduling and layout challenges. In this paper, we tackle the problem of scheduling, applying our speedups from TELS. We leave the design of the layouts of distillation factories for future work.

When executing an algorithm in the core using TELS on PP sets of size $k$, we denote the time to execute the algorithmic PP set as $T_{\text{PBC}}$. The distillation factory will simultaneously be working to distill at least $k$ new magic states for the next algorithmic PP set. It does this in time $T_{\text{magic}}$. General quantum algorithms operate on timescales much longer than $T_{\text{magic}}$, and so it is important to ensure the core is never idle and waiting for magic states. This situation, where the core is idling, is called a magic-state bottleneck. To avoid this bottleneck, we would like to ensure
\begin{equation}
\label{Cliff-bottleneck}
    T_{\text{magic}} \leq  T_{\text{PBC}} .
\end{equation}
The best case scenario is when $T_{\text{magic}} \approx  T_{\text{PBC}}$. As discussed in Ref.~\cite{Chamberland22}, this yields the smallest space-time cost for running algorithms.

We now wish to determine how many distillation tiles are required to satisfy the above condition. This can vary depending on the size of the PP set executed in the core and its relative speed-up. The number of tiles required also depends on the scheduling algorithm used in the factory, especially since some distillation tiles may detect errors and have to restart before producing magic states that can be used by the core. We first show how a simple algorithm can lead to large time costs when a distillation tile fails. Next we show how to use round robin scheduling to minimize additional time costs due to distillation tile failures. Note that the round robin scheduling algorithm can also be used when distilling lower level magic states in a concatenated distillation protocol.

Consider a situation where \cref{Cliff-bottleneck} is satisfied and $k$ is greater than or equal to the number of magic state storage cells (green cells in \cref{fig:15to1layouts}) in the factory. This implies each distillation tile takes less time to produce magic states than $T_{\text{PBC}}$. Hence a simple factory schedule would be to  start distillation on all tiles when a new PP set is beginning to be executed in the core. However, if the core is only marginally slower than the factory and a distillation tile rejects, the core will pause and wait for new magic states to be produced. Such a situation is undesirable since the core would now need to wait by a time $T_m$, where $T_m$ is the worst-case time needed to produce distilled magic states by a distillation tile.

In an attempt to reduce the core waiting time due to the rejection of a distillation tile (either due to TELS or the distillation algorithm), we suggest a round-robin approach. We assume that we have a distillation tile where, in the case where no errors are detected during the magic state distillation protocol, the tile produces $l$ magic states in time $T_m$ using a deterministic algorithm (TELS distillation is adaptive, but we consider the worst case time). The probability that a tile detects an error on an input magic state, or that the TELS protocol detects a timelike failure during lattice surgery is $p_D$.  If $D$ distillation tiles are used with round robin scheduling, the average time to distill $k$ magic states is
\begin{widetext}
\begin{align}
    T_{\text{magic}} = & \frac{T_m k}{l D} + {k/l \choose 1} p_D (1-p_D) ^{k/l-1} \frac{T_m}{D}   \quad+  {{k/l+1} \choose 2} p_D^2 (1-p_D)^{k/l-1} \frac{2 T_m}{D} + ... \nonumber \\
    = & \frac{T_m k}{l D} + \bigg (\frac{p_D (1-p_D)^{k/l-1} T_m}{D} \bigg ) \times \sum_{j=1}^{\infty} j {{k/l+j-1} \choose j}  p_D^{j-1} \nonumber \\
    = & \frac{T_m k}{l D} + \bigg (\frac{p_D (1-p_D)^{k/l-1} T_m}{D}\bigg ) \frac{k}{l} (1-p_D)^{-k/l-1} \nonumber \\
    = & \frac{T_m k (1-p_D+p_D^2)}{l D (1-p_D)^2} .
\end{align}
\end{widetext}
In \cref{fig:factoryscheduling}, we show an example of the round robin scheduling algorithm with $3$ distillation tiles that each produce $2$ distilled magic states in time $T_m$. If the core executes PP sets with size $8$, the $8$ required magic states are distilled and prepared in time $\frac{T_m k}{l D}$, if no errors are detected.

From this we may solve for the number of distillation tiles required when $T_{\text{magic}} <  T_{\text{PBC}}$ with $D$ distillation tiles given by
\begin{equation}
D= \frac{T_m k (1-p_D+p_D^2)}{l T_{\text{magic}} (1-p_D)^2} .
\end{equation}

The shortcoming of this calculation is that it only applies to constant-time distillation tiles. If the tile takes adaptive time, such as in the magic state distillation protocol of \cref{sec:Cliff}, where there are a non-trivial number of extra measurements, it is unclear what the most efficient scheduling algorithm is. In this case, we may still upper bound the total time $T_m$ of a magic state distillation tile thus making the round robin scheduling algorithm applicable. However, adapting a scheduling algorithm to the type of distillation tile used could allow for a more precise calculation of the time cost of the factory, which could possibly reduce the required number of distillation tiles.

\section{Conclusion}
\label{sec:Conclusion}

The most promising route to performing large-scale quantum computation on a surface code is to use Pauli-based computation with lattice surgery. For full fault tolerance and logical accuracy that scales with the code distance, each multi-qubit Pauli measurement implemented via lattice surgery requires multiple rounds of syndrome measurements to ensure the exponential suppression of timelike logical failures. In a recent paper, temporally encoded lattice surgery (TELS) was suggested to reduce the average time per lattice surgery measurement~\cite{Chamberland22}. In our paper, we make new developments for TELS protocols in two areas. First, we show an improved scheme for temporally encoded lattice surgery. Second, we apply TELS to magic state distillation factories and estimate their resulting space-time costs.

In the old scheme for TELS,  temporally encoded measurements are used to detect lattice surgery logical timelike failures. If an error is found, the original lattice surgery measurements are performed. In the new scheme, if a logical timelike failure is detected, the temporally encoded measurements are repeated. The average time per Pauli measurement is decreased but only if there is a high probability of detecting a logical timelike failure. We additionally reduce the time by combining error correction and error detection in the classical code. Low-weight errors are corrected using a lookup table, and other non-trivial syndromes signal detection events.

Next we applied TELS to the PP measurements that constitute magic state distillation circuits, like in \cref{fig:Haddist}. TELS can greatly reduce the time cost of distillation, but will require more logical qubits to be held in memory, requiring more space. We devised a new protocol to implement magic state distillation circuits derived from triorthgonal quantum codes using TELS, which yields a distilled magic state up to a Clifford frame. We show that Clifford frames are not too problematic for algorithms that use these states (since additional ancillas to perform $Y$ measurements may be required). In devising our TELS protocols for magic state factories, we provide implementations of the $15$-to-$1$, $116$-to-$12$, $114$-to-$14$ and $125$-to-$3$ distillation algorithms using different classical codes for TELS. Finally, we show how to use a round robin scheduling algorithm to schedule the operation of multiple distillation tiles in a distillation factory. 

For future work, adapting TELS protocols to magic state distillation protocols implemented in the Pauli frame which require auto-corrected $T$ gadgets could lead to greater space-time improvements. Furthermore, we believe that applying TELS protocols to the magic state distillation schemes considered in Ref.~\cite{Litinski19magic} could lead to distillation tiles with even smaller space-time costs than those obtained in that paper.

\section{Acknowledgments}
We thank Alexander Dalzell for useful discussions. P.P. thanks the AWS Center for Quantum Computing for the opportunity to undertake a summer research internship.

\appendix

\section{Malignant set counting}
\label{app:malignantsets}

An $[n,k,d]$ binary classical error-correcting code encodes $k$ logical bits of information into $n \geq k$ physical bits, with distance $d$. During error detection/correction, all errors of weight less than $d$ are detected. However \textbf{some} of the weight-$d$ errors are not detected. These errors are called malignant sets as they can cause erroneous flips of the logical bits. The task of computing how many of the ${n \choose d}$ weight-$d$ bit strings are malignant is computationally hard. The deterministic method is to evaluate the weights of all ${n \choose d}$ bit strings. But this takes time that is exponential in the problem size. 

For larger codes, we searched for faster methods to estimate the number of malignant fault sets. The first was a Monte Carlo simulation. The second method modelled the malignancy of weight-$d$ errors using a Bernoulli random variable. Finally, a third method used the MacWilliams identity. 

\subsection{Monte-Carlo sampling}
\label{subapp:monte}

For physical bit error rate $p$, the logical bit error rate of an $[n,k,d]$ code is $p_L = \sum_{j=d}^{n-d} l_j p^j (1-p)^{n-j}$, where $l_j$ is the number of malignant sets of weight $j$. At sufficiently low $p$, $p_L$ is approximately the first term of the polynomial, $l_d p^d (1-p)^{n-d}$. 

We can estimate $l_d$ using Monte Carlo simulations in two steps:
\begin{enumerate}
    \item For different, small values of $p$, compute $p_L$ by sampling errors and evaluating the fraction of them that are malignant. An $n$-bit error sample $e$ is obtained by sampling each bit from a Bernoulli random variable with probability $p$. The error $e$ is malignant if $H e = 0$, where $H$ is the parity check matrix of the code.
    \item Perform a least squares fit of the obtained values with the polynomial $p^d (1-p)^{n-d}$. The coefficient of the fit is the Monte Carlo approximation of $l_d$. 
\end{enumerate} 

At sufficiently low $p$, many of the error samples will be trivial. Hence a lot of time is wasted evaluating these samples. For large $d$, this problem becomes worse. The probability of observing a weight-$d$ error scales as $p^d$, implying that the errors that actually may be malignant are rarely ever observed.

\subsection{Modelling malignancy with the Bernoulli distribution}
\label{subapp:bernoulli}

Since we only care to check whether a weight-$d$ error is malignant or not, it is faster to sample only from the set of weight-$d$ errors. Let an error be sampled by choosing $d$ out of $n$ locations at random without replacement. We can now model the malignancy of a weight-$d$ error using a Bernoulli random variable: a weight-$d$ error sample is malignant with probability $p$. We can estimate $p$ with high confidence by checking for malignancy on many samples. Finally, $l_d = p {n \choose d}$.

\subsection{MacWilliams identity}
\label{subapp:macwilliams}

For a code $C$ with $k$ codeword generators, there exist $n-k$ vectors spanning the nullspace (kernel), $C^{\perp}$.  If the weights of the $\abs {C^\perp}= 2^{n-k}$ bit strings can be enumerated, then the weights of the codewords of $C$ can be evaluated using the MacWilliams identity:
\begin{equation}
    W_j^{C} = \frac{1}{\abs C^\perp} \sum_{i=0}^n W_i^{C^\perp} K_j(i,n),
\end{equation}
for $j= 0,1,\mathellipsis , n$. Here $W_i^C$ is the number of codewords of $C$ of weight $i$ and $K_j(i,n)$ is the Krawtchouk polynomial 
\begin{equation}
    K_j(i,n) = \sum_{l=0}^j (-1)^l {i \choose l} {n-i \choose j-l}.
\end{equation}

Then $W_d^C$ is the number of malignant fault sets of weight-$d$.

A short Mathematica script can enumerate the weights of  $2^{36} = 68{,}719{,}476{,}736$ codewords in just under $24$ hours.

\section{Construction of classical codes}
\label{app:codeconstruction}

\subsection{Cyclic codes defined using polynomials}

Binary cyclic codes can be constructed using cyclic shifts of polynomials defined over finite fields. To understand why, first note the isomorphic map between the field $\mathbb{F}_2^n$ and polynomials of degree $<$ $n$ with coefficients in $\mathbb{F}_2$.  For example, in $\mathbb{F}_2^4$, the polynomial $x^3 + 0 x^2 + x + 1$ corresponds to the bit string $1011$, where the bit at position $i\in \{ 0,1,\mathellipsis, n-1 \}$ (right-to-left) is the coefficient of the term $x^i$. An $[n,k,d]$ code is cyclic if the $k$ codewords can be generated by cyclic shifts of a generator polynomial, $g(x)$. Cyclic shifts of $g(x)$ are obtained by multiplying $g(x)$ with $\{ 1,x,x^2, \mathellipsis , x^{k-1} \}$.

\subsubsection{Single Error Detect code}
\label{subsubsec:SED}

The Single Error Detect code with parameters $[\alpha +1, \alpha, 2]$ can be generated by taking cyclic shifts of the generating polynomial $g(x)= x+1$ over the field $\mathbb{F}_2^{\alpha+1}$. For example the codewords of the $[4,3,2]$ code are the rows of $G$ below.
\begin{equation}
    G = \begin{bmatrix} 
0011 \\
0110 \\ 
1100 \\
\end{bmatrix}.
\end{equation}
The parity check matrix $H$ is the nullspace of $G$, i.e. the span of all vectors in $\mathbb{F}_2^n$ that are orthogonal to elements of $G$. For the above code,
\begin{equation}
     H = \begin{bmatrix} 
1111 \\
\end{bmatrix}.
\end{equation}

We also display the codeword generator matrix for the $[12,11,2]$ code which we use in \cref{sec:15to1} for magic state distillation.
\begin{equation}
    G = \begin{bmatrix} 
000000000011 \\
000000000110 \\
000000001100 \\
000000011000 \\
000000110000 \\
000001100000 \\
000011000000 \\
000110000000 \\
001100000000 \\
011000000000 \\
110000000000 
\end{bmatrix}.
\label{eq:CyclicG12}
\end{equation}

\begin{table*}
    \centering
    \begin{tabular}{c c}
         Code & Generator polynomial \\
         \hline
         $[7,4,3]$ & $1011 = x^3+x+1$ \\
         $[15,11,3]$ & $10011$ \\
         $[31,26,3]$ & $ 100101 $ \\
         $[43,36,3]$ & $ 10101011 $ \\
         $[49,43,3]$ & $ 1000011 $ \\
         $[63,57,3]$ & $ 1000011 $ \\
         $[85,77,3]$ & $ 100011101 $ \\
         $[127,120,3]$ & $ 10000011 $ \\[.25cm]
         
         $[15,7,5]$ & $ 111010001 $ \\
         $[31,21,5]$ & $ 11101101001 $ \\
         $[43,29,5]$ & $ 100111110100011 $ \\
         $[49,37,5]$ & $ 1010100111001 $ \\
         $[63,51,5]$ & $ 1010100111001 $ \\
         $[85,69,5]$ & $ 10110111101100011 $ \\
         $[127,113,5]$ & $ 101010001111101 $ \\[.25cm]
         
         $[15,5,7]$ & $ 10100110111$ \\
         $[31,16,7]$ & $ 1000111110101111 $ \\
         $[43,22,7]$ & $ 1010010100110010100001 $ \\
         $[49,31,7]$ & $ 1111000001011001111 $ \\
         $[63,45,7]$ & $ 1111000001011001111 $ \\
         $[85,61,7]$ & $ 1101110111010000110110101 $ \\
         $[127,106,7]$ & $ 1010010011000000011011 $ \\[.25cm]
         
         $[49,25,9]$ & $ 1110110110010011101110111 $ \\
         $[63,39,9]$ & $ 1110110110010011101110111 $ \\
         $[85,53,9]$ & $ 111101110010110110100001011111101$ \\
         $[127,99,9]$ & $ 11000101001010111100100111111 $ \\[.25cm]
         
         $[43,15,10]$ & $ 11111110001001100100000101011 $ \\[.25cm]
         
         $[31,11,11]$ & $ 101100010011011010101 $ \\
         $[49,22,11]$ & $ 1000011011101000000100010011 $ \\
         $[63,36,11]$ & $ 1000011011101000000100010011 $ \\
         $[85,45,11]$ & $ 10011001101111101110100111010110100010001$ \\
         $[127,92,11]$ & $ 111000010001110010101001101101010111 $ \\
    \end{tabular}
    \caption{BCH codes and associated generator polynomials. The codewords generators are  cyclic shifts of the generator polynomial.}
    \label{tab:BCHpolynomials}
\end{table*}

\begin{table*}
    \centering
    \begin{tabular}{c c c}
         $u$ & Code & Generator Polynomial  \\
         \hline
        $ 3$ & $[9,2,6]$ & $ 10111101 = x^7+x^5+x^4+x^3+x^2+1$ \\
        $ 4$ & $[17,9,5]$ & $ 100111001$ \\
        $ 5$ & $[33,22,6]$ & $ 101001100101 $ \\
        $ 6$ & $[65,53,5]$ & $ 1000111110001 $ \\
        $ 7$ & $[129,114,6]$ & $ 1001010000101001$ \\
    \end{tabular}
    \caption{Zetterberg codes with associated generator polynomials}
    \label{tab:Zettpolynomials}
\end{table*}

\subsubsection{Golay code}
\label{subsubsec:Golay}

The $[23,12,7]$ Golay code is a cyclic code generated by the polynomial $x^{11}+x^9+x^7+x^6+x^5+x+1$ over $\mathbb{F}_2^{23}$. 

\begin{equation}
G = \begin{bmatrix} 

000000000001 0 1 0 1 1 1 0 0 0 1 1 \\
00000000001 0 1 0 1 1 1 0 0 0 1 1 0 \\
0000000001 0 1 0 1 1 1 0 0 0 1 1 00 \\
000000001 0 1 0 1 1 1 0 0 0 1 1 000 \\
00000001 0 1 0 1 1 1 0 0 0 1 1 0000 \\
0000001 0 1 0 1 1 1 0 0 0 1 1 00000 \\
000001 0 1 0 1 1 1 0 0 0 1 1 000000 \\
00001 0 1 0 1 1 1 0 0 0 1 1 0000000 \\
0001 0 1 0 1 1 1 0 0 0 1 1 00000000 \\
001 0 1 0 1 1 1 0 0 0 1 1 000000000 \\
01 0 1 0 1 1 1 0 0 0 1 1 0000000000 \\
1  0 1 0 1 1 1 0 0 0 1 1 00000000000 \\
\end{bmatrix}.
\end{equation}

\subsubsection{BCH codes}
\label{subsubsec:BCH}

Bose-Chaudhuri–Hocquenghem (BCH) codes are a well-studied family of classical cyclic codes constructed using polynomials over finite fields. Due to the flexible nature of the construction of these codes, codes of different distances can be defined for the same code size. In \cref{tab:BCHpolynomials}, we show the generating polynomials for the BCH codes that were considered in this paper.

\subsubsection{Zetterberg codes}

Zetterberg codes are binary cyclic codes defined as $[2^u+1,2^u+1-2u,5 \leq d \leq 6]$ codes for even $u$. For odd $u$, we obtain the parameters $[2^u+1,2^u-2u,6]$. These codes are quasi-perfect: the distance between two codewords is $5 \leq d \leq 6$.

In this paper we consider Zetterberg codes for $u\in \{3,4,5,6,7\}$. The codewords are defined by taking cyclic shifts of polynomials shown in \cref{tab:Zettpolynomials}. Note that all the polynomials are palindromic. For a chosen $u$, other codes with the same parameters may be defined using different palindromic polynomials of the same degree. For a more detailed description of the construction, consider~\cite{Jing10}.

\subsection{Reed-Muller and polar codes}

Binary Reed-Muller codes are $[ 2^m, k ,2^{m-r} ]$ codes for $r \leq m$ where 
\begin{equation}
    k= 2^m-\sum_{i=0}^{m-r-1} {m \choose i} = \sum_{i=0}^{r} {m \choose i}.
\end{equation}
To determine the codewords of the $(r,m)$-Reed-Muller code, start with the m-fold tensor product of the generator matrix $\begin{bmatrix} 1 & 1 \\ 0 & 1\end{bmatrix}$.
Remove the $\sum_{i=0}^{m-r-1} {m \choose i}$ rows with fewer than $d$ $1$'s. The $k$ remaining rows denote the codewords. In this paper, we look at the family of $m=r+1$ Single Error Detect codes, $m=r+2$ distance-$4$ Extended Hamming codes, and $m=r+3$ distance-$8$ codes.

Polar codes are $2^m$-bit codes that were initially developed for communication systems to tackle analog noise. The binary codes constructed using this formalism can be used against discrete noise too. The method of construction is the same as that of the Reed-Muller codes, but allows for codes with fewer encoded bits. After removing low-weight codewords to create a Reed-Muller code, remove extra codewords (lowest weight first) until there are exactly as many encoded bits as required.

\section{Speedups offered by different codes}
\label{app:speedups}

\begin{figure*}
    \centering
    \hspace{-1mm}
    \subfloat[\label{fig:delt15}]{\includegraphics[width=.98\textwidth]{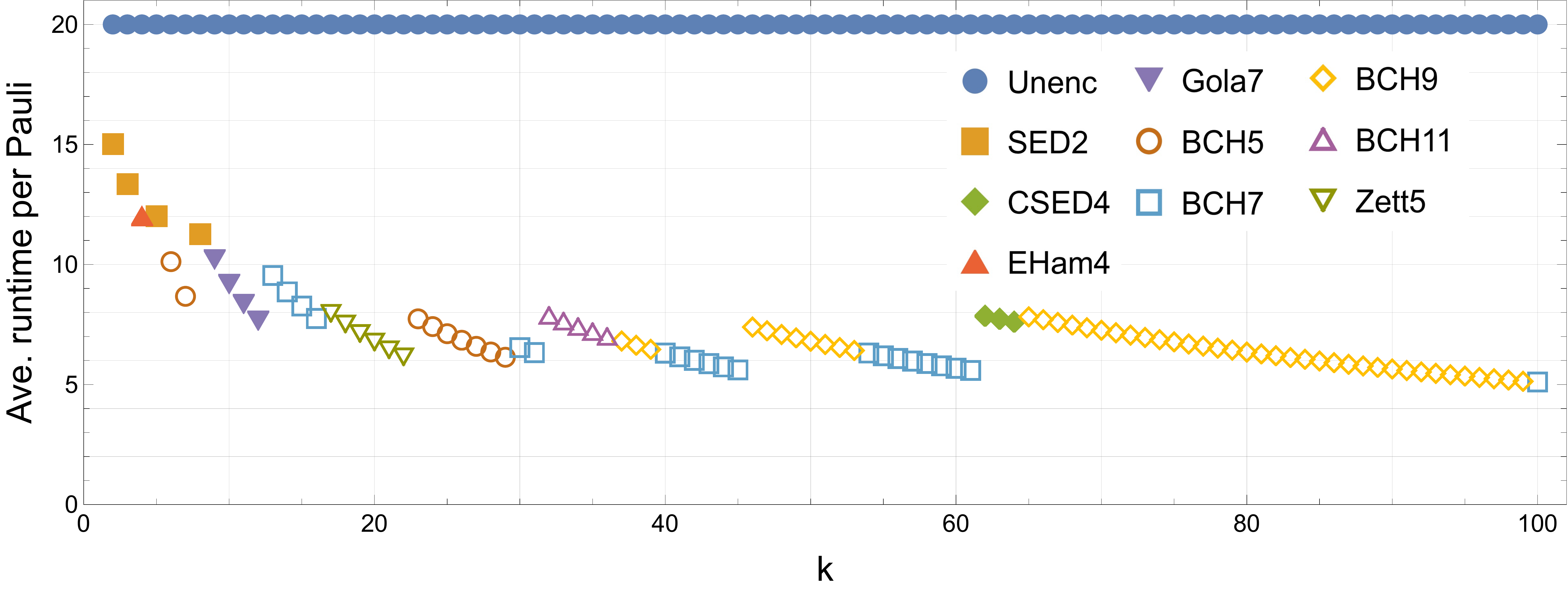}}
    \hspace{0.1cm}
    \subfloat[\label{fig:delt20}]{\includegraphics[width=.98\textwidth]{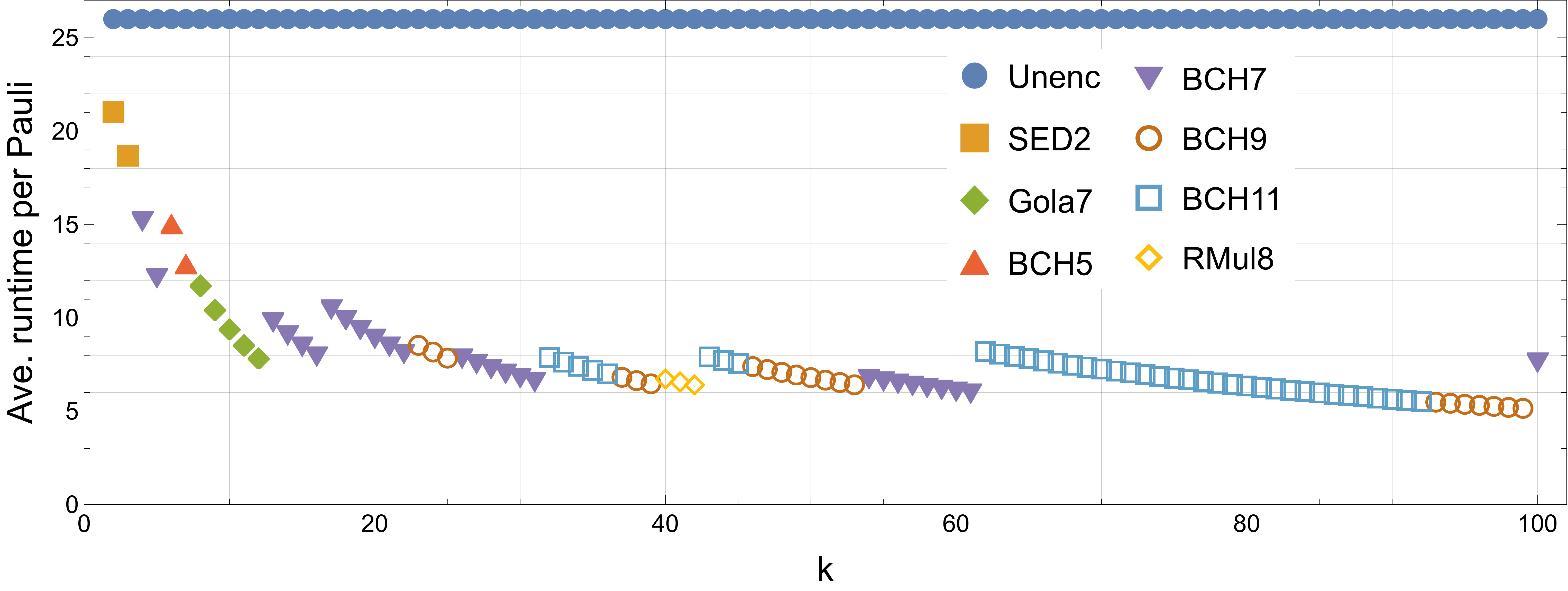}}
    \hspace{0.1cm}
    \subfloat[\label{fig:delt25}]{\includegraphics[width=.98\textwidth]{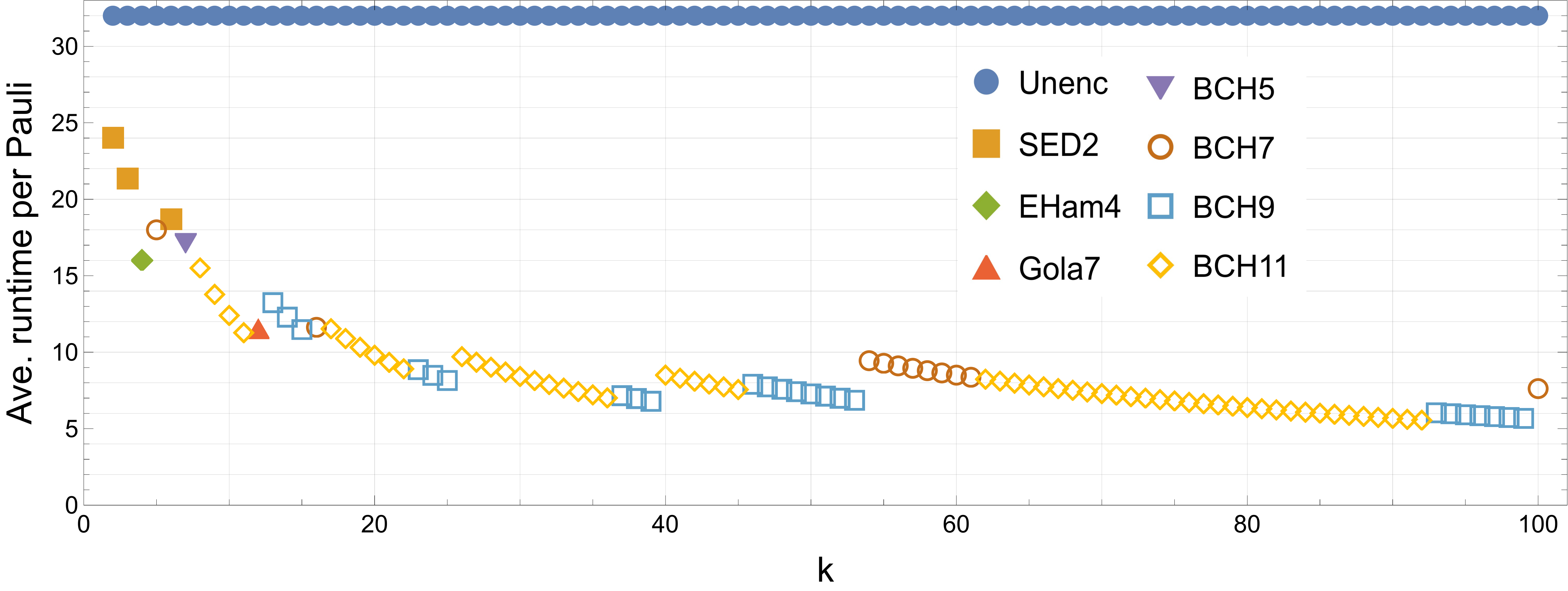}}
    \caption{We show the classical codes achieving the lowest average runtime per Pauli for $k \in \{ 2,3, \mathellipsis, 100\}$ at $p=10^{-3}$ and for (a)~$\delta= 10^{-15}$, (b)~$\delta= 10^{-20}$ and (c)~$\delta= 10^{-25}$. We set the routing space area $A = 100$. }
    \label{fig:p3delt152025}
\end{figure*}

\begin{figure*}
    \centering
    \hspace{-1mm}
    \subfloat[\label{fig:delt15p4}]{\includegraphics[width=.98\textwidth]{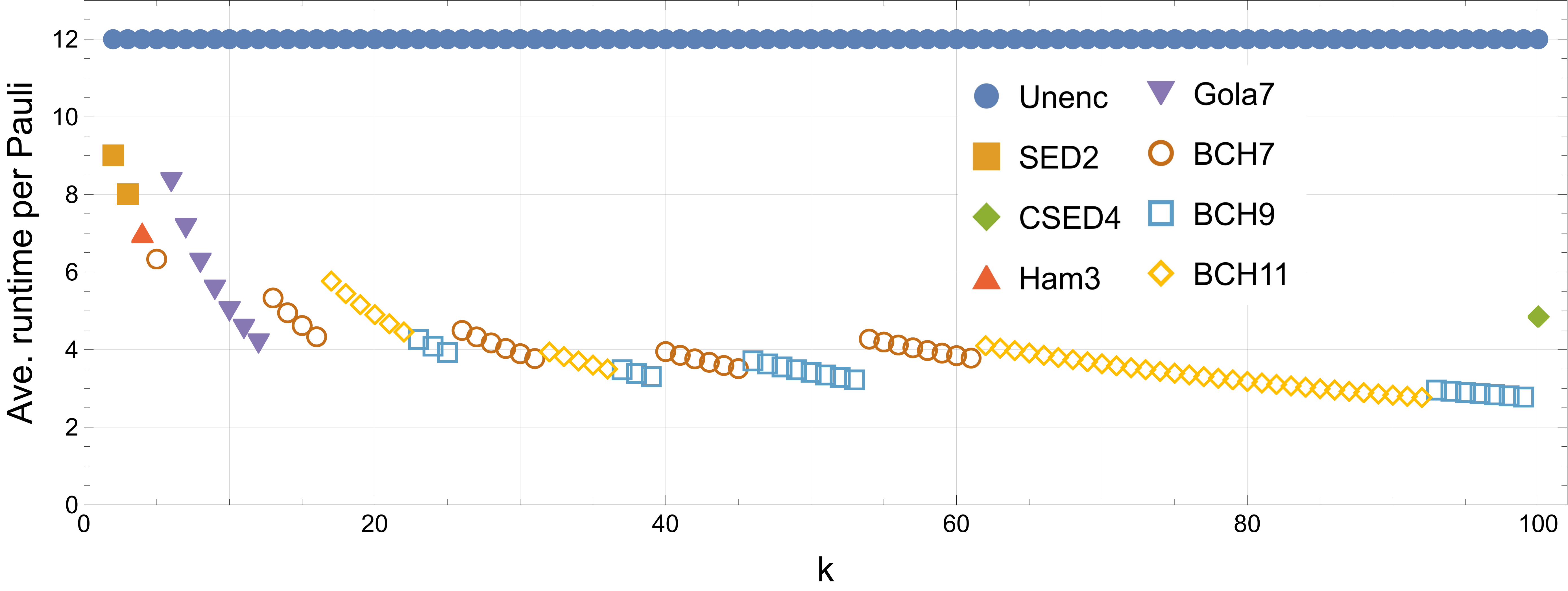}}
    \hspace{0.1cm}
    \subfloat[\label{fig:delt20p4}]{
    \includegraphics[width=.98\textwidth]{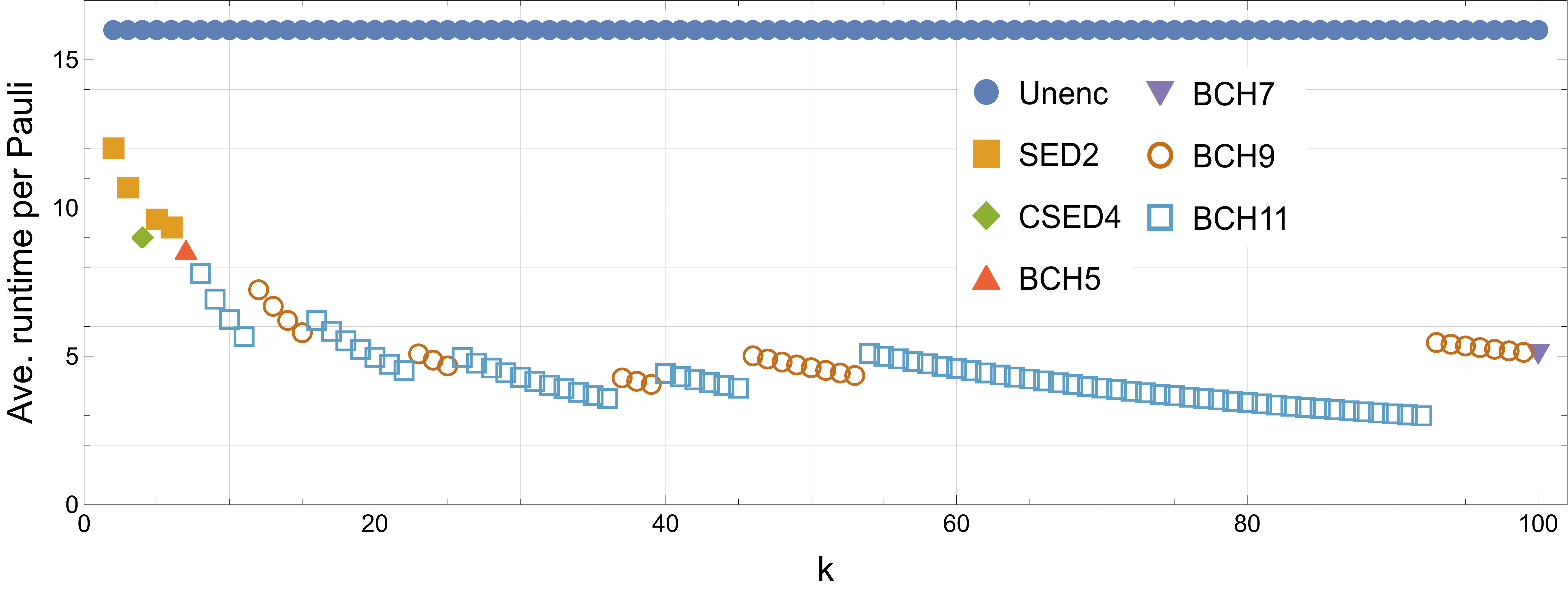}}
    \caption{We show the classical codes achieving the lowest average runtime per Pauli for $k \in \{ 2,3, \mathellipsis, 100\}$ at $p=10^{-4}$ and for (a)~$\delta= 10^{-15}$ and (b)~$\delta= 10^{-20}$. We set the routing space area $A = 100$.}
    \label{fig:p4delt1520}
\end{figure*}

In \cref{fig:p3delt152025}, we show the lowest average runtime per Pauli for temporally encoded lattice surgery of $k \in \{ 2,3, \mathellipsis, 100\}$ measurements for $p=10^{-3}$. We also indicate which code achieves the lowest average runtime per Pauli for each $k$. Note that this paper only considered a limited number of classical codes for TELS. It may be possible for other codes to perform better than the ones outlined here.
\cref{fig:p4delt1520} shows the best classical codes for $p=10^{-4}$ and $\delta = 10^{-15}$ and $\delta = 10^{-20}$ respectively.

In \cref{tab:k2to50}, we show the best average speedup due to a TELS code for $k \in \{ 2,3, \mathellipsis, 100\}$. These speedups are computed with respect to performing the $k$ measurements sequentially at the regular measurement distance $d_m$. These speedups are computed for various regimes: $p = 10^{-3}$, $\delta \in \{ 10^{-10}, 10^{-15}, 10^{-20}, 10^{-25}\}$ and $p = 10^{-4}$, $\delta \in \{ 10^{-15}, 10^{-20}\}$.

\begin{table*}[]
    \centering
    \caption{The best lattice surgery speedup for $k \in \{ 1,2, \mathellipsis 100\}$, and associated classical code achieving it, in different noise regimes $p$, and for different target logical error rates $\delta$.}
    \begin{tabular}{c c c c c c c}
        $k$ & $p=10^{-3}$ & $p=10^{-3}$& $p=10^{-3}$ & $p=10^{-3}$ & $p=10^{-4}$ & $p=10^{-4}$ \\
        &  $\delta = 10^{-10}$& $\delta = 10^{-15}$& $\delta = 10^{-20}$& $\delta = 10^{-25}$& $\delta = 10^{-15}$& $\delta = 10^{-20}$ \\
        \hline
        $2$ & SED2, $1.167$ & SED2, $1.333$ & SED2, $1.238$ & SED2, $1.333$ & SED2, $1.333$  & SED2, $1.333$ \\
        $3$ & SED2, $1.312$ & SED2, $1.5  $ & SED2, $1.393$   & SED2, $1.5$ & SED2, $1.5$ & SED2, $1.5$ \\
        $4$ & EHam4, $1.739$ & EHam4, $1.666$ & BCH7, $1.713$& EHam4, $2$ & Ham3, $1.714$ & CSED4, $1.778$ \\
        $5$ & SED2, $1.458$ & SED2, $1.667$ & BCH7, $2.141$   & BCH7, $1.778$ & BCH7, $1.895$ & SED2, $1.667$ \\
        $6$ & SED2, $1.5$ & BCH5, $1.977$ & BCH5, $1.733$     & SED2, $1.714$ & Gola7, $1.441$ & SED2, $1.714$ \\
        $7$ & BCH5, $1.633$ & BCH5, $2.306$ & BCH5, $2.022$ & BCH5, $1.867$ & Gola7, $1.681$ & BCH5, $1.866$ \\
        $8$ & CSED4, $1.728$ & SED2, $1.778$ & Gola7, $2.22$ & BCH11, $2.065$ & Gola7, $1.922$ & BCH11, $2.053$ \\
        $9$ & CSED4, $1.944$ & Gola7, $1.956$ & Gola7, $2.498$ & BCH11, $2.323$ & Gola7, $2.162$ & BCH11, $2.31$ \\
        $10$ & EHam4, $2.16$ & Gola7, $2.174$ & Gola7, $2.775$ & BCH11, $2.581$ & Gola7, $2.402$ &  BCH11, $2.566$  \\
        $11$ & EHam4, $2.376$ & Gola7, $2.391$ & Gola7, $3.053$ & BCH11, $2.839$ & Gola7, $2.642$ & BCH11, $2.823$  \\
        $12$ & Gola7, $1.826$ & Gola7, $2.608$ & Gola7, $3.331$ & Gola7, $2.783$ & Gola7, $2.882$ & BCH9, $2.209$  \\
        $13$ & CSED4, $1.785$ & BCH7, $2.096$ & BCH7, $2.66 $ & BCH9, $2.417$ & BCH7, $2.251$ & BCH9, $2.393$ \\
        $14$ & CSED4, $1.922$ & BCH7, $2.257$ & BCH7, $2.865$ & BCH9, $2.603$ & BCH7, $2.424$ & BCH9, $2.577$ \\
        $15$ & CSED4, $2.059$ & BCH7, $2.419$ & BCH7, $3.069$ & BCH9, $2.789$ & BCH7, $2.597$ & BCH9, $2.761$ \\
        $16$ & CSED4, $2.196$ & BCH7, $2.58$ & BCH7, $3.274$ & BCH7, $2.753$ & BCH7, $2.771$ & BCH11, $2.577$ \\
        $17$ & BCH5, $1.919$ & Zett5, $2.51$ & BCH7, $2.484$ & BCH11, $2.775$ & BCH11, $2.082$ & BCH11, $2.738$ \\
        $18$ & BCH5, $2.032$ & Zett5, $2.657$ & BCH7, $2.63$ & BCH11, $2.939$ & BCH11, $2.204$ & BCH11, $2.899$ \\
        $19$ & BCH5, $2.145$ & Zett5, $2.805$ & BCH7, $2.777$ & BCH11, $3.102$ & BCH11, $2.326$ & BCH11, $3.06$ \\
        $20$ & BCH5, $2.257$ & Zett5, $2.953$ & BCH7, $2.923$ & BCH11, $3.265$ & BCH11, $2.449$ & BCH11, $3.221$ \\
        $21$ & BCH5, $2.37$ & Zett5, $3.1$ & BCH7, $3.069$ & BCH11, $3.429$ & BCH11, $2.571$ & BCH11, $3.382$ \\
        $22$ & EHam4, $2.346$ & Zett5, $3.248$ & BCH7, $3.215$ & BCH11, $3.592$ & BCH11, $2.694$ & BCH11, $3.543$ \\
        $23$ & Pol4, $2.453$ & BCH5, $2.586$ & BCH9, $3.049$ & BCH9, $3.613$ & BCH9, $2.814$ & BCH9, $3.149$ \\
        $24$ & Pol4, $2.56$ & BCH5, $2.698$ & BCH9, $3.181$ & BCH9, $3.77$ & BCH9, $2.937$ & BCH9, $3.286$ \\
        $25$ & Pol4, $2.666$ & BCH5, $2.81$ & BCH9, $3.314$ & BCH9, $3.927$ & BCH9, $3.059$ & BCH9, $3.423$ \\
        $26$ & EHam4, $2.773$ & BCH5, $2.923$ & BCH7, $3.319$ & BCH11, $3.298$ & BCH7, $2.67$ & BCH11, $3.23$ \\
        $27$ & BCH5, $2.196$ & BCH5, $3.035$ & BCH7, $3.446$ & BCH11, $3.425$ & BCH7, $2.773$ & BCH11, $3.354$ \\
        $28$ & BCH5, $2.278$ & BCH5, $3.148$ & BCH7, $3.574$ & BCH11, $3.551$ & BCH7, $2.876$ & BCH11, $3.478$ \\
        $29$ & BCH5, $2.359$ & BCH5, $3.26$ & BCH7, $3.702$ & BCH11, $3.678$ & BCH7, $2.978$ & BCH11, $3.603$ \\
        $30$ & BCH7, $2.143$ & BCH7, $3.059$ & BCH7, $3.829$ & BCH11, $3.805$ & BCH7, $3.081$ & BCH11, $3.727$ \\
        $31$ & BCH7, $2.214$ & BCH7, $3.161$ & BCH7, $3.957$ & BCH11, $3.932$ & BCH7, $3.184$ & BCH11, $3.851$ \\
        $32$ & BCH5, $2.284$ & BCH11, $2.54$ & BCH11, $3.302$ & BCH11, $4.059$ & BCH11, $3.047$ & BCH11, $3.975$ \\
        $33$ & BCH5, $2.355$ & BCH11, $2.619$ & BCH11, $3.405$ & BCH11, $4.186$ & BCH11, $3.143$ & BCH11, $4.1$ \\
        $34$ & BCH5, $2.427$ & BCH11, $2.698$ & BCH11, $3.508$ & BCH11, $4.312$ & BCH11, $3.238$ & BCH11, $4.224$ \\
        $35$ & BCH5, $2.498$ & BCH11, $2.778$ & BCH11, $3.611$ & BCH11, $4.439$ & BCH11, $3.333$ & BCH11, $4.348$ \\
        $36$ & BCH5, $2.57$ & BCH11, $2.857$ & BCH11, $3.714$ & BCH11, $4.566$ & BCH11, $3.428$ & BCH11, $4.472$ \\
        $37$ & BCH5, $2.641$ & BCH9, $2.937$ & BCH9, $3.813$ & BCH9, $4.471$ & BCH9, $3.447$ & BCH9, $3.747$ \\
        $38$ & BCH9, $2.111$ & BCH9, $3.016$ & BCH9, $3.916$ & BCH9, $4.592$ & BCH9, $3.541$ & BCH9, $3.849$ \\
        $39$ & BCH9, $2.167$ & BCH9, $3.095$ & BCH9, $4.019$ & BCH9, $4.713$ & BCH9, $3.634$ & BCH9, $3.95$ \\
        $40$ & BCH7, $2.222$ & BCH7, $3.171$ & RMul8, $3.863$ & BCH11, $3.765$ & BCH7, $3.038$ &  BCH11, $3.623$\\
        $41$ & BCH7, $2.278$ & BCH7, $3.25$ & RMul8, $3.96$ & BCH11, $3.859$ & BCH7, $3.114$ & BCH11, $3.713$ \\
        $42$ & BCH7, $2.333$ & BCH7, $3.329$ & RMul8, $4.059$ & BCH11, $3.953$ & BCH7, $3.19$ & BCH11, $3.804$ \\
        $43$ & BCH7, $2.389$ & BCH7, $3.409$ & BCH11, $3.288$ & BCH11, $4.047$ & BCH7, $3.266$ & BCH11, $3.895$ \\
        $44$ & BCH7, $2.444$ & BCH7, $3.488$ & BCH11, $3.365$ & BCH11, $4.141$ & BCH7, $3.342$ & BCH11, $3.985$ \\
        $45$ & BCH7, $2.5$ & BCH7, $3.567$ & BCH11, $3.441$ & BCH11, $4.235$ & BCH7, $3.418$ & BCH11, $4.076$ \\
        $46$ & BCH5, $2.553$ & BCH9, $2.706$ & BCH9, $3.517$ & BCH9, $4.05$ & BCH9, $3.235$ & BCH9, $3.191$ \\
        $47$ & BCH5, $2.608$ & BCH9, $2.765$ & BCH9, $3.594$ & BCH9, $4.138$ & BCH9, $3.306$ & BCH9, $3.26$ \\
        $48$ & BCH5, $2.664$ & BCH9, $2.824$ & BCH9, $3.67$ & BCH9, $4.226$ & BCH9, $3.376$ & BCH9, $3.33$ \\
        $49$ & BCH5, $2.719$ & BCH9, $2.882$ & BCH9, $3.747$ & BCH9, $4.314$ & BCH9, $3.446$ & BCH9, $3.399$ \\
    \end{tabular}
    \label{tab:k2to50}
\end{table*}

\begin{table*}[]
    \centering
    \begin{tabular}{c c c c c c c}
        $50$ & BCH5, $2.775$ & BCH9, $2.941$ & BCH9, $3.823$ & BCH9, $4.402$ & BCH9, $3.517$ & BCH9, $3.468$ \\
        $51$ & BCH5, $2.83$ & BCH9, $3$ & BCH9, $3.9$ & BCH9, $4.49$ & BCH9, $3.587$  & BCH9, $3.538$ \\
        $52$ & Zett5, $2.797$ & BCH9, $3.059$ & BCH9, $3.976$ & BCH9, $4.578$ & BCH9, $3.657$ &BCH9, $3.607$ \\
        $53$ & Zett5, $2.85$ & BCH9, $3.118$ & BCH9, $4.053$ & BCH9, $4.666$ & BCH9, $3.728$ & BCH9, $3.676$ \\
        $54$ & EHam4, $2.808$ & BCH7, $3.17$ & BCH7, $3.862$ & BCH7, $3.388$ & BCH7, $2.809$ & BCH11, $3.141$ \\
        $55$ & EHam4, $2.86$ & BCH7, $3.228$ & BCH7, $3.934$ & BCH7, $3.451$ & BCH7, $2.861$ & BCH11, $3.199$ \\
        $56$ & EHam4, $2.912$ & BCH7, $3.287$ & BCH7, $4.006$ & BCH7, $3.514$ & BCH7, $2.913$ & BCH11, $3.257$ \\
        $57$ & EHam4, $2.964$ & BCH7, $3.346$ & BCH7, $4.077$ & BCH7, $3.576$ & BCH7, $2.965$ & BCH11, $3.315$ \\
        $58$ & BCH7, $2.388$ & BCH7, $3.405$  & BCH7, $4.149$ & BCH7, $3.639$ & BCH7, $3.017$ & BCH11, $3.374$ \\
        $59$ & BCH7, $2.429$ & BCH7, $3.463$  & BCH7, $4.22 $ & BCH7, $3.702$ & BCH7, $3.069$ & BCH11, $3.432$ \\
        $60$ & BCH7, $2.47$ & BCH7, $3.522$ &   BCH7, $4.292$ & BCH7, $3.765$ & BCH7, $3.122$ & BCH11, $3.49$ \\
        $61$ & BCH7, $2.512$ & BCH7, $3.581$ &  BCH7, $4.363$ & BCH7, $3.827$ & BCH7, $3.174$ & BCH11, $3.548$ \\
        $62$ & BCH5, $2.548$ & CSED4, $2.548$ & BCH11, $3.173$ & BCH11, $3.887$ & BCH11, $2.926$ & BCH11, $3.606$ \\
        $63$ & BCH5, $2.589$ & CSED4, $2.589$ & BCH11, $3.224$ & BCH11, $3.95 $ & BCH11, $2.973$ & BCH11, $3.664$ \\
        $64$ & BCH5, $2.63$ & CSED4, $2.63 $ &  BCH11, $3.276$ & BCH11, $4.013$ & BCH11, $3.02 $ & BCH11, $3.723$ \\
        $65$ & BCH5, $2.671$ & BCH9, $2.559$  & BCH11, $3.327$ & BCH11, $4.076$ & BCH11, $3.067$ & BCH11, $3.781$ \\
        $66$ & BCH5, $2.712$ & BCH9, $2.598$ & BCH11, $3.378$ & BCH11, $4.138$ & BCH11, $3.114$ & BCH11, $3.839$ \\
        $67$ & BCH5, $2.753$ & BCH9, $2.638$ & BCH11, $3.429$ & BCH11, $4.201$ & BCH11, $3.162$ & BCH11, $3.897$ \\
        $68$ & BCH5, $2.794$ & BCH9, $2.677$ & BCH11, $3.48 $ & BCH11, $4.264$ & BCH11, $3.209$ & BCH11, $3.955$ \\
        $69$ & BCH5, $2.835$ & BCH9, $2.717$ & BCH11, $3.531$ & BCH11, $4.326$ & BCH11, $3.256$ & BCH11, $4.013$ \\
        $70$ & CSED4, $2.265$ & BCH9, $2.756$ & BCH11, $3.583$ & BCH11, $4.389$ & BCH11, $3.303$ & BCH11, $4.072$ \\
        $71$ & CSED4, $2.297$ & BCH9, $2.795$ & BCH11, $3.634$ & BCH11, $4.452$ & BCH11, $3.35$ &  BCH11, $4.13$ \\
        $72$ & CSED4, $2.329$ & BCH9, $2.835$ & BCH11, $3.685$ & BCH11, $4.514$ & BCH11, $3.397$ & BCH11, $4.188$ \\
        $73$ & CSED4, $2.362$ & BCH9, $2.874$ & BCH11, $3.736$ & BCH11, $4.577$ & BCH11, $3.445$ & BCH11, $4.246$ \\
        $74$ & CSED4, $2.394$ & BCH9, $2.913$ & BCH11, $3.787$ & BCH11, $4.64 $ & BCH11, $3.492$ & BCH11, $4.304$ \\
        $75$ & CSED4, $2.427$ & BCH9, $2.953$ & BCH11, $3.839$ & BCH11, $4.703$ & BCH11, $3.539$ & BCH11, $4.362$ \\
        $76$ & CSED4, $2.459$ & BCH9, $2.992$ & BCH11, $3.89 $ & BCH11, $4.765$ & BCH11, $3.586$ & BCH11, $4.421$ \\
        $77$ & CSED4, $2.491$ & BCH9, $3.031$ & BCH11, $3.941$ & BCH11, $4.828$ & BCH11, $3.633$ & BCH11, $4.479$ \\
        $78$ & CSED4, $2.524$ & BCH9, $3.071$ & BCH11, $3.992$ & BCH11, $4.891$ & BCH11, $3.681$ & BCH11, $4.537$ \\
        $79$ & CSED4, $2.556$ & BCH9, $3.11 $ & BCH11, $4.043$ & BCH11, $4.953$ & BCH11, $3.728$ & BCH11, $4.595$ \\
        $80$ & CSED4, $2.588$ & BCH9, $3.15 $ & BCH11, $4.094$ & BCH11, $5.016$ & BCH11, $3.775$ & BCH11, $4.653$ \\
        $81$ & CSED4, $2.621$ & BCH9, $3.189$ & BCH11, $4.146$ & BCH11, $5.079$ & BCH11, $3.822$ & BCH11, $4.711$ \\
        $82$ & BCH11, $2.26 $ & BCH9, $3.228$ & BCH11, $4.197$ & BCH11, $5.141$ & BCH11, $3.869$ & BCH11, $4.77$ \\
        $83$ & BCH11, $2.287$ & BCH9, $3.268$ & BCH11, $4.248$ & BCH11, $5.204$ & BCH11, $3.917$ & BCH11, $4.828$ \\
        $84$ & BCH11, $2.315$ & BCH9, $3.307$ & BCH11, $4.299$ & BCH11, $5.267$ & BCH11, $3.964$ & BCH11, $4.886$ \\
        $85$ & BCH11, $2.343$ & BCH9, $3.346$ & BCH11, $4.35 $ & BCH11, $5.33 $ & BCH11, $4.011$ & BCH11, $4.944$ \\
        $86$ & BCH11, $2.37 $ & BCH9, $3.386$ & BCH11, $4.402$ & BCH11, $5.392$ & BCH11, $4.058$ & BCH11, $5.002$ \\
        $87$ & BCH11, $2.398$ & BCH9, $3.425$ & BCH11, $4.453$ & BCH11, $5.455$ & BCH11, $4.105$ & BCH11, $5.06$  \\
        $88$ & BCH11, $2.425$ & BCH9, $3.465$ & BCH11, $4.504$ & BCH11, $5.518$ & BCH11, $4.152$ & BCH11, $5.119$  \\
        $89$ & BCH11, $2.453$ & BCH9, $3.504$ & BCH11, $4.555$ & BCH11, $5.58 $ & BCH11, $4.2. $ & BCH11, $5.177$  \\
        $90$ & BCH11, $2.48 $ & BCH9, $3.543$ & BCH11, $4.606$ & BCH11, $5.643$ & BCH11, $4.247$ & BCH11, $5.235$  \\
        $91$ & BCH11, $2.508$ & BCH9, $3.583$ & BCH11, $4.657$ & BCH11, $5.706$ & BCH11, $4.294$ & BCH11, $5.293$  \\
        $92$ & BCH11, $2.535$ & BCH9, $3.622$ & BCH11, $4.709$ & BCH11, $5.768$ & BCH11, $4.341$ & BCH11, $5.351$  \\
        $93$ & BCH9, $2.563$ & BCH9, $3.661$ & BCH9, $4.738$ & BCH9, $5.302$ & BCH9, $4.057$ & BCH9, $2.929$  \\
        $94$ & BCH9, $2.591$ & BCH9, $3.701 $ & BCH9, $4.789$ & BCH9, $5.359$ & BCH9, $4.101$ & BCH9, $2.961$  \\
        $95$ & BCH9, $2.618$ & BCH9, $3.74 $ & BCH9, $4.84$ & BCH9, $5.416$ & BCH9, $4.144$ & BCH9, $2.992$  \\
        $96$ & BCH9, $2.646$ & BCH9, $3.78$ & BCH9, $4.891$ & BCH9, $5.473$ & BCH9, $4.188$ & BCH9, $3.024$  \\
        $97$ & BCH9, $2.673$ & BCH9, $3.819$ & BCH9, $4.942$ & BCH9, $5.53$ & BCH9, $4.232$ & BCH9, $3.055$ \\
        $98$ & BCH9, $2.701$ & BCH9, $3.858$ & BCH9, $4.993$ & BCH9, $5.587$ & BCH9, $4.275$ & BCH9, $3.087$ \\
        $99$ & BCH9, $2.728$ & BCH9, $3.898$ & BCH9, $5.043$ & BCH9, $5.644$ & BCH9, $4.319$ & BCH9, $3.118$ \\
        $100$ & BCH7, $2.755$ & BCH7, $3.919$ & BCH7, $3.412$ & BCH7, $4.199$ & CSED4, $2.477$ & BCH7, $3.15$ \\
    \end{tabular}
    \label{tab:k51to100}
\end{table*}

\section{Clifford frames of distilled magic states}
\label{app:CliffpartTraceproof}

In circuits like \cref{fig:Haddist}, where non-Clifford gates are implemented using Pauli measurements, we prove that the Clifford frame of the distilled magic states are powers of $X_{\pi/4}$.
After the temporally encoded non-Clifford gates, the Clifford frame consists of a sequence of Clifford rotations which are tensor products of $X$ and $\Id$. We first observe the effect of an $(X \otimes X)_{\pi/4}$ gate on an input $\ket{T_X} \otimes \ket{\psi}$ state, where $\ket{\psi} = \alpha \ket 0 + \beta \ket 1$ is some arbitrary state. If we can determine the effect of the $(X \otimes X)_{\pi/4}$ rotation on the subsystem of the distilled magic state, we can determine the final Clifford frame of the distilled magic state.
\begin{align}
    (X \otimes &X)_{\pi/4} \ket{T_X} \otimes \ket{\psi} =   \nonumber \\
    & \begin{bmatrix} 
 \frac{1}{\sqrt{2}} &0 &0 &\frac{-i}{\sqrt{2}}  \\
 0 &\frac{1}{\sqrt{2}} &\frac{-i}{\sqrt{2}} &0  \\
 0 &\frac{-i}{\sqrt{2}} &\frac{1}{\sqrt{2}} &0  \\
 \frac{-i}{\sqrt{2}} &0 &0 &\frac{1}{\sqrt{2}}
\end{bmatrix} \cdot \frac{1}{\sqrt{2}}
\begin{bmatrix} 
 (1+ e^{\frac{i \pi}{4}}) \alpha \\
 (1+ e^{\frac{i \pi}{4}}) \beta  \\
 (1- e^{\frac{i \pi}{4}}) \alpha \\
 (1- e^{\frac{i \pi}{4}}) \beta
\end{bmatrix}.
\end{align}
When we trace out the subsystem that started as $\ket {\psi}$, we obtain the following state on the subsystem of the magic state (up to a global phase)
\begin{align}
    tr_{\ket {\psi}}((X \otimes &X)_{\pi/4} \ket{T_X} \otimes \ket{\psi}) =   \nonumber \\
& \begin{bmatrix} 
 1+ e^{\frac{i \pi}{4}} - i (1- e^{\frac{i \pi}{4}}) \\
 1- e^{\frac{i \pi}{4}} - i (1+ e^{\frac{i \pi}{4}})
\end{bmatrix}.
\end{align}
This is equivalent to $X_{\pi/4} \ket{T_X}$, which is 
\begin{equation}
    X_{\pi/4} \ket{T_X} = 
    \begin{bmatrix} 
 1+ e^{\frac{i \pi}{4}} - i (1- e^{\frac{i \pi}{4}}) \\
 1- e^{\frac{i \pi}{4}} - i (1+ e^{\frac{i \pi}{4}})
\end{bmatrix}. 
\end{equation}

This shows that if there is one Clifford operator in the Clifford frame with $X$ support on the magic state qubit, it essentially results in an $X_{\pi/4}$ Clifford frame update on the magic state. However with more than one Clifford correction in the Clifford frame, the total rotation accumulated on the magic state qubit is the product of $X_{\pi/4}$ rotations for all the Cliffords in the Clifford frame that contain an $X$ operator on the support of the magic state qubit.

\section{Choice of codewords for the Golay code for TELS of a $15$-to-$1$ distillation protocol}
\label{app:golaycodechoice}

In the $15$-to-$1$ distillation protocol of \cref{sec:15to1}, we implemented a TELS protocol for the non-Clifford measurements using the $[23,12,7]$ Golay code. We remove one codeword, since we only need to perform $11$ Pauli measurements in the PP set, and permute some columns (reordering the resulting Pauli measurements) to get the following codeword generator matrix
\begin{equation}
    G = \begin{bmatrix} 

 1 1 1 1 1 1 1 0 0 0 0 0 0 0 0 0 0 0 0 0 0 0 0 \\
 0 0 0 1 1 0 1 1 1 1 1 0 0 0 0 0 0 0 0 0 0 0 0 \\
 0 1 1 0 1 0 0 0 0 1 1 1 1 0 0 0 0 0 0 0 0 0 0 \\
 0 0 0 1 0 0 0 0 1 1 0 1 1 1 1 0 0 0 0 0 0 0 0 \\
 0 0 1 0 1 1 0 0 0 0 0 1 0 1 1 1 0 0 0 0 0 0 0 \\
 0 0 0 1 0 1 1 0 0 1 0 0 0 1 0 1 1 0 0 0 0 0 0 \\
 0 0 0 0 1 1 1 0 0 0 1 1 0 0 0 0 1 1 0 0 0 0 0 \\
 0 0 0 0 0 0 1 0 0 1 1 0 1 1 0 0 0 1 1 0 0 0 0 \\
 0 0 0 0 0 1 0 0 0 0 1 1 1 0 1 0 0 0 1 1 0 0 0 \\
 0 0 0 0 0 0 1 0 0 0 0 0 1 1 1 1 0 0 0 1 1 0 0 \\
 0 0 0 0 0 0 0 0 0 0 0 0 0 1 1 0 1 1 0 1 1 1 1
\end{bmatrix}.
\end{equation}
Now, after the seventh measurement, the cell holding the magic state associated with the first row of $G$ can be reset and used to inject a new magic state that will only be required for the $14$th Pauli measurement (first column (left-to-right) with a $1$ in the last row).

Note that both the columns and rows of G may be permuted; permuting columns reorders the new sequence of Pauli measurements; permuting rows merely swaps codeword generators. It may be possible to find an algorithm that iteratively applies a permutation rule convention to find a codeword matrix that allows magic states to occupy the fewest number of cells in a distillation tile.

\vspace{-.2cm} 
\section{Procedure for determining code distances of distillation tiles}
\label{app:algoSTcosts}

Here we describe the procedure used to determine the spacelike distances $d_x$ and $d_z$ and timelike distances $d_m$ for lattice surgery in the distillation tiles of \cref{sec:MStilelayouts}. First, note that when performing distillation in the Clifford frame using TELS-assisted lattice surgery, we will need to execute two PP sets. The first PP set performs the non-Clifford gates using $\ket{T_X}$ resource states (using a classical $[n_1,k_1,d_1]$ code), and the second PP set performs a set of Pauli measurements associated with conjugating Clifford corrections through single-qubit logical measurements (using a classical $[n_2,k_2,d_2]$ code).

We first set $\delta^{(M)}$ as the error budget per magic state that is output from a distillation protocol. Logical errors may accumulate on the distilled magic states by different mechanisms. Even with noiseless lattice surgery, errors on the input magic states may may cause the final magic state to be logically wrong. This depends entirely on the choice of quantum code, here $\llbracket n_{\text{dist}}, k_{\text{dist}}, d_{\text{dist}} \rrbracket$. The logical error rate (per output magic state) of a distillation protocol with noiseless gates is
\begin{equation}
    p_L^{(M)} = \frac{l_{\text{dist}}}{k_{\text{dist}}} \Big(\frac{p(1+\eta)}{3 \eta} \Big)^{d_{\text{dist}}},
\end{equation}
according to the analysis in Sec. $1$ of Ref.\cite{Litinski19magic}, using the biased circuit-level noise model of \cref{subsec:PauliBasedReview}. Here, $l_{\text{dist}}$ is the number of weight-$d_{\text{dist}}$ fault sets that can cause a logical error.

Given the error budget $\delta^{(M)}$ and the logical error rate with noiseless gates $p_L^{(M)}$, we may now upper bound the logical error rate due to noisy lattice surgery measurements,
\begin{equation}
\delta = (\delta^{(M)} - p_L^{(M)}) k_{\text{dist}}.
\end{equation}
Note that we multiply by $k_{\text{dist}}$ since $\delta^{(M)}$ is the error budget per output magic state, but $\delta$ is the error budget of lattice surgery for the entire distillation protocol. Logical errors due to lattice surgery may occur due to spacelike or timelike errors. 

For each PP set, the logical error rate due to lattice surgery is
\begin{equation}
    p_{\text{PP1}}(d_m, n) = p_{L,X}(d_m , n) + p_{L,Z}(d_m , n) + p_{L}(p_m(d_m , n)), 
\end{equation}
where  $p_{L,X}$ and $p_{L,Z}$ are the spacelike contributions and $p_{L}$ is the timelike failure rate of TELS described in \cref{sec:NewTELS}. Here $d_m$ and $n$ refer to the lattice surgery measurement distance and the number of Pauli measurements respectively. Although we express $p_{\text{PP1}}$ as a function of two variables $d_m$ and $n$, the spacelike distance $d_x$ and $d_z$ are also input variables. The important point is that $d_m$ and $n$ are the only two variables that are different for each PP set. 

In Ref.~\cite{Chamberland22}, Eqs. $3-6$ denote the logical error rates of an $X \otimes X$ lattice surgery measurement. We obtain equations for $p_{L,X}$, $p_{L,Z}$, and $p_m$ by modifying the above equations as shown below
\begin{align}
    p_m(d_m , n) =& 0.01634 n A (21.93p)^{\tfrac{d_m+1}{2}}, \\ 
    p_{L,Z}(d_m , n) =& 0.03148 T N d_x (28.91p)^{\tfrac{d_z+1}{2}}, \\ 
    p_{L,X}(d_m , n) =& 0.0148 T \frac{F}{d_x} (0.762p)^{\tfrac{d_x+1}{2}}, 
\end{align}
where $A$ is the area of the routing space (in units of $d_x$ and $d_z$), $T$ is the average time taken to execute the parallelizable Pauli set (from \cref{sec:NewTELS}), $N$ is the maximum number of logical qubits that are concurrently used during any lattice surgery measurement in a TELS protocol, and $F$ is a pessimistic estimate of the maximum area used during a lattice surgery measurement (routing space $+$ logical qubits associated in the measurement). In \cref{app:constants}, we show equations for $F,A,N$ and $\text{Space}$ in terms of $d_x$ and $d_z$ for each of the different distillation layouts suggested in \cref{sec:MStilelayouts}. The time to complete the entire distillation protocol is given in \cref{subsubsec:clifftimecost}.

If we use measurement distance $d_m'$ for the first PP set and measurement distance $d_m''$ for the second PP set, then the objective is to find a set of parameters $\{d_x, d_z, d_m', d_m''\}$ that minimizes the space-time cost of a distillation factory, while ensuring the following equation is satisfied,
\begin{equation}
\label{eq:conditiondelta}
p_{\text{PP1}}(d_m', n_1) + p_{\text{PP2}}(d_m'', n_2) < \delta.
\end{equation}

Note that \cref{eq:conditiondelta}  ensures that the probability of a single logical failure event is less than $\delta$. Two independent logical failure events occur with probability $\sim \delta^2$, so we omit these higher order events from \cref{eq:conditiondelta}.

\section{Constants used to determine spacetime costs of distillation tiles}
\label{app:constants}

In this section we list the constants that are used to determine the minimum space-like and time-like distances for the distillation layouts described in \cref{app:algoSTcosts}. Distillation protocols that do not use TELS perform auto-corrected non-Clifford gate gadgets for the entire protocol, and hence contain one value each for the number of logical qubits, $N$, routing space area, $A$, and full area of lattice surgery, $F$. For protocols that perform TELS using the Clifford frame distillation circuit of \cref{sec:Cliff}, there are two PP sets. We display two sets of values for $N,A$, and $F$ to account for the changes between the execution of the first and second PP sets.

\vspace{13cm} 
\hbox{} 

\begin{table}[]
    \begin{tabular}{l}
    $15$-to-$1$ distillation - No temporal encoding \\
    \hline
    Space $= (2 d_z + 2 d_x + 2) (5(d_x+1))$\\
    $N = 7$ \\
    $A =(d_x+1)(d_z + 4(d_x+2))$\\
    $F = \text{Space} - 2 d_z (d_x + 1)$\\
    \hline
    $15$-to-$1$ distillation - No temp. encoding, parallelized \\
    \hline
    Space $= (2 d_z + 4 d_x + 4) (5(d_x+1))$\\
    $N = 9$ \\
    $A =(d_x+1)(d_z + 4(d_x+2))$\\
    $F = \text{Space} - 3 d_z (d_x + 1)$\\
    \hline
    $15$-to-$1$ distillation - SED2 \\
    \hline
    Space $= (2 d_z + d_x + 3) (5(d_x+1))$\\
    $N = 7$ \\
    $A = 5 (d_x+1)(d_x+3)$\\
    $F = \text{Space} - 3 d_z (d_x + 1)$\\
    $N2 = 5$ \\
    $A2 = A + 4 d_z (d_x+1)$\\
    $F2 = \text{Space} - d_z (d_x + 1)$\\
    \hline
    $15$-to-$1$ distillation - SED2, parallellized \\
    \hline
    Space $= (2 d_z + 3 d_x + 3) (7(d_x+1))$\\
    $N = 8$ \\
    $A =  (d_x+1)(2 d_z + 14 d_x+3)$\\
    $F = \text{Space} - 4 d_z (d_x + 1)$\\
    $N2 = 5$ \\
    $A2 = A + 4 d_z (d_x+1)$\\
    $F2 = \text{Space} - d_z (d_x + 1)$\\
    \hline
    $15$-to-$1$ distillation - BCH3 \\
    \hline
    Space $= (2 d_z + d_x + 3) (6(d_x+1))$\\
    $N = 10$ \\
    $A = 6 (d_x+1)(d_x+3)$\\
    $F = \text{Space} - 2 d_z (d_x + 1)$\\
    $N2 = 5$ \\
    $A2 = A + 4 d_z (d_x+1)$\\
    $F2 = \text{Space} - d_z (d_x + 1)$\\
    \hline
    $15$-to-$1$ distillation - BCH3, parallellized \\
    \hline
    Space $= (2 d_z + 3 d_x + 3) (8(d_x+1))$\\
    $N = 11$ \\
    $A =  (d_x+1)(2 d_z + 16 d_x+3)$\\
    $F = \text{Space} - 3 d_z (d_x + 1)$\\
    $N2 = 5$ \\
    $A2 = A + 4 d_z (d_x+1)$\\
    $F2 = \text{Space} - d_z (d_x + 1)$\\
    \hline
    \end{tabular}
    \caption{Constants associated with layouts of distillation tiles described in \cref{sec:MStilelayouts}. These constants are used to determine minimum spacelike and timelike distances using the procedure in \cref{app:algoSTcosts}.}
    \label{tab:constants}
\end{table}

\begin{table}[]
    \begin{tabular}{l}
    $15$-to-$1$ distillation - Golay \\
    \hline
    Space $= (2 d_z + d_x + 3) (8(d_x+1))$\\
    $N = 15$ \\
    $A = 8 (d_x+1)(d_x+3)$\\
    $F = \text{Space} -  d_z (d_x + 1)$\\
    $N2 = 5$ \\
    $A2 = A + 4 d_z (d_x+1)$\\
    $F2 = \text{Space} - d_z (d_x + 1)$\\
    \hline
$15$-to-$1$ distillation - Golay, parallellized \\
    \hline
    Space $= (2 d_z + 3 d_x + 3) (9(d_x+1))$\\
    $N = 15$ \\
    $A =  (d_x+1)(2 d_z + 18 d_x+3)$\\
    $F = \text{Space} - d_z (d_x + 1)$\\
    $N2 = 5$ \\
    $A2 = A + 4 d_z (d_x+1)$\\
    $F2 = \text{Space} - d_z (d_x + 1)$\\
    \hline
$116$-to-$12$ distillation - No temporal encoding \\
    \hline
    Space $= \max(2(d_z + d_x), 4(d_x+1)) (22d_x + d_z +23)$\\
    $N = 31$ \\
    $A =(d_x+1)(25d_x+4)$\\
    $F = \text{Space} - 13 d_z (d_x + 1)$\\
    \hline
    $116$-to-$12$ distillation - No temp. encoding, parallelized \\
    \hline
    Space $= \max(2(d_z + 3d_x), 4(d_x+1)) (23d_x + 2d_z +25)$\\
    $N = 33$ \\
    $A =(d_x+1) (30 d_x + \max(2(d_z + 3d_x), 4(d_x+1)))$\\
    $F = \text{Space} - 14 d_z (d_x + 1)$\\
    \hline
    $116$-to-$12$ distillation - Zett5, parallelized \\
    \hline
    Space $= (2 d_z + 3d_x + 3) (31(d_x+1))$\\
    $N = 46$ \\
    $A = (d_x+1)(2 d_z + 51 d_x+3)$\\
    $F = \text{Space} - 14 d_z (d_x + 1)$\\
    $N2 = 29$ \\
    $A2 = A + 19 d_z (d_x+1)$\\
    $F2 = \text{Space} - 12 d_z (d_x + 1)$\\
    \hline
    $116$-to-$12$ distillation - BCH9, parallellized \\
    \hline
    Space $= (2 d_z + 3 d_x + 3) (38(d_x+1))$\\
    $N = 58$ \\
    $A =  (d_x+1)(2 d_z + 65 d_x+3)$\\
    $F = \text{Space} - 15 d_z (d_x + 1)$\\
    $N2 = 29$ \\
    $A2 = A + 32 d_z (d_x+1)$\\
    $F2 = \text{Space} - 12 d_z (d_x + 1)$\\
    \hline
     $114$-to-$14$ distillation - No temporal encoding \\
    \hline
    Space $= \max(2(d_z + d_x), 4(d_x+1)) (23d_x + d_z + 24)$\\
    $N = 31$ \\
    $A =(d_x+1)(26d_x+4)$\\
    $F = \text{Space} - 15 d_z (d_x + 1)$\\
    \hline
    \end{tabular}
\end{table}

\begin{table}[]
    \begin{tabular}{l}
    $114$-to-$14$ distillation - No temp. encoding, parallelized \\
    \hline
    Space $= \max(2(d_z + 3d_x), 4(d_x+1)) (24d_x + 2d_z + 26)$\\
    $N = 33$ \\
    $A =(d_x+1) (30 d_x + \max(2(d_z + 3d_x), 4(d_x+1)))$\\
    $F = \text{Space} - 16 d_z (d_x + 1)$\\
    \hline
    $114$-to-$14$ distillation - Zett5, parallelized \\
    \hline
    Space $= (2 d_z + 3d_x + 3) (32(d_x+1))$\\
    $N = 46$ \\
    $A = (d_x+1)(2 d_z + 51 d_x+3)$\\
    $F = \text{Space} - 16 d_z (d_x + 1)$\\
    $N2 = 29$ \\
    $A2 = A + 19 d_z (d_x+1)$\\
    $F2 = \text{Space} - 14 d_z (d_x + 1)$\\
    \hline
    $114$-to-$14$ distillation - BCH7, parallellized \\
    \hline
    Space $= (2 d_z + 3 d_x + 3) (35(d_x+1))$\\
    $N = 52$ \\
    $A =  (d_x+1)(2 d_z + 57 d_x+3)$\\
    $F = \text{Space} - 16 d_z (d_x + 1)$\\
    $N2 = 29$ \\
    $A2 = A + 25 d_z (d_x+1)$\\
    $F2 = \text{Space} - 14 d_z (d_x + 1)$\\
    \hline
    $125$-to-$3$ distillation - No temporal encoding \\
    \hline
    Space $= \max(2(d_z + d_x), 4(d_x+1)) (18d_x + d_z +19)$\\
    $N = 31$ \\
    $A =(d_x+1)(20d_x+4)$\\
    $F = \text{Space} - 4 d_z (d_x + 1)$\\
    \hline
    $125$-to-$3$ distillation - No temp. encoding, parallelized \\
    \hline
    Space $= \max(2(d_z + 3d_x), 4(d_x+1)) (20d_x + 2d_z  +22)$\\
    $N = 33$ \\
    $A =(d_x+1) (29 d_x + \max(2(d_z + 3d_x), 4(d_x+1)))$\\
    $F = \text{Space} - 5 d_z (d_x + 1)$\\
    \hline
    $125$-to-$3$ distillation - BCH7, parallelized \\
    \hline
    Space $= (2 d_z + 3d_x + 3) (30(d_x+1))$\\
    $N = 52$ \\
    $A = (d_x+1)(2 d_z + 58 d_x+3)$\\
    $F = \text{Space} - 6 d_z (d_x + 1)$\\
    $N2 = 29$ \\
    $A2 = A + 26 d_z (d_x+1)$\\
    $F2 = \text{Space} - 3 d_z (d_x + 1)$\\
    \hline
    $125$-to-$3$ distillation - BCH9, parallellized \\
    \hline
    Space $= (2 d_z + 3 d_x + 3) (33(d_x+1))$\\
    $N = 56$ \\
    $A =  (d_x+1)(2 d_z + 64 d_x+3)$\\
    $F = \text{Space} - 5 d_z (d_x + 1)$\\
    $N2 = 29$ \\
    $A2 = A + 32 d_z (d_x+1)$\\
    $F2 = \text{Space} - 3 d_z (d_x + 1)$\\
    \hline
    \end{tabular}
\end{table}

\pagebreak

\bibliography{q}

\end{document}